\documentclass[aps,prd,twoside,twocolumn,superscriptaddress,floatfix,nofootinbib]{revtex4}
\usepackage{amsmath}
\usepackage{bm}
\usepackage{xcolor}

\newcommand\beq{\begin{equation}}
\newcommand\eeq{\end{equation}}
\newcommand\beqn{\begin{eqnarray}}
\newcommand\eeqn{\end{eqnarray}}

\newcommand{\ba}{\begin{eqnarray}}
\newcommand{\ea}{\end{eqnarray}}
\newcommand{\be}{\begin{equation}}
\newcommand{\ee}{\end{equation}}

\newcommand\lsim{\mathrel{\rlap{\lower4pt\hbox{\hskip1pt$\sim$}}
        \raise1pt\hbox{$<$}}}
\newcommand\gsim{\mathrel{\rlap{\lower4pt\hbox{\hskip1pt$\sim$}}
        \raise1pt\hbox{$>$}}}

\newcommand{\jcap}{{J.~Cosm.~Astrop.~Phys.}}
\newcommand{\araa}{{Annu.~Rev.~Astron.~Astrophys.}}
\newcommand{\aap}{{Astron.~Astrophys.}}
\newcommand{\apjl}{{Astrophys.~J.~Lett.}}
\newcommand{\apjs}{{Astrophys.~J.~Supp.}}

\newcommand{\mnras}{{Mon.~Not.~R.~Astron.~Soc.}}
\usepackage{graphicx}
\usepackage[large]{subfigure}
\usepackage{amssymb, amsmath}
\usepackage[amssymb]{SIunits}
\usepackage{epstopdf}
\usepackage{aas_macros}
\usepackage{natbib}

\begin{document}

\title{Foreground Biases on Primordial Non-Gaussianity Measurements from the CMB Temperature Bispectrum: Implications for {\em Planck} and Beyond}
\author{J.~Colin Hill\footnote{jch@ias.edu}}
\affiliation{School of Natural Sciences, Institute for Advanced Study, Princeton, NJ, USA 08540}
\affiliation{Center for Computational Astrophysics, Flatiron Institute, New York, NY, USA 10003}
\begin{abstract}
The cosmic microwave background (CMB) temperature bispectrum is currently the most precise tool for constraining non-Gaussianity (NG) in the primordial curvature perturbations.  The {\em Planck} temperature data tightly constrain the amplitude of local-type NG: $f_{\rm NL}^{\rm loc} = 2.5 \pm 5.7$.  In this paper, we compute previously-neglected foreground biases in temperature-based $f_{\rm NL}^{\rm loc}$ measurements.  We consider signals from the integrated Sachs-Wolfe (ISW) effect, gravitational lensing, the thermal (tSZ) and kinematic Sunyaev-Zel'dovich (kSZ) effects, and the cosmic infrared background (CIB).  In standard analyses, a significant foreground bias arising from the ISW-lensing bispectrum is subtracted from the $f_{\rm NL}^{\rm loc}$ measurement.  However, a number of other terms sourced by the ISW, lensing, tSZ, kSZ, and CIB fields are also present in the temperature bispectrum.  We compute the dominant biases on $f_{\rm NL}^{\rm loc}$ arising from these signals, focusing on ``squeezed'' bispectrum shapes.  Most of the biases are non-blackbody in nature, and are thus reduced by multifrequency component separation methods; however, recent analyses have found that extragalactic foregrounds are present at non-negligible levels in the {\em Planck} component-separated maps.  Moreover, the {\em Planck} FFP8 simulations do not include the correlations amongst components that are responsible for these biases.  We compute the biases for individual {\em Planck} frequencies, finding that some are comparable to the statistical error bar on $f_{\rm NL}^{\rm loc}$, even for the main CMB channels (100, 143, and 217 GHz).  For future experiments, they can greatly exceed the statistical error bar (considering temperature data only).  Alternatively, the foreground contributions can be marginalized over, but without strong priors this leads to a non-negligible increase in the error bar on $f_{\rm NL}^{\rm loc}$.  A full assessment for {\em Planck} and other experiments will require calculations in tandem with component separation, ideally using simulations.  We also compute these biases for equilateral and orthogonal NG, finding large effects for the latter.  Similar calculations must be performed for trispectrum NG.  We conclude that the search for primordial NG using \emph{Planck} data may not yet be over.
\end{abstract}
\maketitle

\section{Introduction}
\label{sec:intro}

Primordial non-Gaussianity (NG) is a key probe of the physics thought to have generated all structure in our Universe during its earliest moments.  The simplest models of inflation (i.e., single-field, slow-roll) predict negligibly small departures from Gaussianity in the primordial curvature perturbations~\cite{Acquavivaetal2003,Maldacena2003}, but a rich spectrum of non-Gaussian signatures can be produced in more complex inflationary scenarios or non-inflationary early-Universe models~(see, e.g.,~Refs.~\cite{Bartolo2004,Chen2010} for reviews).  A key quantity of interest is the amplitude of the bispectrum of curvature perturbations in the so-called ``squeezed'' limit (in which one wavenumber is much smaller than the other two, i.e., $k_1 \ll k_2, k_3$), conventionally denoted as $f_{\rm NL}^{\rm loc}$~\cite{Komatsu-Spergel2001}.  Single-field, slow-roll inflation predicts that $f_{\rm NL}^{\rm loc}$ vanishes exactly~(e.g.,~\cite{Pajer2013}), modulo small, higher-order corrections due to the nonlinearity of gravity~\cite{Cabass2017}.   A detection of non-zero $f_{\rm NL}^{\rm loc}$ would rule out essentially all single-field models of inflation~\cite{Maldacena2003,Creminelli-Zaldarriaga2004,Bravo2018}.  We focus on local-type NG here; discussion and results for additional ``shapes'' of NG (equilateral or orthogonal), which also contain a wealth of information about the physics of the early Universe~\cite{Bartolo2004,Chen2010}, can be found in the appendices.

The most powerful observable for constraining primordial NG in current data sets is the bispectrum of temperature fluctuations in the cosmic microwave background (CMB)~\cite{Hinshawetal2013,Bennettetal2013,Planck2015params,Planck2015NG}.  The well-understood, linear physics responsible for the CMB anisotropy permits an essentially direct mapping of the primordial curvature perturbations, thus allowing NG templates to be directly fit to CMB maps.  The most stringent current constraint on local NG is derived from {\em Planck} data in this manner, yielding $f_{\rm NL}^{\rm loc} = 2.5 \pm 5.7$ (temperature data only) or $f_{\rm NL}^{\rm loc} = 0.8 \pm 5.0$ (temperature and polarization data)~\cite{Planck2015NG}.

Nevertheless, a number of complex problems must be surmounted in order to extract robust NG constraints from CMB data.  In this paper, we focus on one such problem: extragalactic foreground contamination in CMB temperature maps.  Prior to NG analysis, maps of the microwave sky at multiple frequencies must be combined to extract a map of the CMB anisotropy from the multitude of other sky signals, a process known as ``component separation''~(see Refs.~\cite{Planck2013compsep,Planck2015compsep,Planck2018compsep} for an overview of the {\em Planck} CMB component separation methods).  The goal of such techniques is to minimize the contributions from non-CMB contaminants (including both foregrounds and noise) while preserving the CMB signal.  However, such methods are generally imperfect, and some level of foreground residuals will propagate into the final map.  These residuals must be well-understood in order for robust NG constraints to be obtained.

We focus in this paper on extragalactic foreground contributions to local-type NG estimators in CMB temperature maps, as detailed further below.  Most of these contributions are non-blackbody in frequency dependence, and can therefore be reduced (or, in some cases, completely removed~\cite{Remazeillesetal2011}) via multifrequency component separation.  However, the extent to which such reduction for extragalactic foregrounds has occurred in the {\em Planck} component-separated CMB temperature maps is presently unclear, and evidence has recently accumulated that some small-scale foregrounds may be present at non-negligible levels~\cite{MH2018,Chenetal2018}.  A crucial cross-check could come from a polarization-only NG analysis, where the only extragalactic foreground is that due to point source emission, which is well-understood and simple to remove.  However, as can be immediately seen in the results quoted above, the {\em Planck} $f_{\rm NL}^{\rm loc}$ constraints are strongly dominated by information in the CMB temperature field.  Upcoming experiments, including the {\em Simons Observatory}\footnote{\url{http://www.simonsobservatory.org}} ({\em SO})~\cite{SO2018} and {\em CMB-S4}~\cite{S4}, may be sufficiently sensitive to allow independent measurements from temperature and polarization with comparable error bars.  At present, CMB temperature dominates the information content in NG constraints.

Moreover, some contaminants to NG estimators cannot be removed via multifrequency component separation, as they possess the same blackbody frequency dependence as the CMB itself.  Chief amongst these is the CMB temperature bispectrum sourced by the correlation between the integrated Sachs-Wolfe (ISW) effect~\cite{SW1967,RS1968} and the gravitational lensing potential by which CMB photons are deflected (see Ref.~\cite{Lewis-Challinor2006} for a review of CMB lensing).  The ISW effect is the change in the temperature of CMB photons due to the decay (or enhancement) of late-time gravitational potentials (e.g., as a result of dark energy).  Both the ISW effect and gravitational lensing do not alter the blackbody spectrum of the CMB.  In addition, both fields trace the late-time gravitational potential of the Universe.  Finally, since CMB lensing couples previously-independent spherical harmonic coefficients of the CMB, the ISW-lensing correlation produces a non-zero bispectrum in the CMB temperature field~\cite{Smith-Zaldarriaga2011,Hansonetal2009,Lewisetal2011,Junk-Komatsu2012}.  This ISW-lensing bispectrum has a non-zero projection onto the bispectrum shape of local-type NG, thereby producing a non-negligible, irreducible bias to estimates of $f_{\rm NL}^{\rm loc}$ from CMB temperature maps.  For {\em Planck}, this bias is substantial and must be subtracted to obtain unbiased constraints: $\Delta f_{{\rm NL},Planck}^{{\rm loc, ISW-}\phi} = 7.6$ (see \S\ref{sec:ISW-phi}).  The ISW-lensing bispectrum also produces biases on equilateral and orthogonal NG estimates (particularly the latter), as discussed in the appendices.

In this paper, we point out the existence of additional foreground biases in CMB temperature-derived $f_{\rm NL}^{\rm loc}$ constraints.  To our knowledge, the ISW-related biases presented here have not been computed elsewhere, with the exception of the ISW-lensing bias.  Other biases involving CMB lensing have received some attention, albeit limited~\cite{Serra-Cooray2008,Curtoetal2015}.  In general, $f_{\rm NL}^{\rm loc}$ biases are generated in the CMB temperature field by bispectra involving the ISW effect, CMB lensing, the cosmic infrared background (CIB), and the thermal (tSZ) and kinematic Sunyaev-Zel'dovich (kSZ) effects.\footnote{Bispectra involving point sources also generate biases, but are smaller in magnitude, particularly for \emph{Planck}, and have been considered in previous analyses.}  The CIB refers to the cumulative infrared emission of dusty, star-forming galaxies over cosmic time, which has a broad redshift kernel peaking around $z \approx 2$ (similar to the CMB lensing kernel).  The tSZ effect is the inverse-Compton scattering of CMB photons off hot, free electrons, producing a shift in the photon spectrum to higher energies, and thus leaving a non-blackbody spectral distortion in the CMB~\cite{Zeldovich-Sunyaev1969,Sunyaev-Zeldovich1970}.  The kSZ effect is the Doppler-boosting of CMB photons scattering off electrons that have a non-zero line-of-sight (LOS) velocity in the CMB rest frame~\cite{Sunyaev-Zeldovich1972,Sunyaev-Zeldovich1980,Ostriker-Vishniac1986}.  Along with the ISW and CMB lensing signals, these fields all trace the large-scale structure of the universe in some way.  The resulting correlations generate bispectra in CMB temperature maps.  While these bispectra are interesting in their own right for astrophysical and cosmological reasons~\cite{Wilsonetal2012,Hill-Sherwin2013,Bhattacharyaetal2012,Crawfordetal2014,Hill2014PDF,Planck2015ymap,Coultonetal2017}, here we focus on their role in biasing measurements of local-type primordial NG, akin to the bias due to the ISW-lensing bispectrum described above.  Our goal is to consider all extragalactic foreground bispectra that have strong contributions in the squeezed limit, i.e., to provide a complete assessment of relevant biases for local-type NG.  Because this set of terms is not exhaustive for other shapes of primordial NG (although it contains some that are nevertheless non-negligible), we relegate calculations for equilateral and orthogonal NG to appendices, deferring a complete bias assessment for these shapes to future work.

Note that rather than treating these foreground contributions as biases, it is possible to include the relevant bispectrum templates in the NG analysis and marginalize over their amplitudes, thereby mitigating the biases at the cost of increasing the error bars on the primordial NG parameters~(e.g.,~\cite{Calabrese2010}).  We consider this approach for each foreground contribution throughout the paper (assuming no priors are placed on the amplitudes of any bispectra).  In general, for $f_{\rm NL}^{\rm loc}$ we find that marginalizing over lensing-related foregrounds (i.e., the lensing-ISW, lensing-tSZ, or lensing-CIB bispectra) leaves the error bar on $f_{\rm NL}^{\rm loc}$ nearly unchanged.  However, marginalizing over ISW-related bispectra (i.e., ISW-tSZ-tSZ, ISW-CIB-CIB, ISW-tSZ-CIB, or ISW-kSZ-kSZ) generally increases the error bar on $f_{\rm NL}^{\rm loc}$ by a non-negligible amount, e.g., $\approx 50$\% for {\em Planck}.  This increase is simply due to the high correlation coefficient of the ISW-related bispectra with the local-type bispectrum (all of these bispectra peak in the squeezed limit).  Therefore, precise theoretical calculations of these signals are important, so that strong priors can be placed on the relevant amplitudes, thereby mitigating the error bar increase on $f_{\rm NL}^{\rm loc}$.

\begin{table*}[ht]
\begin{center}
  Biases on $f_{\rm NL}^{\rm loc}$ for {\em Planck} $\langle TTT \rangle$: $\ell_{\rm max} = 1590$, $\sigma(f_{\rm NL}^{\rm loc}) = 5.0/\sqrt{f_{\rm sky}}$ \vspace{2pt} \\
  \begin{tabular}{| c | c | c | c | c | c | c |}
    \hline 
     Frequency & ISW-$\phi$ (\S\ref{sec:ISW-phi}) & tSZ-$\phi$ (\S\ref{sec:tSZ-phi}) & CIB-$\phi$ (\S\ref{sec:CIB-phi}) & ISW-tSZ-tSZ (\S\ref{sec:ISW-tSZ-tSZ}) & ISW-CIB-CIB (\S\ref{sec:ISW-CIB-CIB}) & ISW-kSZ-kSZ (\S\ref{sec:ISW-kSZ-kSZ}) \\
     \hline \hline
    100 GHz & 7.6 & $-1.2$ & 0.9 & $-4.5$ & $\approx 0$ & $-0.1$ \\ \hline
    143 GHz & 7.6 & $-0.8$ & 1.4 & $-2.1$ & $\approx 0$ & $-0.1$ \\ \hline
    217 GHz & 7.6 & $\approx 0$ & 4.3 & $\approx 0$ & $-0.4$ & $-0.1$ \\ \hline
  \end{tabular}
  \\ \vspace{12pt} Biases on $f_{\rm NL}^{\rm loc}$ for {\em SO/CMB-S4} $\langle TTT \rangle$: $\ell_{\rm max} = 3000$, $\sigma(f_{\rm NL}^{\rm loc}) = 2.6/\sqrt{f_{\rm sky}}$ \vspace{2pt} \\
    \begin{tabular}{| c | c | c | c | c | c | c |}
    \hline 
     Frequency & ISW-$\phi$ (\S\ref{sec:ISW-phi}) & tSZ-$\phi$ (\S\ref{sec:tSZ-phi}) & CIB-$\phi$ (\S\ref{sec:CIB-phi}) & ISW-tSZ-tSZ (\S\ref{sec:ISW-tSZ-tSZ}) & ISW-CIB-CIB (\S\ref{sec:ISW-CIB-CIB}) & ISW-kSZ-kSZ (\S\ref{sec:ISW-kSZ-kSZ}) \\
     \hline \hline
    100 GHz & 14.9 & $-3.4$ & 2.2 & $-47.1$ & $\approx 0$ & $-1.7$ \\ \hline
    143 GHz & 14.9 & $-2.4$ & 3.3 & $-22.5$ & $ -1.2 $ & $-1.7$ \\ \hline
    217 GHz & 14.9 & $\approx 0$ & 10.2 & $\approx 0$ & $-10.0$ & $-1.7$ \\ \hline
  \end{tabular}
  \caption{Summary of extragalactic foreground biases on $f_{\rm NL}^{\rm loc}$ measurements from the CMB temperature bispectrum.  The top panel gives the biases for an experiment that is cosmic variance-limited to a maximum multipole $\ell_{\rm max} = 1590$ (i.e., {\em Planck}), while the bottom panel contains the results for an experiment with $\ell_{\rm max} = 3000$ (i.e., {\em Simons Observatory} or {\em CMB-S4}).  These correspond to $\sigma(f_{\rm NL}^{\rm loc}) = 5.0$ and 2.6, respectively, assuming a full-sky measurement (in practice, $f_{\rm sky} \approx 0.75$ for {\em Planck} and $f_{\rm sky} \approx 0.4$ for {\em SO} and {\em CMB-S4}).  We perform these calculations using the measured {\em Planck} bandpasses centered at 100, 143, and 217 GHz, which are the primary CMB channels of {\em Planck} and ground-based experiments, although higher frequencies are considered later in the paper as well.  For brevity, we do not include the ISW-tSZ-CIB bispectrum results in this table, but these biases are computed in \S\ref{sec:ISW-tSZ-CIB}.  Note that no multifrequency cleaning is assumed for the non-blackbody foregrounds.}
  \label{tab:biases}
\end{center}
\end{table*}

The basic conclusions of this paper are summarized in Table~\ref{tab:biases}, which gives the bias on $f_{\rm NL}^{\rm loc}$ sourced by several types of foreground bispectra at the primary CMB channels of {\em Planck} (note that ground-based experiments observe at similar frequencies).  The table includes results for {\em Planck} noise levels, corresponding to an experiment that is cosmic variance-limited to a maximum multipole $\ell_{\rm max} = 1590$, as well as for future experiments that will be cosmic variance-limited to $\ell_{\rm max} = 3000$ (albeit not covering the full sky).  The primary takeaway from these results is that even for the main CMB channels of {\em Planck}, the biases due to these foregrounds are comparable to the $1\sigma$ error bar on $f_{\rm NL}^{\rm loc}$ from the temperature bispectrum (we reach a similar conclusion for $f_{\rm NL}^{\rm orth}$ in Appendix~\ref{app:orth}).  Of course, component separation will reduce the non-blackbody contributions to some extent (and all contributions will be summed in a complex manner in the synthesis of the final CMB map), but a full modeling of this procedure is beyond the scope of this paper and best implemented via simulations.

The other important conclusion from Table~\ref{tab:biases} is that for future experiments aiming to constrain primordial NG from the CMB bispectrum (e.g., {\em SO} and {\em CMB-S4}), non-blackbody foregrounds must be cleaned very precisely to avoid biases from the tSZ and CIB signals.  Furthermore, a non-negligible bias due to the blackbody ISW-kSZ-kSZ bispectrum must be subtracted, which has not been computed in the literature to date.  Given these complexities, future NG constraints may be better off relying on CMB polarization, which is much cleaner than temperature on small scales.

The remainder of this paper is organized as follows.  In \S\ref{sec:motivation}, we motivate this work in the context of current $f_{\rm NL}^{\rm loc}$ measurements and previous calculations.  In \S\ref{sec:fNL}, we provide relevant theoretical background related to primordial NG and formalism for bispectra.  The following two sections contain the main results of the paper: \S\ref{sec:lensing} describes biases related to the correlation of CMB lensing with other secondary fields, while \S\ref{sec:ISW} describes biases generated by the correlation of these fields with the ISW effect.  We discuss the implications of these results in \S\ref{sec:discussion} and conclude in \S\ref{sec:conclusions}.  Appendix~\ref{app:equ} and Appendix~\ref{app:orth} provide analogous calculations and results for equilateral-type and orthogonal-type primordial NG, respectively.

We assume a standard flat $\Lambda$CDM cosmology throughout this paper, adopting the following parameters: matter density $\Omega_m = 0.277$, Hubble constant $H_0 = 70.2$~km/s/Mpc, baryon density $\Omega_b = 0.0459$, spectral tilt $n_s = 0.962$, and amplitude of density fluctuations $\sigma_8 = 0.817$.  Our conclusions are weakly sensitive to the assumed values of these parameters, but we comment below on a few notable exceptions where they are important (related to the tSZ signal).

\section{Motivation and Context}
\label{sec:motivation}

Non-Gaussian signals from extragalactic foregrounds in CMB temperature maps have received significant attention in recent years due to the important biases they can generate in reconstructed CMB lensing maps~(e.g.,~\cite{vanEngelenetal2014,Osborneetal2014,FH2018,MH2018,Prince2018,Schaan2018}).  These biases arise from both non-blackbody (e.g., tSZ and CIB) and blackbody (e.g., kSZ) foregrounds; accordingly, some can be reduced by multifrequency component separation, while some cannot be (although other mitigation methods can be employed).

Similarly, these non-Gaussian signals will generate biases in CMB temperature bispectrum estimates of primordial NG.  To date, the ISW-lensing bispectrum has received significant attention in this context~\cite{Smith-Zaldarriaga2011,Hansonetal2009,Lewisetal2011,Junk-Komatsu2012}, but other biases have been less studied, if at all.  For example, the {\em Planck} 2015 NG analysis considers only the ISW-lensing bispectrum, point source bispectrum, and clustered CIB auto-bispectrum as contaminants to primordial NG measurements~\cite{Planck2015NG}.  An early estimate of the tSZ-lensing bispectrum bias on $f_{\rm NL}^{\rm loc}$ was presented in Ref.~\cite{Serra-Cooray2008}, but it is unclear what observational frequencies were considered in the analysis, and theoretical modeling of the tSZ signal has significantly evolved in the intervening decade.  More recently, the CIB-lensing bispectrum bias on $f_{\rm NL}^{\rm loc}$ was considered in Ref.~\cite{Curtoetal2015}; their results for specific {\em Planck} frequency channels are in qualitative agreement with those presented in \S\ref{sec:CIB-phi} below.  However, to our knowledge, no calculations of ISW-related biases have been presented in the literature to date, beyond the ISW-lensing bias.  Amongst these contributions is that due to the ISW-kSZ-kSZ bispectrum, which is blackbody in frequency dependence and therefore must be subtracted from observational estimates, like the ISW-lensing bias.  We provide a first estimate of this bias on $f_{\rm NL}^{\rm loc}$ in \S\ref{sec:ISW-kSZ-kSZ}.

However, non-blackbody biases are also of significant concern.  Recent analyses indicate that non-negligible extragalactic foreground contamination has leaked into the {\em Planck} component-separated CMB temperature maps, which form the foundation of the {\em Planck} NG analysis.  For example, Fig.~1 of Ref.~\cite{MH2018} demonstrates that the tSZ signal of optically-selected galaxy clusters is present in the {\em Planck} SMICA CMB map with an amplitude nearly identical to that seen in the {\em Planck} 143 GHz map.  Similarly, Ref.~\cite{Chenetal2018} uses cross-correlations with optical galaxy survey data to detect the presence of tSZ residuals in the {\em Planck} NILC CMB map at $54\sigma$ significance.  Their overall estimate is that roughly half of the 143 GHz tSZ signal is present in the NILC map.  While these studies are based on observational estimates, it would be more robust to estimate the leakage via end-to-end simulations containing all relevant sky signals, an analysis which has not yet been performed.

In this context, we note that although the {\em Planck} 2015 NG analysis used the Full Focal Plane 8 (FFP8) simulations to perform end-to-end tests of their analysis pipelines, the FFP8 simulations do not contain the signals responsible for the biases considered in this paper (with the exception of the ISW-lensing bias)~\cite{Planck2015FFP8}.  In particular, the ISW field is generated only as part of the primary CMB map (i.e., as a Gaussian random field), and is not correlated with the tSZ, kSZ, or CIB fields.  Similarly, the CMB lensing field is not correlated with the tSZ, kSZ, or CIB fields, nor is the CIB field correlated with the tSZ or kSZ fields.  The tSZ and kSZ fields are partially correlated with one another, but not with any of the other secondary fields.  The ISW-lensing correlation is generated by the algorithm with which the primary CMB is gravitationally lensed (because the $T-\phi$ correlation is included in {\tt CAMB}~\cite{CAMBref}\footnote{http://www.camb.info} power spectra).  Thus, the NG pipeline verification tests run on the FFP8 simulations do not test for any of the biases considered in this paper (except for the ISW-lensing bias).  This situation could be remedied by using sky simulations in which the extragalactic fields are properly correlated with one another~(e.g.,~\cite{Sehgaletal2010}).

Motivated by the existence of non-negligible extragalactic foreground contamination in the {\em Planck} component-separated CMB temperature maps and the absence of nearly all relevant foreground bispectra in the FFP8 simulations, we consider the role that these effects might have on estimates of primordial NG.  We note that polarization-only analyses would be almost entirely immune to these foreground biases, but the {\em Planck} polarization data are not sufficiently sensitive for such a test ($\sigma(f_{\rm NL}^{\rm loc}) \approx 30-35$ from polarization data alone, in comparison to $\sigma(f_{\rm NL}^{\rm loc}) \approx 5-6$ from temperature data alone~\cite{Planck2015NG}).  We focus on local-type NG in the main text of the paper, but include similar calculations for equilateral- and orthogonal-type NG in the appendices.  We defer foreground bias calculations for trispectrum NG estimators to future work.  The most efficient method for future calculations is likely to simultaneously estimate all such foreground biases via simulations, rather than compute each contribution analytically.\footnote{In addition, NG estimators can be applied to the data that allow one to directly reconstruct the bispectra, rather than fit an overall amplitude to a primordial template bispectrum, as assumed in this work following Ref.~\cite{KSW2005} (see, e.g., the skew-$C_{\ell}$~\cite{Munshi-Heavens2010} or modal estimators~\cite{Fergusson2010} used in Ref.~\cite{Planck2015NG}).  Such reconstructions can then be analyzed to determine the origin of the NG signal(s).}

\section{Primordial Non-Gaussianity and the CMB Bispectrum}
\label{sec:fNL}

In the local model of primordial NG, the primordial potential $\Phi$ (where $\Phi \equiv \frac{3}{5} \zeta$, and $\zeta$ is the adiabatic curvature perturbation) is given by~\cite{Salopek-Bond1990,Ganguietal1994,Komatsu-Spergel2001}
\be
\label{eq.fNLlocdef}
\Phi(\vec{x}) = \Phi_G(\vec{x}) + f_{\mathrm{NL}}^{\rm loc} \left( \Phi_G^2(\vec{x}) - \langle \Phi_G^2 \rangle \right) + \cdots \,,
\ee
where $\Phi_G$ is a Gaussian field and $f_{\rm NL}^{\rm loc}$ is a constant characterizing the lowest-order departure from Gaussianity.  Multi-field inflationary models, such as the curvaton model, or non-inflationary early-Universe scenarios, such as the ekpyrotic/cyclic model, can generate local-type NG~\cite{PNG1,PNG2,PNG3,Ekp1,Ekp2,Ekp3}.  More generally, a detection of $f_{\rm NL}^{\rm loc} \neq 0$ would falsify single-field, slow-roll inflation~\cite{Creminelli-Zaldarriaga2004}.  Current data are consistent with $f_{\rm NL}^{\rm loc} = 0$~\cite{Hinshawetal2013,Giannantonioetal2014,Leistedt2014,Planck2015NG}.  As discussed above, the tightest error bar (by a factor of a few) comes from the {\em Planck} CMB temperature data, thus motivating careful scrutiny of this particular observable.  It will be possible to further shrink the CMB-derived error bar on $f_{\rm NL}^{\rm loc}$ by a factor of $\approx 2-3$, but eventually the cosmic variance (CV) limit will be reached.  Further improvements are then expected to come from large-scale structure data~(e.g.,~\cite{SPHEREx}).

The non-linear coupling in Eq.~\ref{eq.fNLlocdef} generates a non-zero bispectrum in the CMB anisotropy.  We consider only the CMB temperature field in the following.  The angular bispectrum $B_{\ell_1 \ell_2 \ell_3}$ is defined via
\be
\label{eq.bispecdef}
\langle a_{\ell_1 m_1} a_{\ell_2 m_2} a_{\ell_3 m_3} \rangle = B_{\ell_1 \ell_2 \ell_3} \left(\begin{array}{clcr}
\ell_1 & \ell_2 & \ell_3\\
m_1 & m_2 & m_3  \end{array}\right) \,,
\ee
where $a_{\ell m}$ are the spherical harmonic coefficients of the CMB temperature field and $\left(\begin{array}{clcr}
\ell_1 & \ell_2 & \ell_3\\
m_1 & m_2 & m_3  \end{array}\right)$ is the Wigner-$3j$ symbol.  Eq.~\ref{eq.bispecdef} assumes only rotational invariance; if we additionally assume parity invariance (i.e., $B_{\ell_1 \ell_2 \ell_3} = 0$ if $\ell_1 + \ell_2 + \ell_3$ is odd, and thus $B_{\ell_1 \ell_2 \ell_3}$ is invariant under all permutations), then we can define the reduced bispectrum $b_{\ell_1 \ell_2 \ell_3}$ via
\ba
\label{eq.redbispecdef}
B_{\ell_1 \ell_2 \ell_3} & = &  \sqrt{\frac{(2\ell_1+1)(2\ell_2+1)(2\ell_3+1)}{4\pi}} \nonumber \\
& & \times \left(\begin{array}{clcr}
\ell_1 & \ell_2 & \ell_3\\
0 & 0 & 0  \end{array}\right) b_{\ell_1 \ell_2 \ell_3} \,.
\ea

In the local model of primordial NG, the real-space coupling in Eq.~\ref{eq.fNLlocdef} produces a non-zero Fourier-space bispectrum in the primordial potential~\cite{Komatsu-Spergel2001}:
\ba
\label{eq.fNLloc3pt}
\langle \tilde{\Phi}(\vec{k}_1) \tilde{\Phi}(\vec{k}_2) \tilde{\Phi}(\vec{k}_3) \rangle = 2 (2\pi)^3 \delta^{(3)}(\vec{k}_1+\vec{k}_2+\vec{k}_3) f_{\rm NL}^{\rm loc} \times \nonumber \\
P_{\Phi}(k_2) P_{\Phi}(k_3) + 2 \,\, {\rm perm.} \,,
\ea
where $\tilde{\Phi}(\vec{k})$ is the Fourier transform of the primordial potential, $\delta^{(3)}$ is the 3D Dirac delta function, and $P_{\Phi}(k)$ is the power spectrum of the primordial potential.  This non-zero bispectrum then yields a non-zero angular bispectrum in the CMB anisotropy, $B_{\ell_1 \ell_2 \ell_3}^{\rm loc}$.  The CMB temperature bispectrum can be computed straightforwardly via integrals involving the radiation transfer function, spherical Bessel functions, and $P_{\Phi}(k)$ (see, e.g., Ref.~\cite{Komatsu-Spergel2001} for explicit formulae).  The parameter $f_{\rm NL}^{\rm loc}$ characterizes the amplitude of the bispectrum, as seen in Eq.~\ref{eq.fNLloc3pt}.  This bispectrum peaks in the squeezed limit, in which one of the wavenumbers is much smaller than the other two (e.g., $k_1 \ll k_2, k_3$)~\cite{Komatsu-Spergel2001,Maldacena2003}.

For weak NG, assuming full-sky CMB temperature data that is CV-limited up to a multipole $\ell_{\rm max}$ (with homogeneous noise properties), we can define the Fisher matrix element for two bispectra $B,B'$~(e.g.,~\cite{Hansonetal2009}):\footnote{Fisher matrices for all calculations presented in this paper are available upon request from the author.}
\be
\label{eq.FishBBp}
F(B,B') = \frac{1}{6} \sum_{\ell_1 \ell_2 \ell_3}^{\ell_{\rm max}} \frac{B_{\ell_1 \ell_2 \ell_3} B'_{\ell_1 \ell_2 \ell_3}}{C_{\ell_1}^{TT} C_{\ell_2}^{TT} C_{\ell_3}^{TT}} \,,
\ee
where $C_{\ell}^{TT}$ is the lensed primary CMB power spectrum (see Ref.~\cite{Junk-Komatsu2012} for useful comments on implementing the sum in Eq.~\ref{eq.FishBBp}).  The error bar on the amplitude of a bispectrum $B$ is then given by the square root of $1/F(B,B)$.  For example, the error bar on $f_{\rm NL}^{\rm loc}$ (in the Gaussian approximation) is
\be
\sigma(f_{\rm NL}^{\rm loc}) = \sqrt{1/F(B^{\rm loc},B^{\rm loc})} \,.
\label{eq.fNLerror}
\ee
Similarly, the marginalized error on the amplitude is given by the square root of the relevant element of the inverted Fisher matrix (assumed here to be a simple $2\times2$ matrix with elements for $B^{\rm loc}$, a contaminating bispectrum $B^{\rm cont}$, and their cross-term):
\be
\sigma(f_{\rm NL}^{\rm loc, marg.}) = \sqrt{F^{-1}(B^{\rm loc},B^{\rm loc})} \,.
\label{eq.fNLerrormarg}
\ee
Finally, the bias on the minimum-variance estimator for $f_{\rm NL}^{\rm loc}$ due to a contaminating bispectrum $B^{\rm cont}$ is given by a ratio of Fisher matrix elements~(e.g.,~\cite{Hansonetal2009}):
\be
\Delta f_{\rm NL}^{{\rm loc},{\rm cont}} = \frac{F(B^{\rm loc},B^{\rm cont})}{F(B^{\rm loc},B^{\rm loc})} \,.
\label{eq.fNLbias}
\ee

\section{CMB Lensing-Related Biases}
\label{sec:lensing}

We first compute biases to $f_{\rm NL}^{\rm loc}$ associated with the CMB lensing field.  These biases arise from bispectra with a form identical to that of the standard lensing-ISW bispectrum~\cite{Smith-Zaldarriaga2011,Hansonetal2009}:
\be
B_{\ell_1\ell_2\ell_3}^{X\phi} = f_{\ell_1\ell_2\ell_3} C_{\ell_2}^{X\phi} C_{\ell_3}^{TT} + 5 \,\, {\rm perm.},
\label{eq.Blens}
\ee
where $X \in \left\{{\rm ISW, tSZ, CIB} \right\}$, $C_{\ell}^{X\phi}$ is the cross-power spectrum of $X$ and the CMB lensing potential, and $f_{\ell_1\ell_2\ell_3}$ is a coupling kernel~(e.g.,~\cite{Hansonetal2009,Planck2015NG}):
\ba
f_{\ell_1\ell_2\ell_3} & & = \frac{1}{2} \left[ -\ell_1(\ell_1+1) + \ell_2(\ell_2+1) + \ell_3(\ell_3+1) \right] \times \nonumber \\
& & \sqrt{\frac{(2\ell_1+1)(2\ell_2+1)(2\ell_3+1)}{4\pi}} \left(\begin{array}{clcr}
\ell_1 & \ell_2 & \ell_3\\
0 & 0 & 0  \end{array}\right) \,.
\label{eq.coupling}
\ea
Note that $B_{\ell_1\ell_2\ell_3}^{X\phi}$ vanishes if $X=$ kSZ due to the equal probability of positive or negative LOS velocities (in general, all bispectra involving odd numbers of kSZ fields vanish for this reason).  In Eq.~\ref{eq.Blens}, we use the lensed CMB power spectrum for $C_{\ell}^{TT}$, as this yields a more accurate result for squeezed triangle configurations than using the unlensed spectrum~\cite{Lewisetal2011}.

The remaining quantity to be computed in Eq.~\ref{eq.Blens} is the cross-power spectrum of $X$ and $\phi$.  To lowest order, the CMB lensing potential is a weighted sum of the Newtonian potential along the LOS:
\be
\phi(\hat{n}) = -\frac{2}{c^2} \int_0^{\chi_*} d\chi \left( \frac{\chi_* - \chi}{\chi_* \chi} \right) \Psi (\chi \hat{n}, \chi) \,,
\label{eq.phidef}
\ee
where $\chi(z)$ is the comoving distance to redshift $z$, $\chi_*$ is the comoving distance to the surface of last scattering at $z_* = 1090$, and $\Psi$ is the 3D gravitational potential.  In the following subsections, we detail our computation of $C_{\ell}^{X\phi}$ for $X \in \left\{{\rm ISW, tSZ, CIB} \right\}$, and calculate the associated bias on measurements of $f_{\rm NL}^{\rm loc}$.

\subsection{Lensing-ISW Bias}
\label{sec:ISW-phi}

The late-time ISW effect~\cite{SW1967,RS1968} is generated by the decay of gravitational potentials due to dark energy.  In linear theory, this effect produces positive (negative) CMB temperature fluctuations as CMB photons traverse large-scale overdensities (underdensities) in the late-time matter field.  The fractional temperature shift is given by the LOS integral of the time derivative of the gravitational potential:
\be
\frac{\Delta T^{\rm ISW}(\hat{n})}{T_{\rm CMB}} = \frac{2}{c^2} \int_{\rm LOS} dt \frac{\partial \Psi(\hat{n})}{\partial t} \,,
\label{eq.ISWdef}
\ee
where $T_{\rm CMB}$ is the mean CMB temperature today.  On small scales, nonlinear growth produces a late-time ISW effect (the Rees-Sciama effect~\cite{RS1968}) with the opposite sign to that sourced by dark energy on large, linear scales.  However, this effect is much smaller in amplitude than the linear ISW effect due to dark energy.  It has been shown that the change to the lensing-ISW bispectrum due to the Rees-Sciama effect is essentially undetectable in the CMB, and that linear theory is sufficient for precisely computing the associated bias on $f_{\rm NL}^{\rm loc}$~\cite{Smithetal2009,Junk-Komatsu2012}.  Thus, we only consider the linear ISW effect throughout this paper.\footnote{We compare non-linear and linear-theory predictions for the lensing-ISW biases on equilateral and orthogonal NG in Appendices~\ref{app:equ} and~\ref{app:orth} (see Figs.~\ref{fig:ISWxphi_bias_equ} and~\ref{fig:ISWxphi_bias_orth}).}  We comment on instances where this may not suffice, particularly for the ISW-kSZ-kSZ bispectrum in \S\ref{sec:ISW-kSZ-kSZ}.

\begin{figure}
\includegraphics[width=0.5\textwidth]{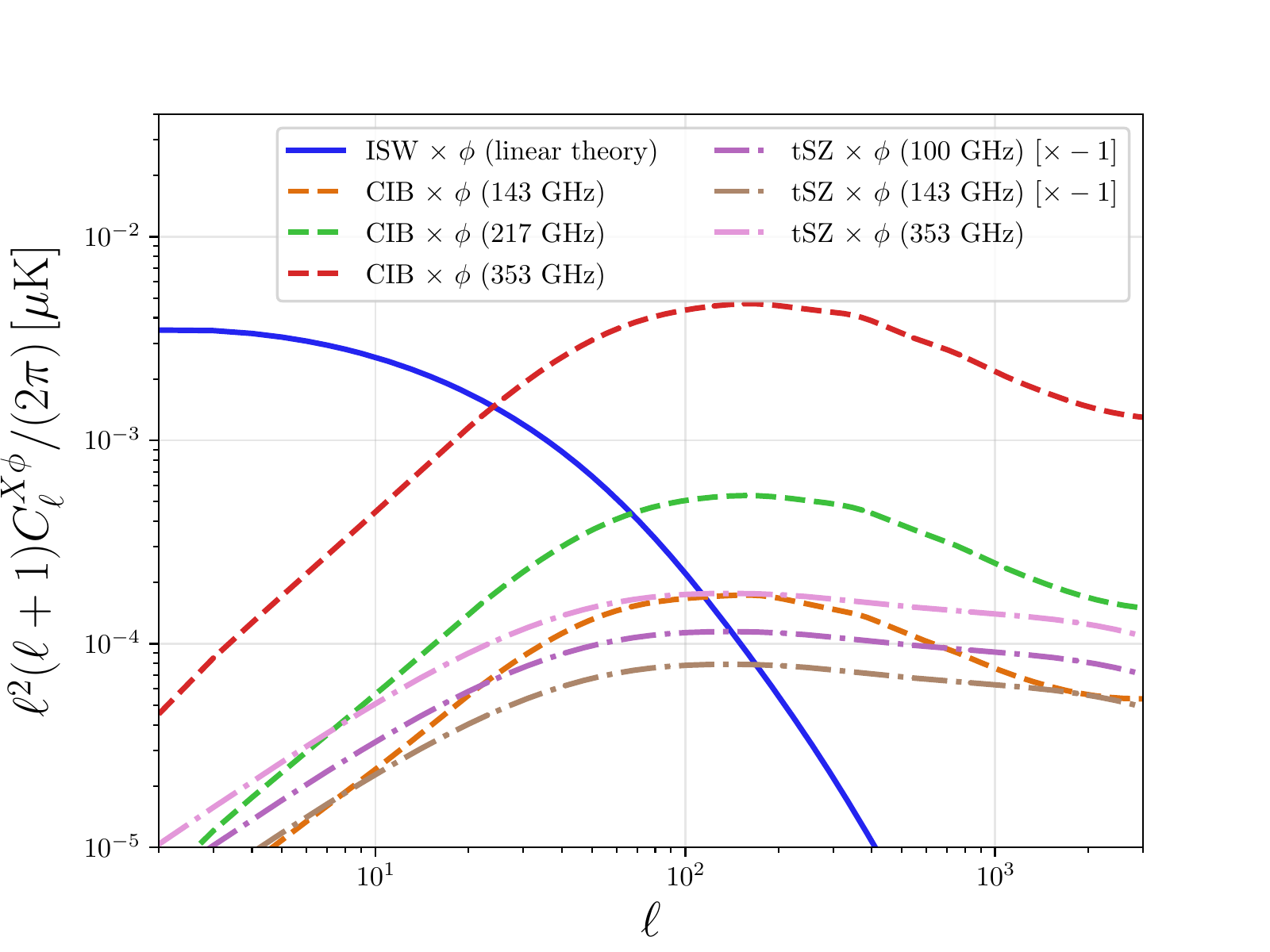}
\caption{\label{fig:phi_xcorr} Cross-power spectra of CMB lensing with other secondary anisotropy fields, $C_{\ell}^{X\phi}$, with $X \in \left\{{\rm ISW, tSZ, CIB} \right\}$.  The lensing-ISW cross-power spectrum (solid blue) is computed in linear theory, i.e., the Rees-Sciama effect is not included, but this has no measurable effect on the associated bias on $f_{\rm NL}^{\rm loc}$.  The lensing-tSZ (dash-dotted) and lensing-CIB cross-power spectra (dashed) are computed via the halo model as described in \S\ref{sec:tSZ-phi} and \S\ref{sec:CIB-phi}, respectively.  The halo model calculations are based on fits to measurements from {\em Planck} data in~\cite{HS2014} (lensing-tSZ) and~\cite{Planck2013CIBxlens} (lensing-CIB).  For clarity, these cross-power spectra are only shown for a subset of the {\em Planck} frequencies.  Note that the tSZ signal is negative (positive) at frequencies below (above) 217 GHz, and vanishes at 217 GHz.}
\end{figure}

In the Limber approximation~\cite{Limber1953}, the CMB lensing-ISW cross-power spectrum is~\cite{Spergel-Goldberg1999,Goldberg-Spergel1999,Verde-Spergel2002}
\be
C_{\ell}^{{\rm ISW} \times \phi} = \frac{2}{c^4} \int dz \left( \frac{\chi_* - \chi}{\chi_*\chi^3} \right) \frac{\partial P_{\Psi}}{\partial z} \bigg\rvert_{k=(\ell+1/2)/\chi} \,,
\label{eq.ISWphi}
\ee
where $P_{\Psi}(k,z)$ is the power spectrum of the 3D gravitational potential at wavenumber $k$ and redshift $z$.  Using the Poisson equation, we can express this result in terms of the linear power spectrum of matter density fluctuations (defined at an arbitrary redshift), $P_{\rm lin}(k)$:
\ba
C_{\ell}^{{\rm ISW} \times \phi} & & = \frac{9 \Omega_m^2 H_0^4}{2 c^4 (\ell+1/2)^2 \chi_*} \int dz \, \chi(z) (\chi_*-\chi(z)) \times \nonumber \\
 & & (1+z) \frac{d}{dz} \left(\frac{D(z)}{a(z)} \right) D(z) P_{\rm lin}\left( \frac{\ell+1/2}{\chi(z)} \right) \,,
\label{eq.ISWphi_simp}
\ea
where $a(z) = 1/(1+z)$ is the scale factor and $D(z)$ is the linear growth factor, normalized in a manner consistent with the redshift at which $P_{\rm lin}$ is defined.  We compute the linear matter power spectrum using {\tt CAMB}.  Note that during matter domination, $D(z) \propto a(z)$, and thus it can be immediately seen from Eq.~\ref{eq.ISWphi_simp} that there are no contributions to $C_{\ell}^{{\rm ISW} \times \phi}$ from this epoch, as expected.

Fig.~\ref{fig:phi_xcorr} shows the lensing-ISW cross-power spectrum (solid blue curve).  The signal falls off steeply with $\ell$ due to the $(\ell+1/2)^{-2}$ dependence in Eq.~\ref{eq.ISWphi_simp}, which arises from the relation between the matter density and gravitational potential in the Poisson equation.  Thus, in Eq.~\ref{eq.Blens}, the multipole associated with $C_{\ell}^{{\rm ISW} \times \phi}$ is generally the long-wavelength mode in squeezed triangle configurations.  Note that the signal is frequency-independent in blackbody CMB temperature units.

\begin{figure}
\includegraphics[width=0.5\textwidth]{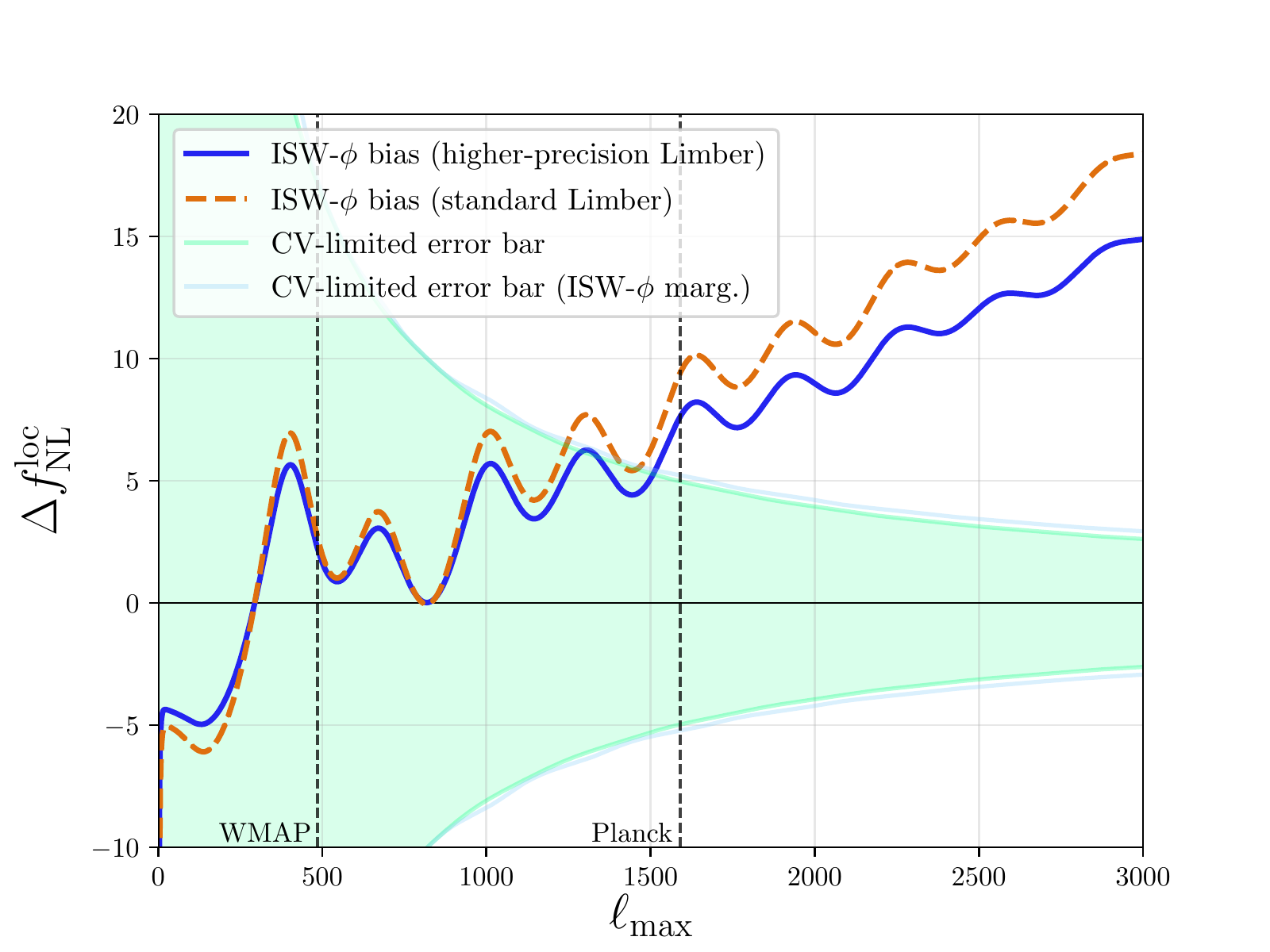} 
\caption{\label{fig:ISWxphi_bias} Bias on $f_{\rm NL}^{\rm loc}$ from the lensing-ISW bispectrum for a CMB temperature measurement that is CV-limited to a maximum multipole $\ell_{\rm max}$.  This bias is blackbody in frequency dependence and cannot be removed via component separation.  The solid blue curve shows the bias when the lensing-ISW cross-power spectrum is computed with a higher-precision implementation of the Limber approximation (i.e., the replacement $\ell \rightarrow \ell + 1/2$), while the dashed orange curve shows the bias for a ``standard'' implementation of the Limber approximation.  The higher-precision calculation gives a bias $\Delta f_{\rm NL}^{\rm loc} = 7.6$ for {\em Planck}, consistent with the {\em Planck} NG analysis~\cite{Planck2015NG}.  The light green shaded region shows the $1\sigma$ uncertainty on $f_{\rm NL}^{\rm loc}$ as a function of $\ell_{\rm max}$ using only information in the CMB temperature bispectrum for a full-sky, CV-limited experiment.  The light blue shaded region shows the $1\sigma$ uncertainty on $f_{\rm NL}^{\rm loc}$ after marginalizing over the lensing-ISW bispectrum amplitude; this marginalization has almost no impact on the sensitivity to $f_{\rm NL}^{\rm loc}$.  The dashed vertical lines indicate the effective $\ell_{\rm max}$ for WMAP9~\cite{Hinshawetal2013} and {\em Planck} 2015~\cite{Planck2015NG}.}
\end{figure}

Fig.~\ref{fig:ISWxphi_bias} shows the bias on $f_{\rm NL}^{\rm loc}$ due to the lensing-ISW bispectrum, computed via Eq.~\ref{eq.fNLbias} for a CMB temperature bispectrum measurement that is CV-limited up to a maximum multipole $\ell_{\rm max}$.  We show the result calculated using Eq.~\ref{eq.ISWphi_simp}, as well as a calculation using a less-precise implementation of the Limber approximation in which all instances of $\ell+1/2$ on the right-hand side of Eq.~\ref{eq.ISWphi_simp} are replaced with $\ell$~\cite{LoVerde-Afshordi2008}.  This choice has a non-negligible impact on the resulting bias on $f_{\rm NL}^{\rm loc}$.  For {\em Planck}, the inferred bias for the higher-precision calculation is $\Delta f_{\rm NL}^{\rm loc} = 7.6$, which agrees with the value used in the {\em Planck} 2015 NG analysis~\cite{Planck2015NG}.  For WMAP9, the bias is $\Delta f_{\rm NL}^{\rm loc} = 2.3$, slightly smaller than the value (2.6) quoted in Ref.~\cite{Hinshawetal2013} (their value is consistent with the ``standard'' Limber calculation).

Fig.~\ref{fig:ISWxphi_bias} also shows the $1\sigma$ uncertainty on $f_{\rm NL}^{\rm loc}$ for a full-sky, CV-limited experiment up to $\ell_{\rm max}$, computed with Eq.~\ref{eq.fNLerror}.  The WMAP9 error bar is $\sigma(f_{\rm NL}^{\rm loc}) = 19.9$, while the {\em Planck} 2015 error bar is $\sigma(f_{\rm NL}^{\rm loc}) = 5.7$ (temperature data only).  Taking into account the sky masks used by WMAP9~\cite{Hinshawetal2013} ($f_{\rm sky} = 0.75$) and {\em Planck}~\cite{Planck2015NG} ($f_{\rm sky} = 0.76$), we infer that $\ell_{\rm max} \approx 485$ for WMAP9 and $\ell_{\rm max} \approx 1590$ for {\em Planck}, which are plotted as dashed vertical lines in Fig.~\ref{fig:ISWxphi_bias}.  Note that these values of $\ell_{\rm max}$ are those appropriate for the component-separated CMB temperature maps used in the WMAP9 and {\em Planck} NG analyses, i.e., they are effective $\ell_{\rm max}$ values that result from a combination of the noise properties of multiple frequency channels.  Focusing on {\em Planck} in particular, the 100, 143, and 217 GHz channels all have individual values of $\ell_{\rm max}$ that are close to that shown in Fig.~\ref{fig:ISWxphi_bias} (and in subsequent plots), and thus we omit the individual channel values for clarity.  However, the 353 and 545 GHz channel sensitivities are lower, and accordingly so are their $\ell_{\rm max}$ values (although we will simply quote {\em Planck}-related biases at the effective {\em Planck} $\ell_{\rm max}$ value given above for brevity).  Note that for a full computation of the combined effect of the frequency-dependent biases computed in subsequent sections, one would have to appropriately take into account the noise properties of each individual {\em Planck} channel, rather than the effective {\em Planck} $\ell_{\rm max}$ value for the component-separated CMB map.

Finally, Fig.~\ref{fig:ISWxphi_bias} also shows the $1\sigma$ uncertainty on $f_{\rm NL}^{\rm loc}$ after marginalizing over the amplitude of the lensing-ISW bispectrum, computed with Eq.~\ref{eq.fNLerrormarg}.  Interestingly, this marginalization hardly increases the error bar on $f_{\rm NL}^{\rm loc}$, even though the bias sourced by the lensing-ISW bispectrum is large.  Mathematically, this is due to the fact that the error bar increase due to marginalization depends only on the correlation coefficient between the two bispectra (i.e., their shapes), which is independent of their amplitudes, while the bias depends explicitly on the amplitudes.  Thus, the error bar increase can be small, even if the bias is large (and vice versa, as we will see later in the paper).  The result shown in Fig.~\ref{fig:ISWxphi_bias} indicates that the amplitude of the lensing-ISW bispectrum could simply be simultaneously fit in the $f_{\rm NL}^{\rm loc}$ analysis and marginalized over, even with no prior on its amplitude.  Nevertheless, since the lensing-ISW bispectrum amplitude (and shape) can be predicted {\em a priori} (up to a small dependence on cosmological parameters), there is no need to pay even this small penalty in the $f_{\rm NL}^{\rm loc}$ error bar, and thus it is sensible to instead subtract the effect as a known bias.  For other foreground bispectra considered later in the paper, this may not be the case, as will be further discussed.

\subsection{Lensing-tSZ Bias}
\label{sec:tSZ-phi}

The tSZ effect is generated by the inverse-Compton scattering of CMB photons off hot, free electrons, which are predominantly located in galaxy groups and clusters.  Neglecting relativistic corrections~(e.g.,~\cite{Nozawaetal2006}), the tSZ signal is characterized by the Compton-$y$ parameter, which is the LOS integral of the electron pressure~\cite{Zeldovich-Sunyaev1969,Sunyaev-Zeldovich1970}:
\be
y(\hat{n}) = \frac{\sigma_T}{m_e c^2} \int d\chi \, a(\chi) \, P_e(\chi \hat{n}, \chi) \,,
\label{eq.tSZdef}
\ee
where $\sigma_T$ is the Thomson scattering cross-section, $m_e c^2$ is the electron rest-mass energy, and $P_e$ is the electron pressure.  The CMB temperature fluctuation due to the tSZ signal at a given frequency $\nu$ is then given by
\be
\frac{\Delta T^{\rm tSZ}(\hat{n})}{T_{\rm CMB}} = g(\nu) y(\hat{n}) \,,
\label{eq.tSZtemp}
\ee
where $g(\nu)$ is the tSZ spectral function:
\be
g(\nu) = x \coth\left( \frac{x}{2} \right) - 4 \,,
\label{eq.tSZspectralfunc}
\ee
with $x \equiv h\nu/(k_B T_{\rm CMB})$.

The cross-correlation between the tSZ and CMB lensing fields was first measured using {\em Planck} data in Ref.~\cite{HS2014}.  We adopt a model consistent with this measurement in the following.  The lensing-tSZ cross-power spectrum, $C_{\ell}^{y\phi}$, can be computed straightforwardly in the halo model~(e.g.,~\cite{Seljak2000,Cooray-Sheth2002}), analogous to the computation of other tSZ statistics~(e.g.,~\cite{Komatsu-Seljak2002,Hill-Pajer2013}).  We summarize the approach here, and refer the reader to Refs.~\cite{Hill-Pajer2013,HS2014} for full details of these calculations.  The total cross-power spectrum is the sum of the one-halo and two-halo terms:
\be
C_{\ell}^{y\phi} = C_{\ell}^{y\phi,1h} + C_{\ell}^{y\phi,2h} \,,
\label{eq.Clyphi}
\ee
where
\be
C_{\ell}^{y\phi,1h} = \int dz \frac{d^2V}{dz d\Omega} \int dM \frac{dn(M,z)}{dM} \tilde{y}_{\ell}(M,z) \tilde{\phi}_{\ell}(M,z) \,,
\label{eq.Clyphi1h}
\ee
and
\ba
C_{\ell}^{y\phi,2h} = & & \int dz \frac{d^2V}{dz d\Omega} P_{\mathrm{lin}}\left(\frac{\ell+1/2}{\chi(z)},z\right) \times \nonumber \\
& & \int dM_1 \frac{dn(M_1,z)}{dM_1} b(M_1,z) \tilde{y}_{\ell}(M_1,z) \times \nonumber \\
& & \int dM_2 \frac{dn(M_2,z)}{dM_2} b(M_2,z) \tilde{\phi}_{\ell}(M_2,z) \,.
\label{eq.Clyphi2h}
\ea
Here, $\frac{d^2V}{dz d\Omega}$ is the comoving volume per steradian, $dn/dM$ is the halo mass function (number of halos per unit mass per unit comoving volume), $P_{\rm lin}(k,z) \equiv D^2(z) P_{\rm lin}(k)$, $b(M,z)$ is the linear halo bias, and $\tilde{y}_{\ell}(M,z)$ and $\tilde{\phi}_{\ell}(M,z)$ are the Fourier transform of the Compton-$y$ and CMB lensing potential profiles, respectively, of a halo of mass $M$ at redshift $z$:
\ba
\tilde{y}_{\ell}(M,z) & = & \frac{\sigma_T}{m_e c^2} \frac{4 \pi
  r_{s,y}}{\ell_{s,y}^2} \times \nonumber \\
  & & \int dx_y \, x_y^2 \frac{\sin((\ell+1/2)
  x_y/\ell_{s,y})}{(\ell+1/2) x_y/\ell_{s,y}} \times \nonumber \\
  & & \, P_{e}(x_y r_{s,y},M,z) \,;
\label{eq.yprof}
\ea
\ba
\tilde{\phi}_{\ell}(M,z) & = & \frac{2}{\ell(\ell+1)} \frac{4 \pi
  r_{s,\phi}}{\ell_{s,\phi}^2} \times \nonumber \\
  & & \int dx_{\phi} \, x_{\phi}^2
\frac{\sin((\ell+1/2) x_{\phi}/\ell_{s,\phi})}{(\ell+1/2)
  x_{\phi}/\ell_{s,\phi}} \times \nonumber \\
  & & \frac{\rho(x_{\phi} r_{s,\phi},M,z)}{\Sigma_{\rm crit}(z)} \,.
\label{eq.phiprof}
\ea
Here, $r_{s,y}$ is a characteristic scale radius of the electron pressure profile, $\ell_{s,y} = a(z)\chi(z)/r_{s,y} = d_A(z)/r_{s,y}$ is the multipole moment associated with this scale, and $x_y \equiv r/r_{s,y}$ is a dimensionless radial variable for the pressure profile.  Analogously, $r_{s,\phi}$ is a characteristic scale radius of the halo density profile $\rho(r,M,z)$, $\ell_{s,\phi} = a(z)\chi(z)/r_{s,\phi} = d_A(z)/r_{s,\phi}$ is the multipole moment associated with this scale, and $x_{\phi} \equiv r/r_{s,\phi}$ is a dimensionless radial variable for the density profile.  The quantity $\Sigma_{\rm crit}(z)$ is the critical surface density for CMB lensing:
\be
\Sigma_{\rm crit}(z) = \frac{c^2 \chi_* (1+z)}{4 \pi G \chi(z) \left(\chi_{*}-\chi(z) \right)} \,.
\label{eq.Sigmacrit}
\ee
We adopt the electron pressure profile fitting function from the hydrodynamic simulations of~\cite{BBPSS2010,BBPS2012b}, the Navarro-Frenk-White (NFW) density profile~\cite{NFW1997}, the concentration-mass relation of~\cite{Duffyetal2008}, and the fitting functions for the halo mass function and linear halo bias of~\cite{Tinkeretal2010} (updated from~\cite{Tinkeretal2008}).  The concentration-mass relation is required in order to convert between mass definitions; we define $M$ to be the virial mass following the definition of~Ref.~\cite{Bryan-Norman1998}.  Further details of this framework can be found in Refs.~\cite{Hill-Pajer2013,HS2014}.

The fiducial mass and redshift limits for all halo model integrals in this paper (e.g., Eqs.~\ref{eq.Clyphi1h} and~\ref{eq.Clyphi2h}) are $10^5 \, M_{\odot}/h < M < 5 \times 10^{15} \, M_{\odot}/h$ and $0.005 < z < 8$, respectively.  The lower redshift limit is imposed to avoid unphysical divergences at $z=0$.  We verify that all calculations are converged with these choices.

The lensing-tSZ cross-power spectrum computed with this model is shown in the dash-dotted curves in Fig.~\ref{fig:phi_xcorr}.  The frequency dependence is evaluated for the {\em Planck} channels using the bandpass-integrated tSZ spectral function values provided in Ref.~\cite{Planck2015ymap} (note that the signal is negative for $\nu < 217$ GHz, positive for frequencies above this, and effectively vanishes for the 217 GHz channel).  The lensing-tSZ cross-power spectrum is comparable to the lensing-ISW cross-power spectrum around $\ell \approx 100$, and is much larger at higher multipoles.

\begin{figure}
\includegraphics[width=0.5\textwidth]{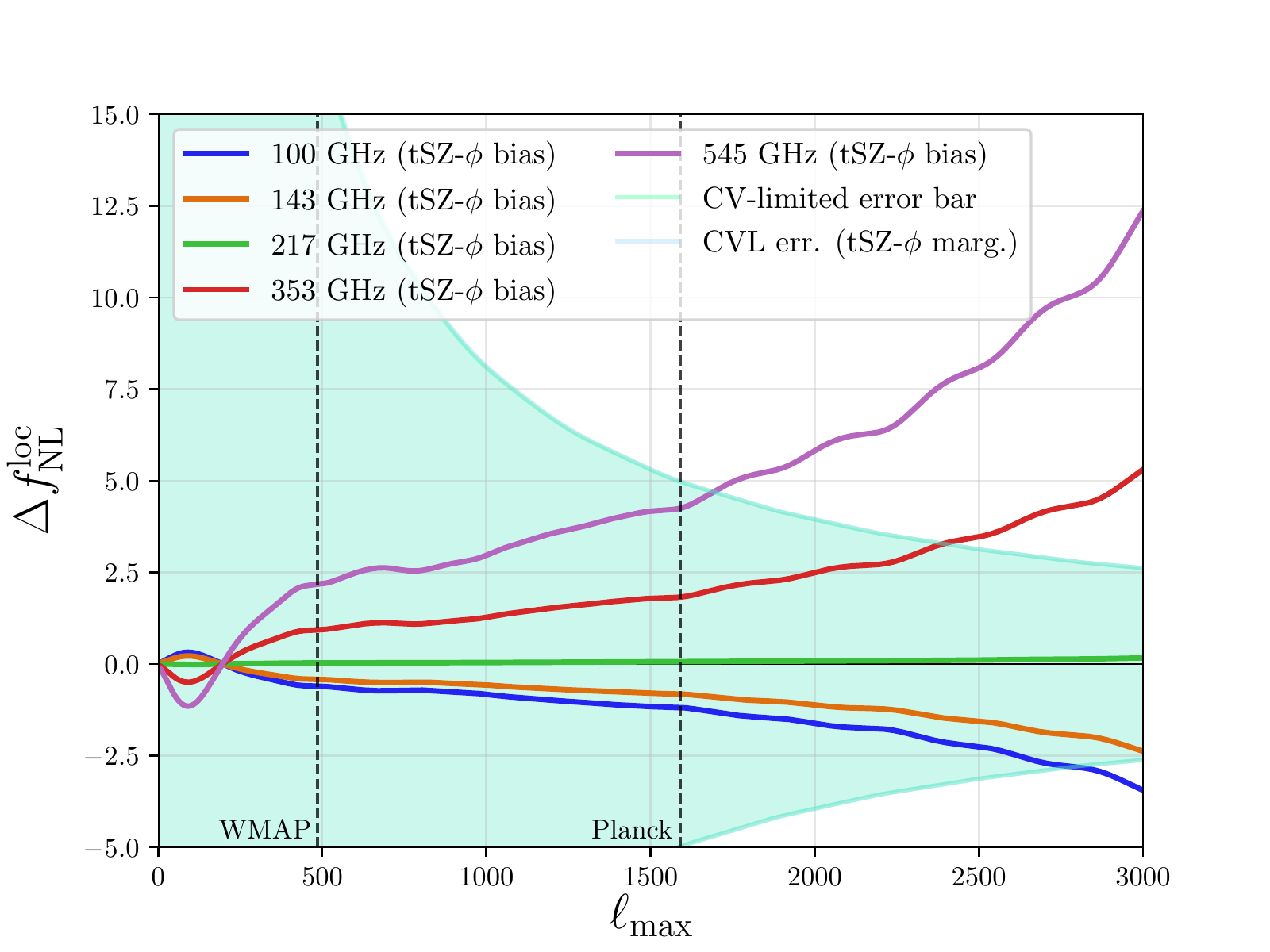} 
\caption{\label{fig:tSZxphi_bias} Bias on $f_{\rm NL}^{\rm loc}$ from the lensing-tSZ bispectrum as a function of $\ell_{\rm max}$, as in Fig.~\ref{fig:ISWxphi_bias}.  The bias is shown for the {\em Planck} HFI channels from 100--545 GHz, with a frequency dependence arising from the tSZ spectral function.  For {\em Planck}, the bias takes values of $\Delta f_{\rm NL}^{\rm loc} = -1.2$ (100 GHz), $\Delta f_{\rm NL}^{\rm loc} = -0.8$ (143 GHz), and $\Delta f_{\rm NL}^{\rm loc} = 1.8$ (353 GHz).  Due to its non-blackbody nature, this bias could be removed via component separation, but recent evidence suggests that non-negligible tSZ signal has leaked into the {\em Planck} component-separated CMB maps~\cite{MH2018,Chenetal2018}.  No multifrequency cleaning is assumed here.  The light green shaded region and dashed vertical lines are identical to those in Fig.~\ref{fig:ISWxphi_bias}.  The light blue shaded region, which is indistinguishable from the green region, shows the $1\sigma$ uncertainty on $f_{\rm NL}^{\rm loc}$ after marginalizing over the lensing-tSZ bispectrum amplitude; this marginalization has no impact on the $f_{\rm NL}^{\rm loc}$ error bar.}
\end{figure}

Fig.~\ref{fig:tSZxphi_bias} shows the bias on $f_{\rm NL}^{\rm loc}$ due to the lensing-tSZ bispectrum, computed via Eqs.~\ref{eq.Clyphi},~\ref{eq.Blens}, and~\ref{eq.fNLbias}.  We show the bias for the {\em Planck} HFI channels from 100--545 GHz, using the bandpass-integrated tSZ spectral function values from Ref.~\cite{Planck2015ymap}.  For {\em Planck}, the bias is of order $|\Delta f_{\rm NL}^{\rm loc}| \approx 1-2$ for these channels, except for the 545 GHz channel, where it is somewhat larger.  This conclusion appears to agree with the results of Ref.~\cite{Serra-Cooray2008} (by comparison to their Fig.~2, after removing the lensing-ISW bias).  For an experiment with $\ell_{\rm max} = 3000$, the bias is comparable to the $1\sigma$ error bar on $f_{\rm NL}^{\rm loc}$ at the dominant CMB channels (100 and 143 GHz).

Fig.~\ref{fig:tSZxphi_bias} also shows the $1\sigma$ uncertainty on $f_{\rm NL}^{\rm loc}$ after marginalizing over the amplitude of the lensing-tSZ bispectrum, computed with Eq.~\ref{eq.fNLerrormarg}.  As in Fig.~\ref{fig:ISWxphi_bias}, this marginalization leaves the error bar on $f_{\rm NL}^{\rm loc}$ essentially unchanged (in fact, the increase after marginalization is even smaller than in the lensing-ISW case).  However, unlike the lensing-ISW case, there is some modeling uncertainty in the lensing-tSZ bispectrum shape (and amplitude) due to ICM astrophysics, and thus additional parameters may have to be marginalized over.  Nevertheless, this result indicates that the correlation between the lensing-tSZ bispectrum and the local bispectrum is quite small.

While marginalization over a lensing-tSZ template may be sensible for a single-frequency $f_{\rm NL}^{\rm loc}$ analysis, this is likely unnecessary for a multifrequency analysis. Unlike the lensing-ISW bias, the lensing-tSZ bias is non-blackbody in frequency dependence, and can therefore be mitigated via multifrequency component separation.  In fact, it can be removed exactly using ``constrained'' component separation methods, in which the frequency channel weights are required to exactly null the tSZ spectral function in Eq.~\ref{eq.tSZspectralfunc}~\cite{Remazeillesetal2011}.  However, such constraints were not applied to the component-separated CMB temperature maps that were used in the {\em Planck} 2015 NG analysis~\cite{Planck2015compsep,Planck2015NG}.  Recent analyses have presented evidence that these maps have non-negligible tSZ contamination~\cite{MH2018,Chenetal2018}.  A precise estimate of the tSZ leakage as a function of angular scale would be needed to convert the frequency-dependent biases in Fig.~\ref{fig:tSZxphi_bias} into a final bias for Planck.  A simpler method would be to perform the NG analysis on a component-separated map in which the tSZ signal has been nulled, although a statistical penalty in signal-to-noise ($S/N$) must be paid accordingly.

Masking individually detected galaxy clusters would reduce the lensing-tSZ bias to some extent, although not by a large amount, as $C_{\ell}^{y\phi}$ is dominated by halos at lower masses and higher redshifts~\cite{HS2014,BHM2015} than are present in the {\em Planck} tSZ catalog~\cite{Planck2015SZcounts,Planck2015SZcosmology}.  Thus, even if such clusters are masked in the {\em Planck} NG analysis, it would not strongly impact the biases shown in Fig.~\ref{fig:tSZxphi_bias}.

Finally, note that $C_{\ell}^{y\phi}$ has a fairly strong dependence on cosmological parameters, particularly $\sigma_8$ and $\Omega_m$~\cite{HS2014}, and thus the associated bias on $f_{\rm NL}^{\rm loc}$ will have a strong dependence as well.\footnote{In fact, the tSZ-related biases computed in this paper also depend on the value of $f_{\rm NL}^{\rm loc}$ itself: increasing (decreasing) $f_{\rm NL}^{\rm loc}$ increases (decreases) the number of massive clusters in the low-redshift universe~(e.g.,~\cite{LVS2011}), and will therefore modify the tSZ-related contributions.}  If we adopt the {\em Planck} 2015 CMB values for these parameters ($\sigma_8 = 0.830$ and $\Omega_m = 0.316$)~\cite{Planck2015params}, the bias would be $\approx 35$\% larger than shown in Fig.~\ref{fig:tSZxphi_bias}, assuming $C_{\ell}^{y\phi} \propto \sigma_8^6 \Omega_m^{1.5}$~\cite{HS2014}.

\subsection{Lensing-CIB Bias}
\label{sec:CIB-phi}

The CIB is sourced by the cumulative emission of dusty, star-forming galaxies over cosmic time.  The emission at different observational frequencies is generated by galaxies at somewhat different redshift ranges, but in general the CIB ``redshift kernel'' has a broad peak around $z \approx 2$, corresponding to the peak in the star formation rate density~\cite{Madau-Dickinson2014,Planck2013CIB,Planck2013CIBxlens,Maniyar2018}.

CIB statistics can be computed in the halo model, analogous to the tSZ calculations above, but with a more complicated prescription for the assignment of infrared flux to halos.  For the lensing-CIB cross-power spectrum, we simply use the best-fit results of the {\em Planck} measurement of this quantity at each of the HFI frequencies~\cite{Planck2013CIBxlens}, in lieu of implementing a detailed model here.\footnote{We thank Olivier Dor\'{e} for providing these fits.}  This guarantees that our calculations are consistent with actual measurements of $C_{\ell}^{{\rm CIB} \times \phi}$.  The cross-power spectra for three of the {\em Planck} HFI frequencies are shown as dashed curves in Fig.~\ref{fig:phi_xcorr}.  The strong frequency dependence of the CIB emission is evident; the cross-power spectrum signal at 353 GHz is nearly an order of magnitude larger than at 217 GHz.

\begin{figure}
\includegraphics[width=0.5\textwidth]{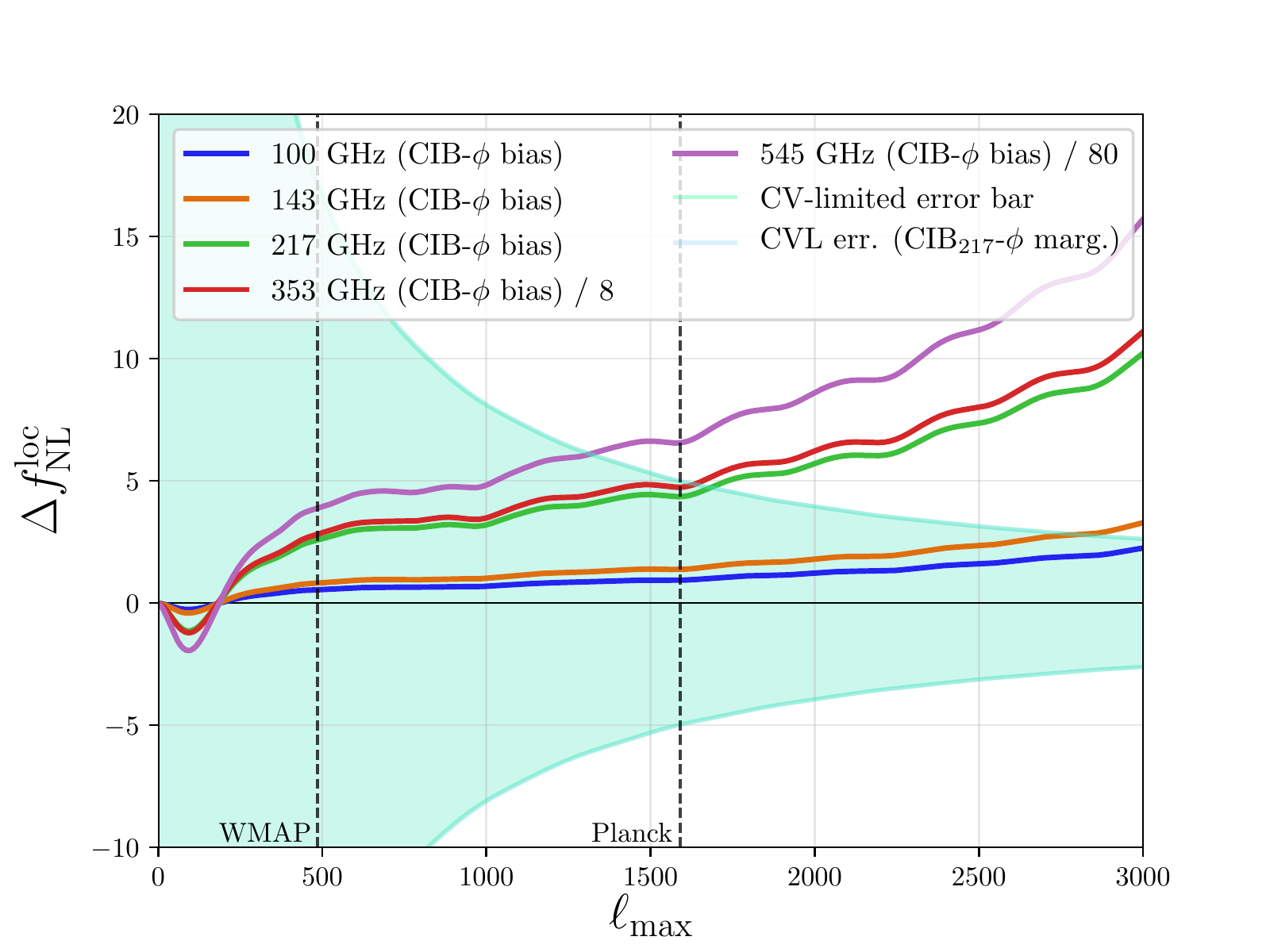} 
\caption{\label{fig:CIBxphi_bias} Bias on $f_{\rm NL}^{\rm loc}$ from the lensing-CIB bispectrum as a function of $\ell_{\rm max}$, as in Fig.~\ref{fig:ISWxphi_bias}.  The bias is shown for the {\em Planck} HFI channels from 100--545 GHz, with a strong frequency dependence due to the increase in dust emission intensity at high frequencies.  Note that the 353 GHz and 545 GHz results are divided by factors of 8 and 80, respectively, to reduce the dynamic range of the plot.  For {\em Planck}, the bias takes values of $\Delta f_{\rm NL}^{\rm loc} = 1.4$ (143 GHz), $\Delta f_{\rm NL}^{\rm loc} = 4.3$ (217 GHz), and $\Delta f_{\rm NL}^{\rm loc} = 38$ (353 GHz).  This bias is non-blackbody in frequency dependence, and is thus reduced by multifrequency component separation techniques (however, it cannot be fully eliminated due to CIB decorrelation).  The extent of this reduction for the {\em Planck} NG analysis is currently unclear.  No multifrequency cleaning is assumed here.  The light green shaded region and dashed vertical lines are identical to those in Fig.~\ref{fig:ISWxphi_bias}.  The light blue shaded region shows the $1\sigma$ uncertainty on $f_{\rm NL}^{\rm loc}$ after marginalizing over the lensing-CIB bispectrum amplitude, evaluated at 217 GHz (as an example); this marginalization has no impact on the $f_{\rm NL}^{\rm loc}$ error bar.}
\end{figure}

Fig.~\ref{fig:CIBxphi_bias} shows the bias on $f_{\rm NL}^{\rm loc}$ due to the lensing-CIB bispectrum, computed using the $C_{\ell}^{{\rm CIB} \times \phi}$ fits from Ref.~\cite{Planck2013CIBxlens} in combination with Eqs.~\ref{eq.Blens} and~\ref{eq.fNLbias}.  We show the bias for the {\em Planck} HFI channels from 100--545 GHz.  The bias is strongly frequency-dependent, a direct result of the strong frequency dependence shown in Fig.~\ref{fig:phi_xcorr}.  However, unlike the lensing-tSZ bias, the lensing-CIB bias has the same sign at all frequencies (at fixed $\ell_{\rm max}$).  At 217 GHz, the bias for {\em Planck} is $\Delta f_{\rm NL}^{\rm loc} = 4.3$, while at 353 GHz, it is nearly an order of magnitude larger.  These results are similar to those presented in Ref.~\cite{Curtoetal2015} (see their Table 3), but up to a factor of 2--3 larger than theirs at some {\em Planck} frequencies.  In this context, we note that our calculation of $C_{\ell}^{{\rm CIB} \times \phi}$ is directly drawn from fits to {\em Planck} measurements as described above, rather than a theoretical model.  For an experiment with $\ell_{\rm max} = 3000$, the lensing-CIB bias is comparable to the $1\sigma$ error bar on $f_{\rm NL}^{\rm loc}$ at 100 or 143 GHz, and is much larger than this at higher frequencies.

Fig.~\ref{fig:CIBxphi_bias} also shows the $1\sigma$ uncertainty on $f_{\rm NL}^{\rm loc}$ after marginalizing over the amplitude of the lensing-CIB bispectrum, computed with Eq.~\ref{eq.fNLerrormarg} (considering only 217 GHz, as an example case).  As in Fig.~\ref{fig:tSZxphi_bias}, this marginalization has no noticeable impact on the $f_{\rm NL}^{\rm loc}$ error bar.  Like the lensing-tSZ case, though, there is some astrophysical modeling uncertainty in the lensing-CIB bispectrum shape (and amplitude), and thus additional parameters may have to be marginalized over.  Nevertheless, this result indicates that the correlation between the lensing-CIB bispectrum and the local bispectrum is quite small.

Like the lensing-tSZ bias, marginalization over the lensing-CIB bispectrum is likely unnecessary for a multifrequency analysis, as the lensing-CIB bias is non-blackbody in frequency dependence and can therefore be reduced via component separation.  However, unlike the lensing-tSZ bias, it cannot be fully eliminated, as the CIB decorrelates across frequency channels to some extent (because the redshift kernel of the emission is different at different frequencies)~\cite{Planck2013CIB,Maketal2017}.  Given the evidence of tSZ leakage into the {\em Planck} component-separated CMB maps, it is plausible that non-negligible CIB leakage is also present, although the extent of such contamination is presently unclear.  A simulation-based analysis is necessary to quantify the total bias on $f_{\rm NL}^{\rm loc}$ resulting from the lensing-CIB cross-correlation.  The most robust route may be a combination of multifrequency cleaning and subsequent marginalization over a lensing-CIB template.

\section{ISW-Related Biases}
\label{sec:ISW}

The ISW effect is a tracer of the late-time gravitational potential, and therefore it is correlated with other such tracers that source secondary CMB anisotropies.  For our purposes here, we are primarily concerned with bispectra involving one ISW ``leg'' and two ``legs'' drawn from the tSZ, CIB, or kSZ fields.  These bispectra have strong contributions in the squeezed limit, as the ISW signal peaks on large scales (i.e., comprising the long-wavelength mode of the triangle) while the other fields peak on small scales (i.e., comprising the two short-wavelength modes of the triangle).  A physical interpretation of these bispectra is that the long-wavelength ISW field is modulating the amplitude of the short-wavelength power spectra of the other fields (e.g., the small-scale tSZ power spectrum).  A long-wavelength overdensity (underdensity) producing a positive (negative) ISW fluctuation will also contain more (fewer) massive halos, thereby corresponding to a higher (lower) amplitude of the small-scale tSZ/CIB/kSZ power spectra.

We compute these bispectra in the halo model, working throughout in the Limber approximation for bispectra~\cite{Buchalter2000,Takada-Jain2004}.  In general, the full bispectrum will contain three-halo, two-halo, and one-halo terms, but here we focus only on the contributions that are expected to dominate in squeezed configurations relevant to $f_{\rm NL}^{\rm loc}$.  The primary such contribution arises from a two-``halo'' term, in which one multipole corresponds to a long-wavelength ISW fluctuation (hence, we use the term ``halo'' loosely here) and the other two multipoles correspond to short-wavelength tSZ/CIB/kSZ fluctuations.  Note that throughout we consider only the linear-theory ISW effect, i.e., the Rees-Sciama effect is neglected.  Thus, there is effectively no one-halo contribution to these bispectra (nonlinear growth would generate a Rees-Sciama one-halo term, but this is much smaller than the linear-theory ISW signal).  We also neglect the three-halo term, which is sourced by the tree-level bispectrum, as it is not expected to contribute strongly to squeezed configurations.  One exception to this may be the three-halo contribution to the ISW-kSZ-kSZ bispectrum (for which we also neglect other potentially important contributions --- see \S\ref{sec:ISW-kSZ-kSZ}).  We defer a full calculation to future work, and focus only on the two-halo contributions in the following.

\subsection{ISW-tSZ-tSZ Bias}
\label{sec:ISW-tSZ-tSZ}

For CMB temperature maps at three frequencies $\nu_1$, $\nu_2$, and $\nu_3$, the two-``halo'' contribution to the ISW-tSZ-tSZ reduced bispectrum is:
\begin{widetext}
\ba
b_{\ell_1\ell_2\ell_3, ({\rm ISW-tSZ-tSZ})}^{({T_{\nu_1} T_{\nu_2} T_{\nu_3}}), 2h} & = & g(\nu_2) g(\nu_3) \int dz \, \mathcal{I}_{\ell_1}(z) \, \left[ P_{\rm lin}\left(\frac{\ell_1+\frac{1}{2}}{\chi(z)}\right) \int dM \, \frac{dn}{dM} b(M,z) \tilde{y}_{\ell_2}(M,z) \tilde{y}_{\ell_3}(M,z) \right. \nonumber \\
& + & P_{\rm lin}\left(\frac{\ell_2+\frac{1}{2}}{\chi(z)}\right) \int dM \, \frac{dn}{dM} b(M,z) \tilde{y}_{\ell_2}(M,z) \int dM' \, \frac{dn}{dM'} b(M',z) \tilde{y}_{\ell_3}(M',z) \frac{M'}{\bar{\rho}_m} \nonumber \\
& + & \left. P_{\rm lin}\left(\frac{\ell_3+\frac{1}{2}}{\chi(z)}\right) \int dM \, \frac{dn}{dM} b(M,z) \tilde{y}_{\ell_3}(M,z) \int dM' \, \frac{dn}{dM'} b(M',z) \tilde{y}_{\ell_2}(M',z) \frac{M'}{\bar{\rho}_m} \right] \nonumber \\
& + & 2 \,\, {\rm perm.} \,,
\label{eq.BIyy}
\ea
\end{widetext}
where
\be
\mathcal{I}_{\ell}(z) = \frac{3 \Omega_m H_0^2}{c^2\left(\ell+\frac{1}{2}\right)^2} \chi^2(z) D(z) \frac{d}{dz} \left( \frac{D(z)}{a(z)} \right) \,,
\label{eq.ISWfac}
\ee
and the additional permutations correspond to cases in which the ISW multipole is either $\ell_2$ or $\ell_3$.  Here, we have made the approximation that the contribution from the internal structure of the halo density profile can be neglected in the ISW factors, i.e., $k_1^{\rm ISW} \rightarrow 0$, so that the Fourier transform of the density profile simply yields a factor of $M$.  This approximation is accurate due to the rapid decline of the ISW signal as $\ell$ increases.  Stated differently, the ISW signal is effectively sourced only by linear modes of the density field.  For computational efficiency, we set the ISW signal to zero above $\ell_{\rm ISW,cut} = 200$ in all of the following calculations.  We verify that our results are converged with this choice, i.e., higher values of $\ell_{\rm ISW,cut}$ do not change the derived bias on $f_{\rm NL}^{\rm loc}$ for any of these bispectra.  Finally, note that the first term in Eq.~\ref{eq.BIyy}, in which both tSZ multipoles belong to the same halo, dominates by a factor of $\gtrsim 10-100$ over the latter two terms, in which the tSZ multipoles are in two distinct halos, except for configurations in which all three multipoles are very small (i.e., on very large scales).  This is precisely analogous to the dominance of the one-halo term over the two-halo term in the tSZ power spectrum for all $\ell \gtrsim 10$~\cite{Komatsu-Kitayama1999,Komatsu-Seljak2002,Hill-Pajer2013}.  Nevertheless, we include all terms in the following calculations.

\begin{figure}
\includegraphics[width=0.5\textwidth]{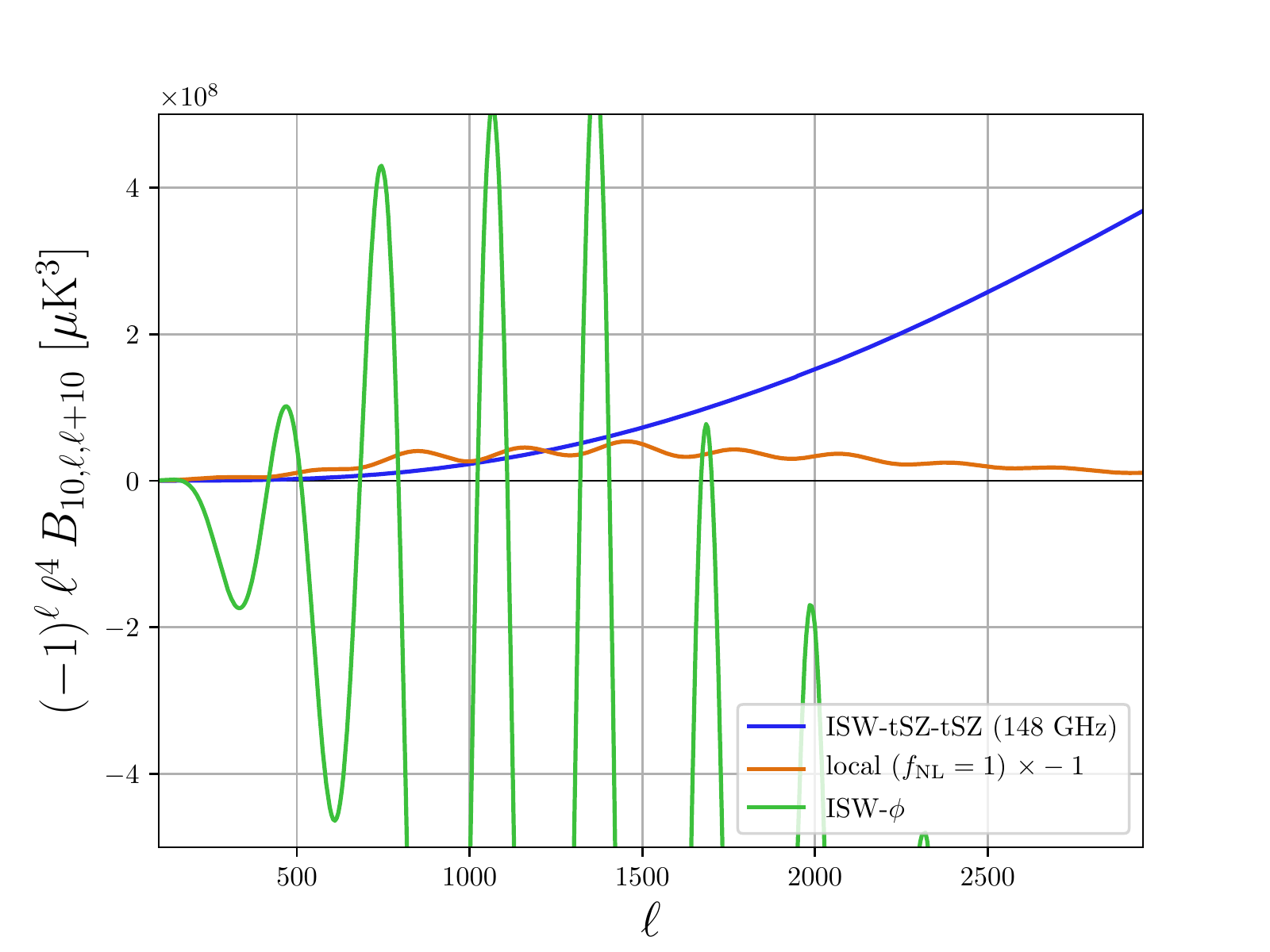}
\caption{\label{fig:bispec_slice} A ``slice'' through the ISW-tSZ-tSZ bispectrum at 148 GHz computed via Eq.~\ref{eq.BIyy} (blue), the local primordial bispectrum with $f_{\rm NL}^{\rm loc}=1$ (orange), and the lensing-ISW bispectrum (green), for squeezed configurations with $\ell_1 = 10$, $\ell_2 \equiv \ell$, and $\ell_3 = \ell+10$.  In contrast to the latter two bispectra, the ISW-tSZ-tSZ bispectrum is a smoothly increasing function of $\ell$.  Note that it peaks in the squeezed limit, as expected.}
\end{figure}

\begin{figure}
\includegraphics[width=0.5\textwidth]{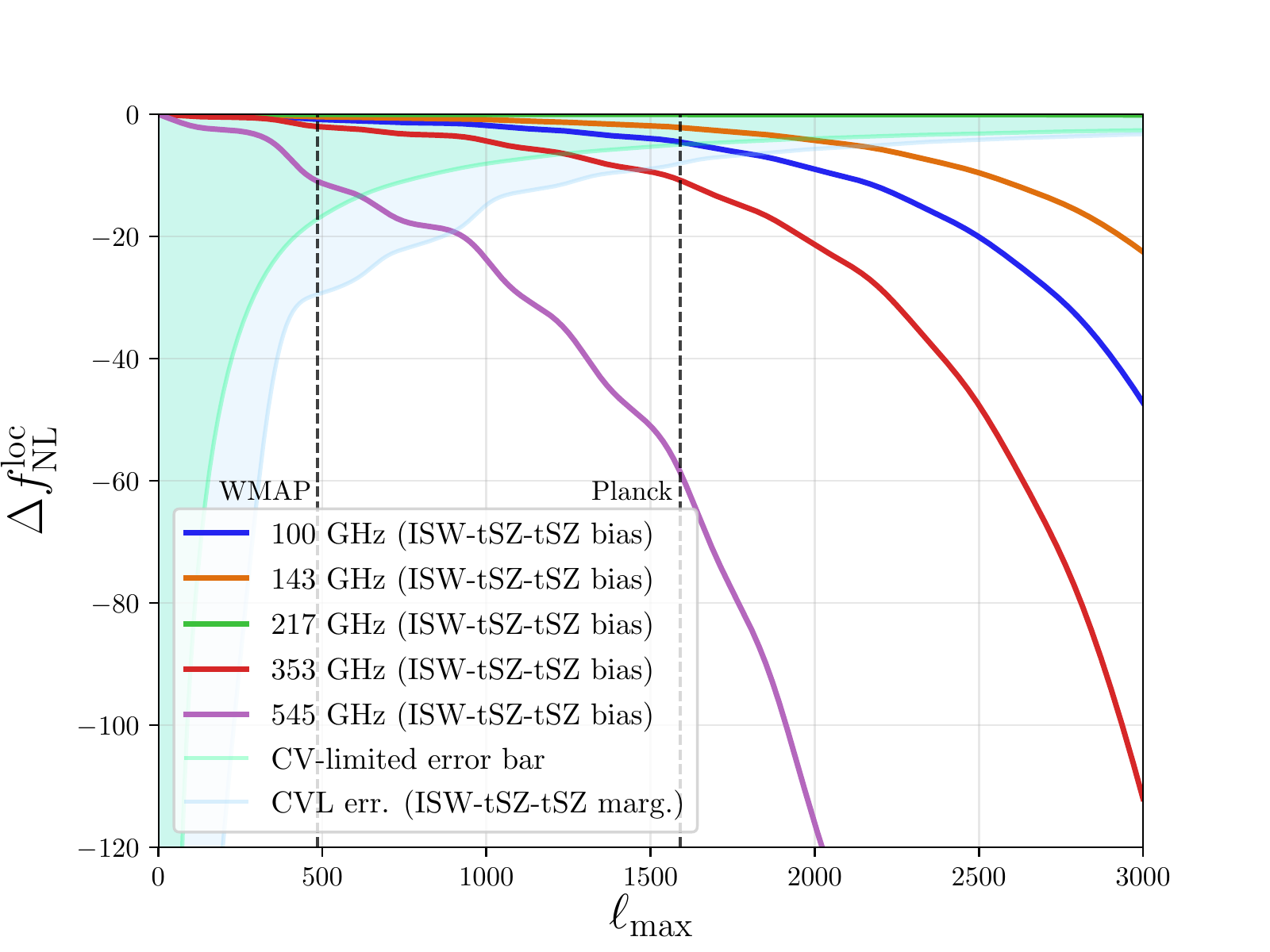} 
\caption{\label{fig:ISWyy_bias} Bias on $f_{\rm NL}^{\rm loc}$ from the ISW-tSZ-tSZ bispectrum as a function of $\ell_{\rm max}$, as in Fig.~\ref{fig:ISWxphi_bias}.  The bias is shown for the {\em Planck} HFI channels from 100--545 GHz, with a frequency dependence arising from the tSZ spectral function.  For {\em Planck}, the bias takes values of $\Delta f_{\rm NL}^{\rm loc} = -4.5$ (100 GHz), $\Delta f_{\rm NL}^{\rm loc} = -2.1$ (143 GHz), and $\Delta f_{\rm NL}^{\rm loc} = -11$ (353 GHz).  Due to its non-blackbody nature, this bias could be removed via component separation, but recent evidence suggests that non-negligible tSZ signal has leaked into the {\em Planck} component-separated CMB maps~\cite{MH2018,Chenetal2018}.  No multifrequency cleaning is assumed here.  The light green shaded region and dashed vertical lines are identical to those in Fig.~\ref{fig:ISWxphi_bias}.  The light blue shaded region shows the $1\sigma$ uncertainty on $f_{\rm NL}^{\rm loc}$ after marginalizing over the ISW-tSZ-tSZ bispectrum amplitude; this marginalization produces a non-negligible increase in the $f_{\rm NL}^{\rm loc}$ error bar ($\approx 60$\% increase for {\em Planck}).}
\end{figure}

We use the same models to compute Eq.~\ref{eq.BIyy} as used for the tSZ calculations described in \S\ref{sec:tSZ-phi}.\footnote{As a cross-check, we also compute the ISW-tSZ cross-power spectrum and obtain results in general agreement with those of Ref.~\cite{CS2016}, although they use a different pressure profile and halo mass function, which will inevitably lead to some differences.  In particular, the agreement is excellent on large scales (within $10$\% at $\ell<10$), but is somewhat discrepant on smaller scales; however, we note that this qualitatively matches the discrepancy between the ISW auto-power spectrum of Ref.~\cite{CS2016} and that of CLASS~\cite{CLASSpaper}, and thus assume that it is related to a numerical issue in their calculation (S.~Bird, priv.~comm.).}  Fig.~\ref{fig:bispec_slice} shows a ``slice'' through the ISW-tSZ-tSZ bispectrum (considering only the contributions in Eq.~\ref{eq.BIyy}), as well as the local bispectrum (with $f_{\rm NL}^{\rm loc} = 1$) and the lensing-ISW bispectrum.  The latter two bispectra display acoustic oscillations arising from the radiation transfer functions, whereas the ISW-tSZ-tSZ bispectrum is smooth (the primary temperature power spectrum does not appear in Eq.~\ref{eq.BIyy}, in contrast to Eq.~\ref{eq.Blens}).  For this particular slice, the ISW-tSZ-tSZ bispectrum becomes comparable in amplitude to the local bispectrum at $\ell \approx 1500$, and is much larger at higher multipoles.  This plot also illustrates the origin of the oscillatory behavior seen for the lensing-ISW bias on $f_{\rm NL}^{\rm loc}$ in Fig.~\ref{fig:ISWxphi_bias} (and to some extent in Figs.~\ref{fig:tSZxphi_bias} and~\ref{fig:CIBxphi_bias}): the lensing-ISW and local bispectra have oscillations that are not exactly in phase, and the lensing-ISW bispectrum furthermore oscillates between positive and negative values.  These effects lead to oscillations in the inner product in the numerator of Eq.~\ref{eq.fNLbias}.  In contrast, the smooth shape of the ISW-tSZ-tSZ bispectrum suggests that the associated bias on $f_{\rm NL}^{\rm loc}$ will be a smoothly increasing function of $\ell_{\rm max}$, which indeed is the case (see Fig.~\ref{fig:ISWyy_bias}).  Finally, while Fig.~\ref{fig:bispec_slice} shows the ISW-tSZ-tSZ bispectrum at an example frequency of 148 GHz, this bispectrum is always positive when evaluated at a single frequency, due to the quadratic tSZ spectral function dependence (when evaluated for a set of different frequencies, it could be negative or positive, but never crosses zero).

Fig.~\ref{fig:ISWyy_bias} shows the bias on $f_{\rm NL}^{\rm loc}$ due to the ISW-tSZ-tSZ bispectrum, computed via Eqs.~\ref{eq.BIyy} and~\ref{eq.fNLbias}.  We show the bias for the {\em Planck} HFI channels from 100--545 GHz, using the bandpass-integrated tSZ spectral function values from Ref.~\cite{Planck2015ymap}.  We do not plot any cross-frequency biases (i.e., involving different values of $\nu_2$ and $\nu_3$ in Eq.~\ref{eq.BIyy}), although these are present and can be of positive or negative sign.  In contrast, the single-frequency biases are always negative, as shown in Fig.~\ref{fig:ISWyy_bias}.  For {\em Planck}, the bias is $\Delta f_{\rm NL}^{\rm loc} = -4.5$ (100 GHz), $\Delta f_{\rm NL}^{\rm loc} = -2.1$ (143 GHz), and $\Delta f_{\rm NL}^{\rm loc} = -11$ (353 GHz).  If a non-negligible fraction of tSZ signal has leaked into the component-separated CMB maps used in the {\em Planck} NG analysis, the ISW-tSZ-tSZ bias could thus yield a shift of order $1\sigma$ in the inferred value of $f_{\rm NL}^{\rm loc}$.  For an experiment with $\ell_{\rm max} = 3000$, the bias is many times larger than the $1\sigma$ error bar on $f_{\rm NL}^{\rm loc}$ at all of the frequencies considered (except 217 GHz, where the tSZ null occurs), including the dominant CMB channels (100 and 143 GHz).

Fig.~\ref{fig:ISWyy_bias} also shows the $1\sigma$ uncertainty on $f_{\rm NL}^{\rm loc}$ after marginalizing over the amplitude of the ISW-tSZ-tSZ bispectrum, computed with Eq.~\ref{eq.fNLerrormarg}.  Unlike marginalizing over the lensing-related foreground bispectra in the previous section, which hardly increases the $f_{\rm NL}^{\rm loc}$ uncertainty, marginalizing over the ISW-tSZ-tSZ bispectrum amplitude significantly increases the error bar on $f_{\rm NL}^{\rm loc}$.  As discussed in \S\ref{sec:ISW-phi}, the error bar increase due to marginalization depends solely on the correlation coefficient between the local bispectrum and the foreground bispectrum, but not on their amplitudes.  The ISW-tSZ-tSZ bispectrum shape is strongly correlated with the local bispectrum template, leading to the significant error bar increase.  For {\em Planck}, assuming no prior is placed on the ISW-tSZ-tSZ bispectrum amplitude, the uncertainty on $f_{\rm NL}^{\rm loc}$ increases by $\approx 60$\%.  Moreover, given astrophysical uncertainties in the tSZ modeling, additional parameters (beyond the amplitude) may have to be marginalized over as well, further increasing the $f_{\rm NL}^{\rm loc}$ error bar.

Fortunately, as discussed in \S\ref{sec:tSZ-phi} for the lensing-tSZ bias, the ISW-tSZ-tSZ bias is non-blackbody in nature, and in fact can be exactly removed via constrained component separation techniques (at a cost in $S/N$).  It can also be modeled and subtracted, although modeling it through the process of (non-constrained) component separation likely requires full simulations.  The bias will also be reduced to some extent by masking known galaxy clusters, but since it is effectively sourced by the small-scale tSZ power spectrum, the bias is mostly generated by clusters that are below the mass threshold for individual detection in {\em Planck} (see, e.g., Refs.~\cite{Hill-Pajer2013,BBPS2012b} for breakdowns of the halo mass and redshift contributions to the tSZ power spectrum).

Amongst the biases considered in this paper, the ISW-tSZ-tSZ bias is the most sensitive to cosmological parameters, as it inherits the strong dependence of the tSZ power spectrum on $\sigma_8$ and $\Omega_m$~(e.g.,~\cite{Komatsu-Seljak2002,Hill-Pajer2013}).  If we adopt the {\em Planck} 2015 CMB values for these parameters~\cite{Planck2015params}, the bias would be $\approx 70$\% larger than shown in Fig.~\ref{fig:ISWyy_bias}, assuming that the bispectrum in Eq.~\ref{eq.BIyy} follows the tSZ power spectrum parameter dependence: $C_{\ell}^{yy} \propto \sigma_8^8 \Omega_m^{3}$~\cite{Hill-Pajer2013,Bollietetal2018}.  In addition, it is sensitive to the modeling of astrophysical processes in the intracluster medium.  Given these parameter and modeling dependences, as well as the significant increase in $\sigma(f_{\rm NL}^{\rm loc})$ after marginalizing over the ISW-tSZ-tSZ bispectrum amplitude, it seems best to simply null the tSZ spectral function via constrained component separation when constructing CMB maps for NG analysis, so that this foreground is not present.

\subsection{ISW-CIB-CIB Bias}
\label{sec:ISW-CIB-CIB}

The two-halo contribution to the ISW-CIB-CIB reduced bispectrum is identical to Eq.~\ref{eq.BIyy}, with the replacement $g(\nu) \tilde{y}_{\ell}(M,z) \rightarrow \tilde{I}_{\ell}^{{\rm CIB},\nu(1+z)}(M,z)$, i.e., the tSZ profile of each halo is replaced by its redshifted infrared emission profile.\footnote{Additional shot noise terms may also be included, depending on the details of the underlying CIB halo model.}  In lieu of implementing a detailed model for the infrared emission of each halo, here we adopt a simpler, approximate approach relying on the very high correlation coefficient between the CIB field (at 100--1000 GHz) and the CMB lensing field~\cite{Planck2013CIBxlens,Holderetal2013,vanEngelenetal2015}.  If the correlation coefficient between these fields were unity at all multipoles, a CIB map would simply be a rescaled version of a CMB lensing map, with an $\ell$-dependent rescaling factor $f_{\ell}$.  Empirically, the correlation coefficient is $\gtrsim 80$\% at all multipoles up to $\ell \approx 2000$ for most {\em Planck} HFI frequencies~\cite{Planck2013CIBxlens}.  For the purpose of approximately estimating biases to $f_{\rm NL}^{\rm loc}$, we consider this sufficiently close to unity to simply approximate $\tilde{I}_{\ell}^{{\rm CIB},\nu(1+z)}(M,z) \approx f_{\ell}(\nu) \tilde{\phi}_{\ell}(M,z)$, where $\tilde{\phi}_{\ell}(M,z)$ is given by Eq.~\ref{eq.phiprof}.  We determine the rescaling factor $f_{\ell}(\nu)$ for a given frequency channel using the lensing-CIB cross-power spectrum results of Ref.~\cite{Planck2013CIBxlens}, i.e., the same fits to $C_{\ell}^{{\rm CIB} \times \phi}$ used in \S\ref{sec:CIB-phi}, in combination with a theoretical calculation of the CMB lensing auto-power spectrum:
\be
f_{\ell}(\nu) = \frac{C_{\ell}^{{\rm CIB}_{\nu} \times \phi}}{C_{\ell}^{\phi\phi}} \,.
\label{eq.CIBrescale}
\ee
Thus, we approximate the ISW-CIB-CIB bispectrum by computing Eq.~\ref{eq.BIyy} with $g(\nu) \tilde{y}_{\ell}(M,z) \rightarrow f_{\ell}(\nu) \tilde{\phi}_{\ell}(M,z)$, for both $\nu_2$ and $\nu_3$.  Clearly this approach will not yield percent-level accuracy, but it suffices to assess the order of magnitude of the bias on $f_{\rm NL}^{\rm loc}$.

\begin{figure}
\includegraphics[width=0.5\textwidth]{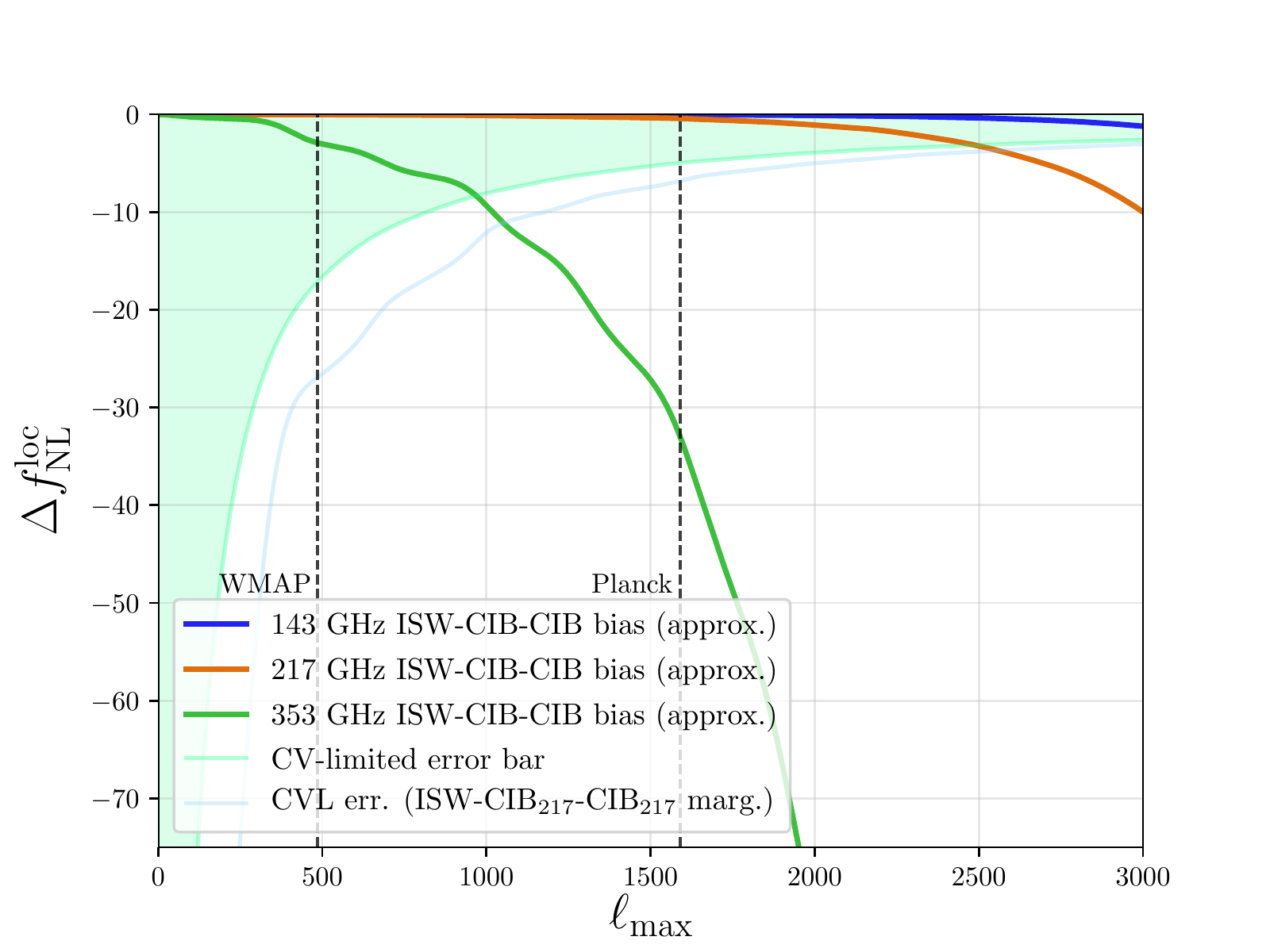} 
\caption{\label{fig:ISWCIBCIB_bias} Bias on $f_{\rm NL}^{\rm loc}$ from the (approximate) ISW-CIB-CIB bispectrum as a function of $\ell_{\rm max}$, as in Fig.~\ref{fig:ISWxphi_bias}.  The bias is shown for only three of the {\em Planck} HFI channels (143, 217, and 353 GHz), due to the approximate nature of the model used in this calculation (see \S\ref{sec:ISW-CIB-CIB}).  For {\em Planck}, the bias takes values of $\Delta f_{\rm NL}^{\rm loc} = -0.4$ (217 GHz) and $\Delta f_{\rm NL}^{\rm loc} = -33$ (353 GHz).  Due to its non-blackbody nature, this bias can be reduced via component separation, although the extent to which this reduction has occurred in the component-separated {\em Planck} CMB maps is currently unclear.  No multifrequency cleaning is assumed here.  The light green shaded region and dashed vertical lines are identical to those in Fig.~\ref{fig:ISWxphi_bias}.  The light blue shaded region shows the $1\sigma$ uncertainty on $f_{\rm NL}^{\rm loc}$ after marginalizing over the ISW-CIB-CIB bispectrum amplitude, evaluated at 217 GHz (as an example); this marginalization produces a non-negligible increase in the $f_{\rm NL}^{\rm loc}$ error bar ($\approx 40$\% increase for {\em Planck}).}
\end{figure}

Fig.~\ref{fig:ISWCIBCIB_bias} shows the bias on $f_{\rm NL}^{\rm loc}$ due to the ISW-CIB-CIB bispectrum, computed via Eqs.~\ref{eq.BIyy},~\ref{eq.CIBrescale}, and~\ref{eq.fNLbias}.  Due to the approximate nature of this calculation, we only show results for three of the {\em Planck} HFI channels (143, 217, and 353 GHz).  We do not plot any cross-frequency biases (i.e., involving different values of $\nu_2$ and $\nu_3$ in Eq.~\ref{eq.BIyy}), although these also exist.  In contrast to the ISW-tSZ-tSZ bias, the ISW-CIB-CIB bias is always negative, since the CIB signal is positive at all frequencies.  For {\em Planck}, the (approximate) ISW-CIB-CIB bias is $\Delta f_{\rm NL}^{\rm loc} = -0.4$ (217 GHz) and $\Delta f_{\rm NL}^{\rm loc} = -33$ (353 GHz).  The steep frequency dependence of this bias is expected due to the strong frequency dependence of the CIB.  The result is thus very sensitive to an assessment of the degree to which CIB emission has leaked into the component-separated {\em Planck} CMB maps.  For an experiment with $\ell_{\rm max} = 3000$, the ISW-CIB-CIB bias is many times larger than the $1\sigma$ error bar on $f_{\rm NL}^{\rm loc}$, even at 217 GHz.

Fig.~\ref{fig:ISWCIBCIB_bias} also shows the $1\sigma$ uncertainty on $f_{\rm NL}^{\rm loc}$ after marginalizing over the amplitude of the ISW-CIB-CIB bispectrum, computed with Eq.~\ref{eq.fNLerrormarg} (considering only 217 GHz, as an example case).  As for the ISW-tSZ-tSZ bispectrum in Fig.~\ref{fig:ISWyy_bias}, marginalizing over the ISW-CIB-CIB bispectrum amplitude noticeably inflates the error bar on $f_{\rm NL}^{\rm loc}$.  The ISW-CIB-CIB bispectrum shape is strongly correlated with the local bispectrum template (although not quite as strongly correlated as the ISW-tSZ-tSZ bispectrum), leading to the significant error bar increase.  For {\em Planck}, assuming no prior is placed on the ISW-CIB-CIB bispectrum amplitude, the uncertainty on $f_{\rm NL}^{\rm loc}$ increases by $\approx 40$\%.  Moreover, given astrophysical uncertainties in our modeling of the CIB signal, additional parameters (beyond the amplitude) may have to be marginalized over as well, further increasing the $f_{\rm NL}^{\rm loc}$ error bar.  Finally, unlike the tSZ signal, the CIB cannot be completely removed via component separation, due to decorrelation across frequencies; thus, some marginalization over residual contributions from the ISW-CIB-CIB signal must be necessary.  In order to avoid a significant increase in $\sigma(f_{\rm NL}^{\rm loc})$, a detailed understanding of the residual CIB emission in the cleaned CMB map is necessary, so that a strong prior can be placed on the residual ISW-CIB-CIB bispectrum before marginalizing.

\subsection{ISW-tSZ-CIB Bias}
\label{sec:ISW-tSZ-CIB}

Using the models described in the previous two subsections, we can readily compute the ISW-tSZ-CIB bispectrum via Eq.~\ref{eq.BIyy}.  We simply replace only one of the tSZ factors in Eq.~\ref{eq.BIyy} with $f_{\ell}(\nu) \tilde{\phi}_{\ell}(M,z)$, rather than both.  Again, we emphasize the approximate nature of the CIB model used here.

\begin{figure}
\includegraphics[width=0.5\textwidth]{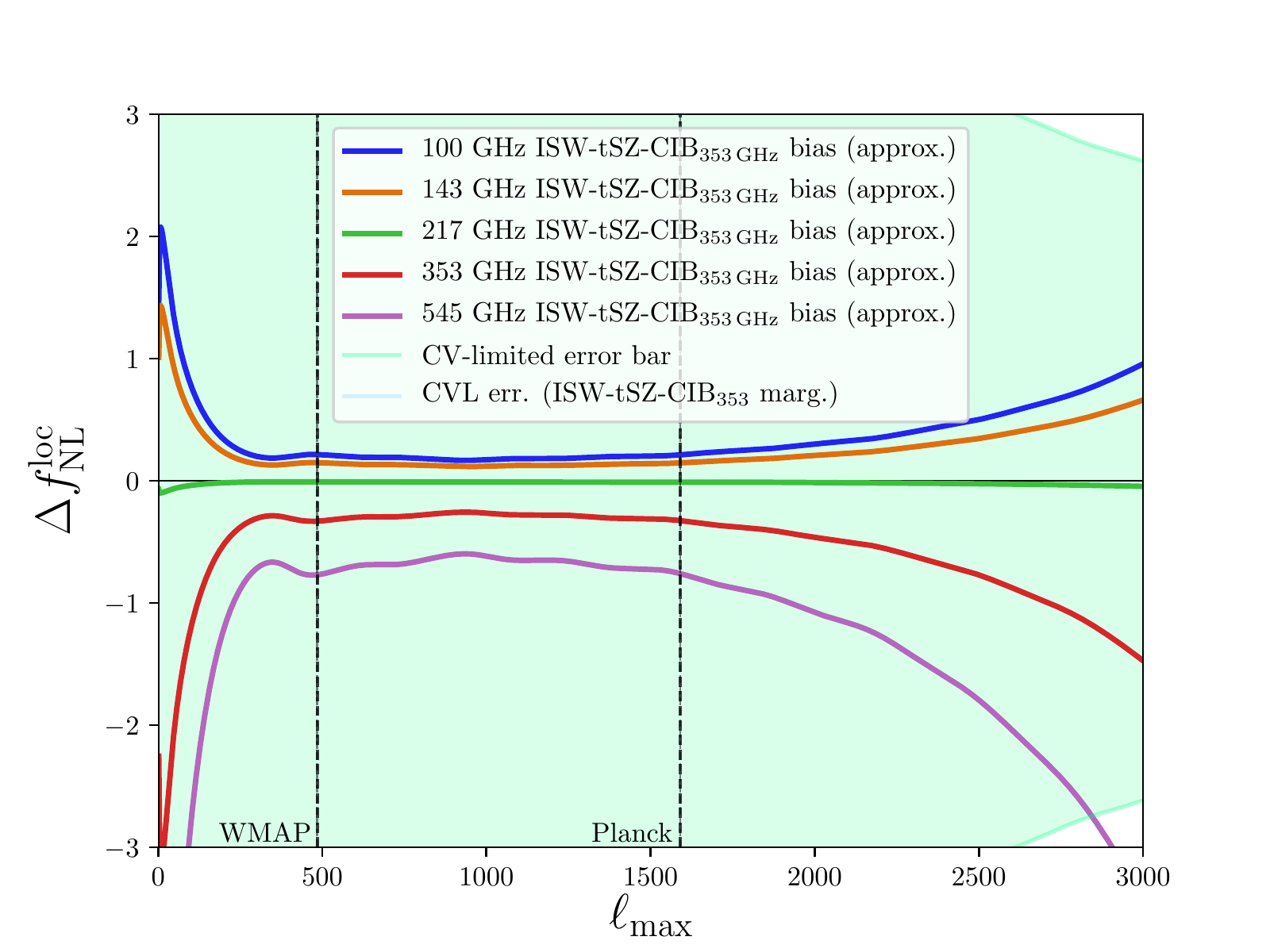} 
\caption{\label{fig:ISWtSZCIB_353_bias} Bias on $f_{\rm NL}^{\rm loc}$ from the (approximate) ISW-tSZ-CIB bispectrum as a function of $\ell_{\rm max}$, as in Fig.~\ref{fig:ISWxphi_bias}.  The bias is shown only for the CIB at 353 GHz, with the tSZ frequency varying over the {\em Planck} HFI channels from 100--545 GHz.  We re-emphasize the approximate nature of the CIB model used in this calculation (see \S\ref{sec:ISW-CIB-CIB}).  For {\em Planck}, the bias takes a value of $\Delta f_{\rm NL}^{\rm loc} = 0.3$ ($100 \times 353$ GHz), which should additionally be multiplied by a combinatorial factor of two.  Note that no multifrequency cleaning is assumed here.  The light green shaded region and dashed vertical lines are identical to those in Fig.~\ref{fig:ISWxphi_bias}.  The light blue shaded region shows the $1\sigma$ uncertainty on $f_{\rm NL}^{\rm loc}$ after marginalizing over the ISW-tSZ-CIB bispectrum amplitude, with the CIB evaluated at 353 GHz; although this is not visible on the plot, the marginalization produces a $\approx 70$\% increase in the $f_{\rm NL}^{\rm loc}$ error bar for {\em Planck}.}
\end{figure}

Fig.~\ref{fig:ISWtSZCIB_353_bias} shows the bias on $f_{\rm NL}^{\rm loc}$ due to the ISW-tSZ-CIB bispectrum, computed via Eqs.~\ref{eq.BIyy},~\ref{eq.CIBrescale}, and~\ref{eq.fNLbias}.  We show results only for a CIB frequency held fixed to 353 GHz, with the tSZ frequency varying over the {\em Planck} HFI channels from 100--545 GHz.  The bias can take on positive or negative values depending on the tSZ frequency considered, due to the behavior of the tSZ spectral function (and it vanishes at 217 GHz, as expected).  For {\em Planck}, the bias is generally small, e.g., $\Delta f_{\rm NL}^{\rm loc} = 0.3$ ($100 \times 353$ GHz), although note that an additional combinatorial factor of two should also be applied beyond this.  Due to the steep frequency dependence of the CIB, the bias will be much smaller when evaluating the CIB at any of the {\em Planck} frequencies below 353 GHz.  Even for an experiment with $\ell_{\rm max} = 3000$, the ISW-tSZ-CIB bias never approaches the $1\sigma$ error bar on $f_{\rm NL}^{\rm loc}$ for the main CMB channels (when considering the CIB at 353 GHz).  The most straightforward explanation for the smaller bias seen here in comparison to Figs.~\ref{fig:ISWyy_bias} and~\ref{fig:ISWCIBCIB_bias} is that the ISW-tSZ-CIB signal is simply smaller than the ISW-tSZ-tSZ and ISW-CIB-CIB signals, since it is suppressed by the tSZ-CIB correlation coefficient.  Note that if the tSZ signal is nulled via constrained component separation as suggested earlier, then this bias will be eliminated (in addition to all other biases involving the tSZ signal).

Although not visible on the plot, Fig.~\ref{fig:ISWtSZCIB_353_bias} also shows the $1\sigma$ uncertainty on $f_{\rm NL}^{\rm loc}$ after marginalizing over the amplitude of the ISW-tSZ-CIB bispectrum, computed with Eq.~\ref{eq.fNLerrormarg} (for the CIB at 353 GHz).  At high $\ell$, the increase in the error bar due to marginalization is small, but at lower $\ell$ values, it can be significant.  For {\em Planck}, if no prior is placed on the ISW-tSZ-CIB bispectrum amplitude, $\sigma(f_{\rm NL}^{\rm loc})$ increases by $\approx 70$\% after marginalization.  As emphasized earlier, this increase depends only on the correlation coefficient between the ISW-tSZ-CIB and local bispectra, and not on their amplitudes.  Without a precise theoretical calculation of the ISW-tSZ-CIB signal (which would allow a strong prior to be placed on its amplitude and therefore this error bar increase to be mitigated), this result strongly motivates the use of tSZ-nulled maps for NG analyses.

\subsection{ISW-kSZ-kSZ Bias}
\label{sec:ISW-kSZ-kSZ}

The final bias that we consider is that due to the cross-bispectrum of the ISW and kSZ effects.  Since bispectra involving odd numbers of kSZ fields vanish, the only such contribution is the ISW-kSZ-kSZ bispectrum.  Like the other bispectra considered in this section, this bispectrum can be thought of as the modulation of the small-scale kSZ power spectrum by a long-wavelength ISW mode.  However, unlike the bispectra involving the tSZ or CIB signals, this bispectrum is blackbody in frequency dependence.  Thus, like the lensing-ISW bias, it cannot be removed by multifrequency component separation techniques.  Its value must be computed and subtracted from any CMB temperature-based estimator for primordial NG, or it must be jointly fit and marginalized over in the NG analysis, at the cost of increased error bars on the primordial NG parameters.

The kSZ effect is generated by the Compton-scattering of CMB photons off free electrons moving with a net LOS velocity with respect to the CMB rest frame~\cite{Sunyaev-Zeldovich1972,Sunyaev-Zeldovich1980,Ostriker-Vishniac1986}.  To lowest order, this produces a Doppler boost in the CMB temperature:
\be
\frac{\Delta T^{\rm kSZ}(\hat{n})}{T_{\rm CMB}} = - \frac{1}{c} \int d\chi \, a(\chi) \, g(\chi) \, \vec{p}_e \cdot {\hat{n}} \,,
\label{eq.kSZdef}
\ee
where $g(\chi) = e^{-\tau} d\tau/d\chi$ is the visibility function, $\tau$ is the optical depth, and $\vec{p}_e = (1+\delta_e) \vec{v}_e$ is the electron momentum.  Here, $\delta_e \equiv (n_e - \bar{n}_e)/\bar{n}_e$ is the electron overdensity field, $n_e$ is the free electron number density, and $\vec{v}_e$ is the electron peculiar velocity field.

As in the previous subsections, we compute the ISW-kSZ-kSZ bispectrum in the halo model, considering only the two-``halo'' contribution that is expected to dominate in the squeezed limit.  However, in this case, our neglect of additional contributions (e.g., the three-halo term) may be less accurate than in the tSZ or CIB cases above.  For a robust assessment of the ISW-kSZ-kSZ bias on $f_{\rm NL}^{\rm loc}$ measurements, a full simulation-based calculation should be undertaken.  We treat the result here as a first estimate of the order of magnitude of the bias, but emphasize that it is likely to be an underestimate.

The ISW-kSZ-kSZ bispectrum is sourced by the ``hybrid bispectrum'' involving one density fluctuation and two LOS electron momenta, $B_{\delta p_{\hat{n}} p_{\hat{n}}}$.  Noting that the latter is $\vec{p}_e \approx \vec{v} \delta_e$ on small scales, we follow Refs.~\cite{Doreetal2004,DeDeoetal2005,Hilletal2016,Ferraroetal2016} in assuming that the term of the form $\langle v v \rangle \langle \delta \delta_e \delta_e \rangle$ dominates the hybrid bispectrum on the scales relevant to our analysis.  Thus, the hybrid bispectrum can be approximated as:
\be
B_{\delta p_{\hat{n}} p_{\hat{n}}} = \frac{1}{3} v_{\rm rms}^2 B_{\rm NL} \,,
\label{eq.Bhybrid}
\ee
where $v_{\rm rms}^2$ is the 3D velocity dispersion and $B_{\rm NL}$ is the nonlinear matter bispectrum (to be more precise, the cross-bispectrum of one matter density fluctuation and two electron density fluctuations).  We compute the velocity dispersion in linear theory, which has been shown to be an excellent approximation~\cite{Hahn2015,Hilletal2016,Ferraroetal2016}:
\be
v_{\rm rms}^2(z) = \frac{1}{2\pi^2} \int dk \left( f(z) a(z) H(z) \right)^2 P_{\rm lin}(k,z) \,,
\label{eq.vrms}
\ee
where $f \equiv d \ln D/d \ln a$ is the growth rate and we have used the continuity equation to relate the linear density and velocity fields.

The remaining ingredient left in the calculation is the matter bispectrum $B_{\rm NL}$.  As in the previous subsections, we consider only the two-``halo'' contribution, as this should dominate squeezed configurations in the ISW-kSZ-kSZ bispectrum.  Putting all of the factors together, the two-halo contribution to the ISW-kSZ-kSZ reduced bispectrum is:
\begin{widetext}
\ba
b_{\ell_1\ell_2\ell_3, ({\rm ISW-kSZ-kSZ})}^{({T_{\nu_1} T_{\nu_2} T_{\nu_3}}), 2h} & = & \int dz \, \left( \frac{v_{\rm rms}^2(z)}{3c^2} \right) \, \mathcal{I}_{\ell_1}(z) \, \left[ P_{\rm lin}\left(\frac{\ell_1+\frac{1}{2}}{\chi(z)}\right) \int dM \, \frac{dn}{dM} b(M,z) \tilde{\tau}_{\ell_2}(M,z) \tilde{\tau}_{\ell_3}(M,z) \right. \nonumber \\
& + & P_{\rm lin}\left(\frac{\ell_2+\frac{1}{2}}{\chi(z)}\right) \int dM \, \frac{dn}{dM} b(M,z) \tilde{\tau}_{\ell_2}(M,z) \int dM' \, \frac{dn}{dM'} b(M',z) \tilde{\tau}_{\ell_3}(M',z) \frac{M'}{\bar{\rho}_m} \nonumber \\
& + & \left. P_{\rm lin}\left(\frac{\ell_3+\frac{1}{2}}{\chi(z)}\right) \int dM \, \frac{dn}{dM} b(M,z) \tilde{\tau}_{\ell_3}(M,z) \int dM' \, \frac{dn}{dM'} b(M',z) \tilde{\tau}_{\ell_2}(M',z) \frac{M'}{\bar{\rho}_m} \right] \nonumber \\
& + & 2 \,\, {\rm perm.} \,,
\label{eq.BIkk}
\ea
\end{widetext}
where $\mathcal{I}_{\ell}(z)$ is given by Eq.~\ref{eq.ISWfac} and $\tilde{\tau}_{\ell}(M,z)$ is the Fourier transform of the optical depth profile of a halo of mass $M$ at redshift $z$:
\ba
\tilde{\tau}_{\ell}(M,z) & = & \sigma_T \frac{4 \pi
  r_{s,\tau}}{\ell_{s,\tau}^2} \int dx_{\tau} \, x_{\tau}^2 \frac{\sin((\ell+1/2)
  x_{\tau}/\ell_{s,\tau})}{(\ell+1/2) x_{\tau}/\ell_{s,\tau}} \nonumber \\
  & & \times \, n_{e}(x_{\tau} r_{s,\tau},M,z) \,.
\label{eq.tauprof}
\ea
In analogy with Eqs.~\ref{eq.yprof} and~\ref{eq.phiprof}, $r_{s,\tau}$ is a characteristic scale radius of the electron number density profile, $\ell_{s,\tau} = a(z)\chi(z)/r_{s,\tau} = d_A(z)/r_{s,\tau}$ is the multipole moment associated with this scale, and $x_\tau \equiv r/r_{s,\tau}$ is a dimensionless radial variable for the electron number density profile.  For simplicity, we assume that the electron number density profile of each halo follows the NFW profile, which is rescaled appropriately from matter density to electron number density assuming a baryon fraction equal to the cosmological value, a mean molecular weight per electron of 1.14 (i.e., a primordial composition of H and He), and a free electron fraction of 0.85.  Our calculation is not particularly sensitive to the details of the profile given the range of angular scales involved in the calculation, but fitting functions from hydrodynamical simulations could be used for improved accuracy~\cite{Battaglia2016tau}.

\begin{figure}
\includegraphics[width=0.5\textwidth]{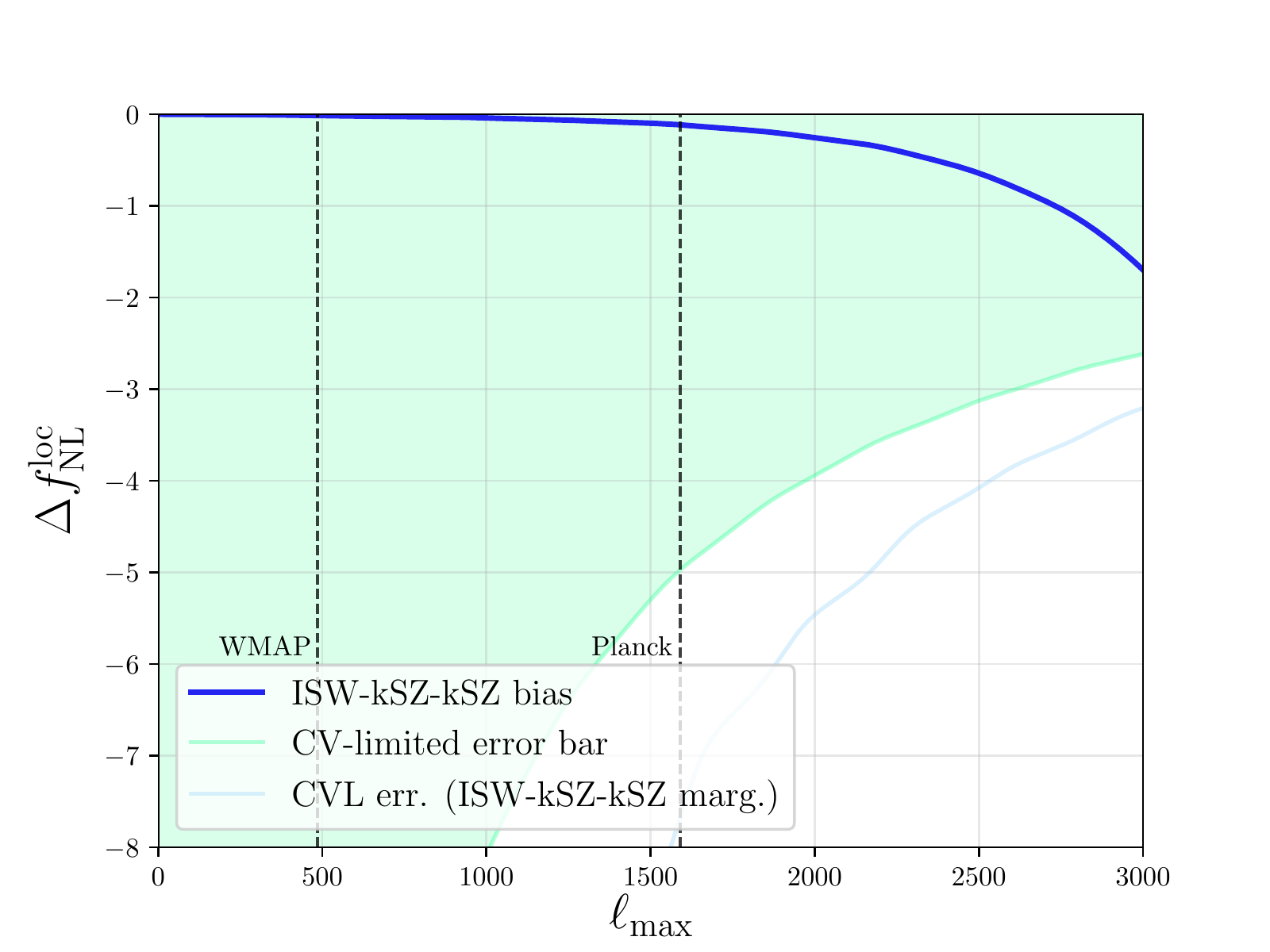} 
\caption{\label{fig:ISWkSZkSZ_bias} Bias on $f_{\rm NL}^{\rm loc}$ from the ISW-kSZ-kSZ bispectrum as a function of $\ell_{\rm max}$, as in Fig.~\ref{fig:ISWxphi_bias}.  The bias is computed considering only the contribution in Eq.~\ref{eq.BIkk}; we caution that this approximation may neglect important contributions, and thus the results shown here should be considered an underestimate.  For {\em Planck}, the bias estimated here is $\Delta f_{\rm NL}^{\rm loc} = -0.1$.  Like the lensing-ISW bias, the ISW-kSZ-kSZ bias is blackbody in frequency dependence and cannot be removed by multifrequency component separation methods.  Thus, it must be computed and subtracted from $f_{\rm NL}^{\rm loc}$ measurements.  The light green shaded region and dashed vertical lines are identical to those in Fig.~\ref{fig:ISWxphi_bias}.  The light blue shaded region shows the $1\sigma$ uncertainty on $f_{\rm NL}^{\rm loc}$ after marginalizing over the ISW-kSZ-kSZ bispectrum amplitude; this marginalization produces a non-negligible increase in the $f_{\rm NL}^{\rm loc}$ error bar ($\approx 50$\% increase for {\em Planck}).}
\end{figure}

Fig.~\ref{fig:ISWkSZkSZ_bias} shows the bias on $f_{\rm NL}^{\rm loc}$ due to the ISW-kSZ-kSZ bispectrum, computed via Eqs.~\ref{eq.BIkk},~\ref{eq.vrms}, and~\ref{eq.fNLbias}.  The bias is frequency-independent in CMB blackbody temperature units.  It is always of negative sign, and therefore correcting for it will increase the inferred value of $f_{\rm NL}^{\rm loc}$.  For {\em Planck}, the bias is $\Delta f_{\rm NL}^{\rm loc} = -0.1$, which is significantly smaller than the statistical error bar.  For an experiment with $\ell_{\rm max} = 3000$, the bias approaches the $1\sigma$ error bar on $f_{\rm NL}^{\rm loc}$.  In this context, we emphasize that this calculation is missing potentially non-negligible contributions, and the true bias could be somewhat larger than estimated here.

There are two main reasons to explain the relative size of the ISW-tSZ-tSZ and ISW-kSZ-kSZ biases.  First, the small-scale tSZ power spectrum appears to be a factor of $\approx 2-4$ larger than the kSZ power spectrum (at frequencies where the tSZ spectral function is near unity, e.g., 150 GHz)~\cite{Sieversetal2013,Georgeetal2015}, although the constraints on this ratio remain weak.  Second, more of the small-scale kSZ power is generated at $z>1$ than the tSZ power~\cite{Shawetal2012,Tracetal2011,BBPS2012b}; thus, the correlation of the kSZ power spectrum with the ISW fluctuations (which predominantly arise at $z<1$) is correspondingly weaker than for the tSZ power spectrum.  In combination, these two effects suppress the ISW-kSZ-kSZ bispectrum by nearly an order of magnitude compared to the ISW-tSZ-tSZ bispectrum, leading to a smaller bias on $f_{\rm NL}^{\rm loc}$.

However, while Eq.~\ref{eq.BIyy} likely includes effectively all relevant contributions to squeezed configurations of the ISW-tSZ-tSZ bispectrum, this may not be true for Eq.~\ref{eq.BIkk} and the squeezed ISW-kSZ-kSZ bispectrum.  In particular, for ``moderately'' squeezed configurations, the approximation in Eq.~\ref{eq.Bhybrid} may not be particularly accurate (analogously, this type of approximation only suffices to compute the kSZ power spectrum accurately on very small scales~\cite{Ma-Fry2002}).  Contributions from the three-halo term may also be non-negligible in the moderately squeezed regime.  Finally, our calculation does not account for correlations that arise between the velocity field and the ISW field due to the change in the growth factor (which sources the velocity field) in the presence of a large-scale void or overdensity.  Thus, Fig.~\ref{fig:ISWkSZkSZ_bias} should only be taken as a very approximate estimate of the order of magnitude of the ISW-kSZ-kSZ bias on $f_{\rm NL}^{\rm loc}$.  A complete calculation, ideally derived from numerical simulations, should be performed to verify the robustness of current NG estimates to this bias.  Simulations with the relevant properties have already been constructed~(e.g.,~\cite{Jubilee1,Jubilee2}), and thus there should be no major obstacle to such a calculation.

Finally, Fig.~\ref{fig:ISWkSZkSZ_bias} also shows the $1\sigma$ uncertainty on $f_{\rm NL}^{\rm loc}$ after marginalizing over the amplitude of the ISW-kSZ-kSZ bispectrum, computed with Eq.~\ref{eq.fNLerrormarg}.  At all $\ell$ values considered, the increase in the error bar due to marginalization is non-negligible.  For {\em Planck}, if no prior is placed on the ISW-kSZ-kSZ bispectrum amplitude, $\sigma(f_{\rm NL}^{\rm loc})$ increases by $\approx 50$\% after marginalization.  As emphasized above, this increase depends only on the correlation coefficient between the ISW-kSZ-kSZ and local bispectra, and not on their amplitudes.  Since the kSZ signal cannot be removed by multifrequency component separation methods, the only option for mitigating this problem in NG analyses is to perform a detailed theoretical calculation of the ISW-kSZ-kSZ bispectrum.  One can then choose whether to place a strong prior on its amplitude (and shape) when jointly analyzing bispectrum templates in the NG analysis (so as not to incur a significant penalty on $\sigma(f_{\rm NL}^{\rm loc})$ when marginalizing), or to directly subtract the theoretically computed bias on $f_{\rm NL}^{\rm loc}$, and not attempt to marginalize at all.  In either case, additional theoretical or simulation work is needed to obtain robust constraints on primordial NG.

\section{Discussion}
\label{sec:discussion}
The results in Figs.~\ref{fig:tSZxphi_bias},~\ref{fig:CIBxphi_bias}, and~\ref{fig:ISWyy_bias}--\ref{fig:ISWkSZkSZ_bias} (summarized in Table~\ref{tab:biases}) suggest that biases due to extragalactic foregrounds may indeed be large enough to be a worry for the {\em Planck} NG constraints, and are clearly a worry for future constraints.\footnote{However, note that polarization is generally free of these foregrounds, and thus offers a robust route forward.}  To take two examples from the preceding sections, for {\em Planck} ($\ell_{\rm max} = 1590$) the lensing-CIB bias is $\Delta f_{\rm NL}^{\rm loc} = 4.3$ at 217 GHz and the ISW-tSZ-tSZ bias is $\Delta f_{\rm NL}^{\rm loc} = -4.5$ at 100 GHz.  If the effective $\ell_{\rm max}$ for {\em Planck} were only slightly larger (e.g., 2000), some of these biases would have likely been noticeable above the statistical uncertainty.  Of course, the non-blackbody biases are reduced to some extent by component separation; for biases that involve two non-blackbody ``legs'' (e.g., ISW-tSZ-tSZ), the reduction is more efficient (e.g., a $50$\% scale-independent reduction of the tSZ signal would suppress the ISW-tSZ-tSZ bias by a factor of four).

The overall, combined effect of the biases computed in this paper is difficult to estimate without performing a detailed calculation that includes the exact weights applied to the frequency maps in the component separation algorithms.  Moreover, in most component separation methods, the weights vary as a function of angular scale and pixel location, and thus the $\ell$-dependences of the non-blackbody foreground bispectra will be modified.  This will affect the associated biases on $f_{\rm NL}^{\rm loc}$, since the shapes of the bispectra will be modified.  For the SMICA component separation method, this calculation could possibly be done analytically, since the weights vary only as a function of $\ell$ (see Fig.~D.1 of~Ref.~\cite{Planck2015compsep}).  However, including these effects analytically is challenging for methods whose weights vary in pixel space.  Thus, numerical simulations are likely a better approach, but this requires the construction of simulations with correlations amongst the relevant fields.

Moreover, the modeling of the secondary anisotropy fields needed to capture these biases via simulations is not as straightforward as calculating the ISW-lensing bias, for which linear theory suffices~\cite{Junk-Komatsu2012}.  The tSZ, kSZ, and CIB fields are all affected by complex baryonic physics.  Some of the biases computed in this paper can likely be modeled at better than $\lesssim 10$\% accuracy given current knowledge, e.g., the ISW-tSZ-tSZ bias, which arises from relatively low-redshift halos whose pressure profiles are well-constrained~(e.g.,~\cite{Arnaudetal2010,Planck2012stack}).  But this may not be true for other contributions, e.g., kSZ-related biases, which depend on the distribution of ionized gas around relatively low-mass halos~(e.g.,~\cite{FH2018}).  Thus, modeling uncertainty for these biases will need to be carefully investigated, and may ultimately have to be included in the final error budget on the primordial NG parameters.  The best approach may be to simply measure as many of the contributions as possible directly from the data; if the $S/N$ is sufficiently high, then the measured signal can be directly used to calculate the bias on primordial NG, as in the lensing-CIB calculation presented in \S\ref{sec:CIB-phi}.  However, for the ISW-related bispectra, the CV-limited $S/N$ on these measurements is likely not high (as for the ISW-lensing bispectrum).  Note that the CV on these bispectra should be propagated into the final error bar on $f_{\rm NL}^{\rm loc}$ if theoretically computed biases are subtracted in the NG analysis.

As discussed throughout the preceding sections, one can instead modify the primordial NG data analysis by simultaneously including the foreground bispectrum templates in the model and marginalizing over their amplitudes.\footnote{This procedure essentially orthogonalizes the primordial NG estimators with respect to the foreground bispectra, analogous to the use of ``bias-hardened'' CMB lensing reconstruction estimators~\cite{Namikawa2013,Namikawa-Takahashi2014}.  Note that performing the primordial NG analysis on component-separated maps in which the tSZ and/or CIB signals have been nulled is one form of such bias-hardening.}  This procedure assumes that the shape of the foreground bispectra are known {\em a priori}, which is not generally the case (see the discussion in the previous paragraph).  Even when marginalizing solely over the amplitudes of the foreground bispectra, the error bar on $f_{\rm NL}^{\rm loc}$ can nevertheless still increase substantially.  The ISW-related bispectra discussed in \S\ref{sec:ISW} are noteworthy in this respect; for {\em Planck}, if no priors are placed on the amplitudes of these bispectra, the error bar on $f_{\rm NL}^{\rm loc}$ increases by $\approx 50$\%.  This increase is due to the fact that these foreground bispectrum templates are highly correlated with the shape of the local-type bispectrum.

\begin{figure}
\includegraphics[width=0.5\textwidth]{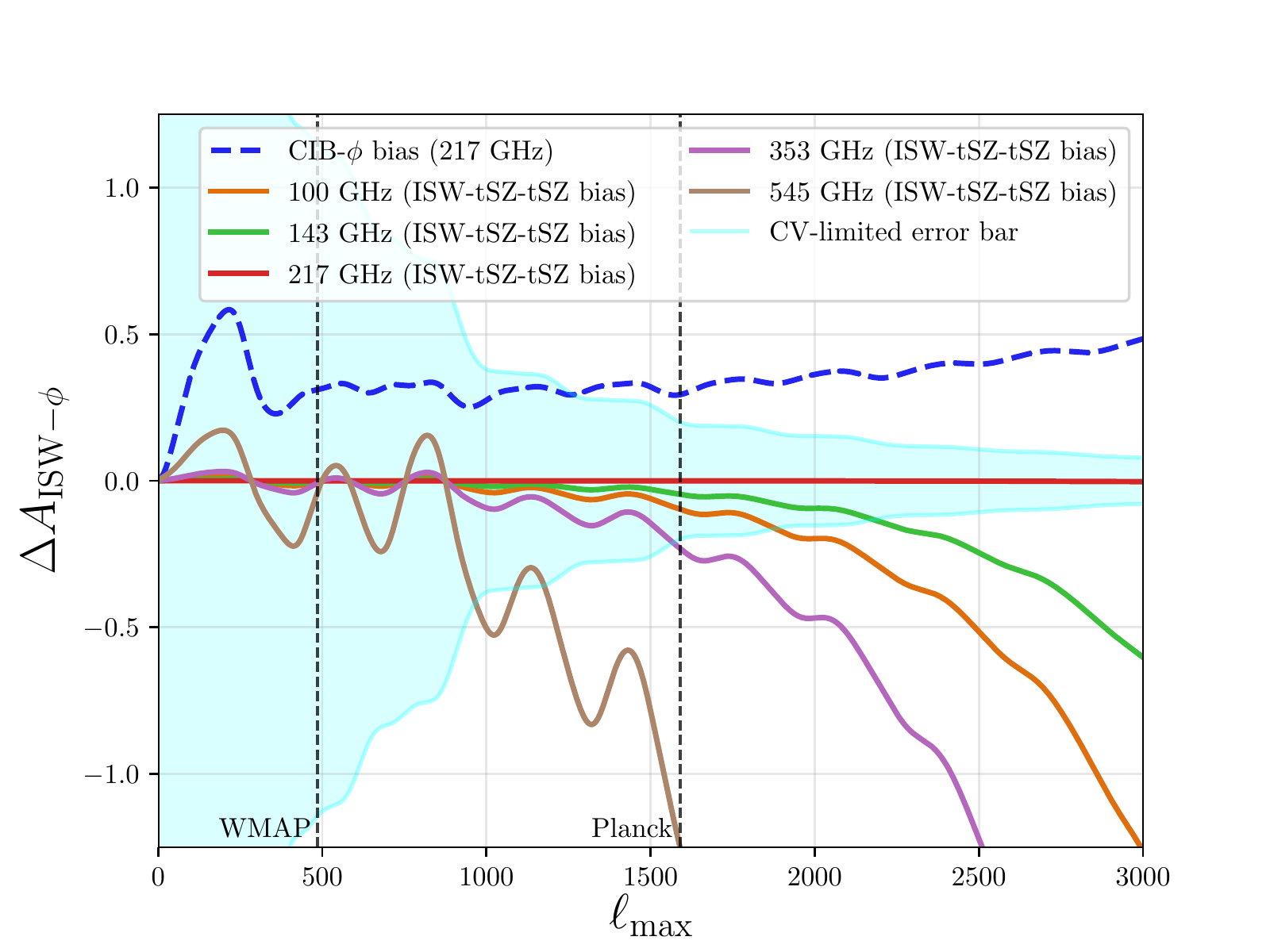}
\caption{\label{fig:A_ISWphi_biases} Bias on the amplitude of the ISW-CMB lensing bispectrum, $A_{{\rm ISW}-\phi}$, from the ISW-tSZ-tSZ bispectrum (solid curves) and the CIB-lensing bispectrum (dashed curve) as a function of $\ell_{\rm max}$, analogous to the $f_{\rm NL}^{\rm loc}$ biases computed earlier in the paper.  For the ISW-tSZ-tSZ bispectrum, the bias is shown for the {\em Planck} HFI channels from 100--545 GHz, with a frequency dependence arising from the tSZ spectral function.  For the CIB-lensing bispectrum, the bias is shown only for the {\em Planck} 217 GHz channel, for clarity.  With no multifrequency-cleaning mitigation, these non-blackbody biases would be marginally detectable in the {\em Planck} measurement of the ISW-CMB lensing bispectrum (disregarding the 545 GHz result).  The light shaded region shows the $1\sigma$ uncertainty on $A_{{\rm ISW}-\phi}$ as a function of $\ell_{\rm max}$ using only information in the CMB temperature bispectrum for a full-sky, CV-limited experiment.  The dashed vertical lines indicate the effective $\ell_{\rm max}$ for WMAP9~\cite{Hinshawetal2013} and {\em Planck} 2015~\cite{Planck2015NG}.}
\end{figure}

A question that is clearly related to the orthogonality of the foreground bispectra and primordial bispectra is the extent to which the foreground biases considered in this paper would also affect the measured amplitude of the ISW-lensing bispectrum, $A_{{\rm ISW}-\phi}$.  If the biases on this amplitude were large, its value could be used as a cross-check for {\em Planck} or other experiments.  Fig.~\ref{fig:A_ISWphi_biases} shows the result of this calculation for two representative examples of the foreground bispectra considered in this paper.  To obtain these results, we simply evaluate Eq.~\ref{eq.fNLbias} with the replacement $B^{\rm loc} \rightarrow B^{{\rm ISW} \times \phi}$.  We consider the ISW-tSZ-tSZ and lensing-CIB bispectra as contaminants for this exercise.  We also compute the Gaussian error bar on $A_{{\rm ISW}-\phi}$ using Eq.~\ref{eq.fNLerror}.  We find that the ISW-lensing bispectrum is likely only a useful diagnostic for extreme foreground contamination, e.g., the ISW-tSZ-tSZ bispectrum evaluated at 545 GHz.  At the main CMB channels, the biases are comparable to or smaller than the error bar on $A_{{\rm ISW}-\phi}$.  Thus, consistency with $A_{{\rm ISW}-\phi} = 1$ is not a robust guarantee against non-negligible foreground biases on $f_{\rm NL}^{\rm loc}$.  In this context, it is interesting to note that in Table 2 of the {\em Planck} 2015 NG analysis~\cite{Planck2015NG}, all of the temperature-based estimators for $A_{{\rm ISW}-\phi}$ return values less than unity (albeit only at $1-1.5\sigma$ significance), perhaps providing a weak indication that residual foregrounds are present.

In this paper, we have only focused on the most obviously relevant terms for $f_{\rm NL}^{\rm loc}$.  There are other foreground bispectra that do not peak in squeezed configurations, but may nonetheless project onto the local template to some extent, e.g., the tSZ-kSZ-kSZ or CIB-kSZ-kSZ bispectra.  A numerical simulation-based approach could simultaneously capture the influence of all contributions.  Moreover, similar foreground biases also exist for the other primordial bispectrum shapes, i.e., the equilateral and orthogonal templates; we provide analogous (but not exhaustive) calculations for these shapes in Appendices~\ref{app:equ} and~\ref{app:orth}.  Finally, similar biases are guaranteed to exist for estimators of primordial NG at the trispectrum level ($g_{\rm NL}$ and $\tau_{\rm NL}$).  These include blackbody contributions that cannot be removed by component separation, such as the ISW-ISW-kSZ-kSZ trispectrum, the lensing-kSZ-kSZ trispectrum, and the kSZ auto-trispectrum.  These terms must be computed and subtracted to obtain unbiased constraints on $g_{\rm NL}$ and $\tau_{\rm NL}$.

We close by noting that the effects considered in this paper are an excellent example of a situation in which component separation should be performed so as to explicitly remove foregrounds that could bias a particular analysis, even at the cost of somewhat increased statistical noise.\footnote{CMB lensing reconstruction is another such example.}  In particular, explicit nulling of the tSZ signal would clearly be beneficial in this instance, and a fiducial CIB spectrum could be nulled as well.  In general, component separation should not be viewed as a homogeneous tool; methods should be adapted and optimized for particular analysis requirements as needed.

Considering these issues more quantitatively, we can estimate the extent to which the tSZ and CIB signals must be removed such that the associated biases on $f_{\rm NL}^{\rm loc}$ are less than some fraction of the statistical error bar, e.g., $0.1\sigma(f_{\rm NL}^{\rm loc})$.  For the lensing-tSZ and ISW-tSZ-tSZ biases, the tSZ signal in the final CMB map must be reduced by a factor of $\approx 3$ relative to its 100 GHz amplitude to satisfy this criterion for {\em Planck} (with ISW-tSZ-tSZ being more stringent); for {\em SO} or {\em CMB-S4} (assuming $f_{\rm sky} = 0.4$), the necessary reduction is a factor of $\approx 10$.  Of course, the tSZ effect can be exactly nulled since its frequency dependence is known from first principles, so these reduction factors are feasible to achieve, but they do place requirements on instrumental systematics, such as relative gain calibration between frequency channels.  For the lensing-CIB and ISW-CIB-CIB biases, the CIB signal in the final CMB map must be reduced by a factor of $\approx 10$ relative to its 217 GHz amplitude to satisfy the bias criterion described above for {\em Planck} (with lensing-CIB being more stringent); for {\em SO} or {\em CMB-S4} (assuming $f_{\rm sky} = 0.4$), the necessary reduction is a factor of $\approx 25$.  Whether such a significant reduction can be achieved in practice is an open question, particularly given our current lack of knowledge about decorrelation of the CIB across frequencies on small scales~\cite{Planck2013CIB,Maketal2017}.  Finally, we re-emphasize that the blackbody biases due to ISW, lensing, and kSZ cannot be removed via multifrequency component separation.

A related issue is the range of angular scales that is most important to clean in order to suppress the non-blackbody biases.  In general, the tSZ- and CIB-related biases are dominated by the smallest-scale modes to which the experiment is sensitive, as these foreground contributions only become comparable to the CMB signal at high-$\ell$.  Thus, these are likely the most important modes to clean.  However, in this regime, component separation algorithms must contend with rapidly increasing noise power spectra.  For algorithms that seek to minimize an overall variance criterion (i.e., with no explicit nulling of any particular signal), this is likely responsible for the leakage of secondary foregrounds --- which are subdominant to the noise --- into the final map.

Thus, explicitly nulling the non-blackbody foregrounds is likely to be worthwhile in primordial NG analyses relying on temperature data, despite the associated penalty in statistical sensitivity that must be paid.  For {\em Planck}, the increase in noise when nulling the tSZ signal is not particularly severe.  Ref.~\cite{MH2018} compares the the noise power spectra of component-separated {\em Planck} CMB maps that do (LGMCA) or do not (SMICA) null the tSZ signal (the methods also have algorithmic differences).  The LGMCA noise power is only $22\%$ larger than the SMICA noise power at $\ell=2000$ (corresponding to a $\lesssim 10$\% decrease in $\ell_{\rm max}$), thus demonstrating that the penalty for nulling tSZ is not large.  The reason for this small increase is that the 217 GHz noise power spectrum is almost as low as the 143 GHz noise power spectrum in {\em Planck}; thus, the 143 GHz channel can be used to remove tSZ, with the 217 GHz channel still available for measuring the CMB, with only a small noise penalty.  For ground-based experiments, the situation is more challenging because of the large atmospheric noise contribution at high frequencies (including 217 GHz).  Thus, a larger statistical penalty may have to be paid.  For the CIB, as long as a high-frequency channel is included that measures the CIB with high $S/N$ (e.g., the 353, 545, or 857 GHz channels in Planck, or a 270 GHz channel from the ground), this signal can be nulled in component separation with little penalty, due to the very steep CIB SED.  However, this requires the assumption of a CIB SED model and the assumption that the CIB is fully correlated across frequencies.  These assumptions are tenable at the $10\%$ level, but may not hold at the $1\%$ level.  Overall, we conclude that for {\em Planck}, nulling the tSZ and CIB signals can likely be done without drastically lowering $\ell_{\rm max}$, i.e., with only a small penalty in the error bar on $f_{\rm NL}^{\rm loc}$ ($\mathcal{O}$(10s)\%).  For {\em SO} and {\em CMB-S4}, this may not be the case, further motivating the use of polarization for primordial NG constraints with these experiments.

\section{Conclusions}
\label{sec:conclusions}

In this paper, we have considered in detail the role of extragalactic foregrounds in biasing measurements of local-type primordial NG from the CMB temperature bispectrum, including contributions that have not been considered previously.  Some of the contributions are non-blackbody in nature, and can thus be reduced by component separation methods, but the extent of this reduction in the {\em Planck} analysis is currently unclear, with evidence suggesting that extragalactic foregrounds have leaked into the {\em Planck} CMB maps~\cite{MH2018,Chenetal2018}.  It is also worth noting that none of these biases are present in the {\em Planck} FFP8 simulations (except for the standard ISW-lensing bias)~\cite{Planck2015FFP8}, which are used to verify the {\em Planck} NG analysis pipelines.  In addition, amongst these biases, only the ISW-lensing contribution is considered in the {\em Planck} 2015 NG analysis~\cite{Planck2015NG}.  For future experiments, the foreground biases are much larger than the statistical error bar on $f_{\rm NL}^{\rm loc}$ and cannot be neglected; moreover, the non-blackbody biases impose stringent  requirements on the component separation accuracy (see \S\ref{sec:discussion}).

Although the largest blackbody contribution (ISW-kSZ-kSZ) is unlikely to be large enough to significantly bias the {\em Planck} constraint on $f_{\rm NL}^{\rm loc}$, it appears possible that residual tSZ and CIB signal could lead to biases that are large enough to shift the inferred value of $f_{\rm NL}^{\rm loc}$ by $\approx 1\sigma$ (we find similar results for orthogonal-type NG in Appendix~\ref{app:orth}).  As discussed in the previous section, the overall foreground bias is sensitive to the component separation details, and thus we do not attempt an estimate here.  A conservative conclusion is that a foreground systematic error bar of order the current statistical error bar should be assigned to the inferred value of $f_{\rm NL}^{\rm loc}$.  Thus, a central value as large as $f_{\rm NL}^{\rm loc} \sim 10$ is still plausible.

Alternatively, instead of treating these foreground contributions as biases, the foreground bispectrum templates can be included in the NG analysis, and their amplitudes can be marginalized over when constraining primordial NG parameters.  For lensing-related bispectra, the marginalization does not noticeably increase the error bar on $f_{\rm NL}^{\rm loc}$, but for ISW-related bispectra, the increase can be substantial.  This increase can be avoided by either performing detailed theoretical calculations that allow strong priors to be placed on the foreground bispectrum amplitudes (and shapes), or by relying on multifrequency mitigation techniques (for non-blackbody foreground bispectra).  A notable case is the ISW-kSZ-kSZ bispectrum, which is blackbody in frequency dependence, and which increases the {\em Planck} error bar on $f_{\rm NL}^{\rm loc}$ by $\approx 50$\% after marginalization.  Additional study of this foreground bispectrum is clearly needed.

Two paths are available for overcoming the non-blackbody biases computed in this paper: (1) the tSZ and (an assumed) CIB spectral dependences can be explicitly nulled in the component separation process, yielding more robust constraints on $f_{\rm NL}^{\rm loc}$ at some cost in statistical constraining power; (2) sky simulations containing all relevant signals, including correlations amongst them, can be processed through the component separation and NG analysis pipelines in order to robustly assess the foreground biases.  Note that some CIB signal will always propagate through the first approach, due to decorrelation across frequencies, but a large fraction can likely be removed explicitly.  Blackbody biases will also persist in the first approach.  The second approach has the advantage of simultaneously capturing the blackbody and non-blackbody biases.  In addition, for trispectrum NG biases, a simulation-based approach may simply be more efficient than computing all of the relevant terms analytically.  Both paths are likely to be of use moving forward.

We conclude that the search for evidence of primordial NG in the {\em Planck} data may not yet be complete.

\begin{acknowledgments}
I am grateful to Kendrick Smith for many helpful interactions, and to the anonymous referee for suggestions that improved the paper.  I also thank David Spergel and Matias Zaldarriaga for useful conversations and comments on the manuscript.  I thank Eiichiro Komatsu for publicly distributing the codes in the Cosmology Routine Library~\cite{CRL}, some of which were beneficial to this work.  JCH is supported by the Friends of the Institute for Advanced Study.
\end{acknowledgments}

\begin{appendix}
\section{Foreground Biases on Equilateral-Type Primordial non-Gaussianity}
\label{app:equ}

In this appendix, we provide foreground bias results for equilateral-type primordial NG analogous to those given in the main text for local-type NG.  Equilateral NG is a unique probe of the physics of the early universe; in the context of inflation, it can be generated by self-interactions of the inflaton field, amongst other possibilities~(e.g.,~\cite{SilversteinTong2004,DBI2004,Creminellietal2006,BaumannGreen2011}).  As the name implies, this bispectrum signal peaks for equilateral triangle configurations in momentum space.  As for local NG, the {\em Planck} 2015 CMB anisotropy data yield the tightest current constraint on the amplitude of equilateral NG, $f_{\rm NL}^{\rm equ}$: $f_{\rm NL}^{\rm equ} = -16 \pm 70$ (temperature data only) or $f_{\rm NL}^{\rm equ} = -4 \pm 43$ (temperature and polarization data)~\cite{Planck2015NG}.  Note that the polarization data (particularly mixed bispectra of temperature and $E$-modes) are somewhat more constraining here than in the local NG analysis, providing a comparable constraint to that derived from temperature alone.  In the following, we compute foreground biases (and foreground-marginalized error bars) for $f_{\rm NL}^{\rm equ}$ inferred from the temperature bispectrum alone.

\begin{figure}
\includegraphics[width=0.5\textwidth]{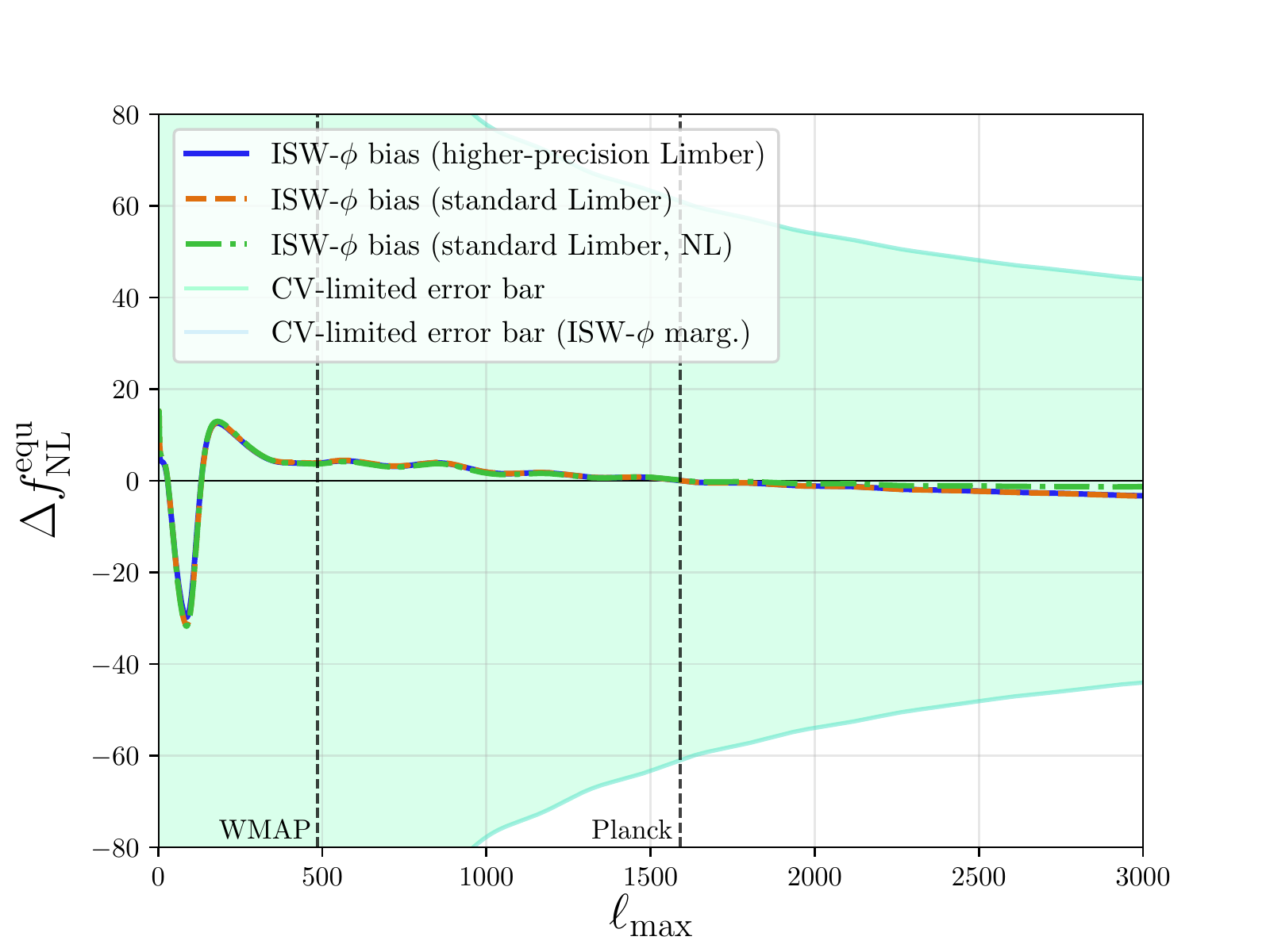} 
\caption{\label{fig:ISWxphi_bias_equ} Bias on $f_{\rm NL}^{\rm equ}$ from the lensing-ISW bispectrum for an experiment that is CV-limited to a maximum multipole $\ell_{\rm max}$, analogous to Fig.~\ref{fig:ISWxphi_bias} for $f_{\rm NL}^{\rm loc}$.  All curves and shaded regions are identical in meaning to those in Fig.~\ref{fig:ISWxphi_bias}, with the exception of the green dot-dashed curve, which shows the effect of using non-linear theory to compute the lensing-ISW bispectrum.}
\end{figure}

\begin{figure}
\includegraphics[width=0.5\textwidth]{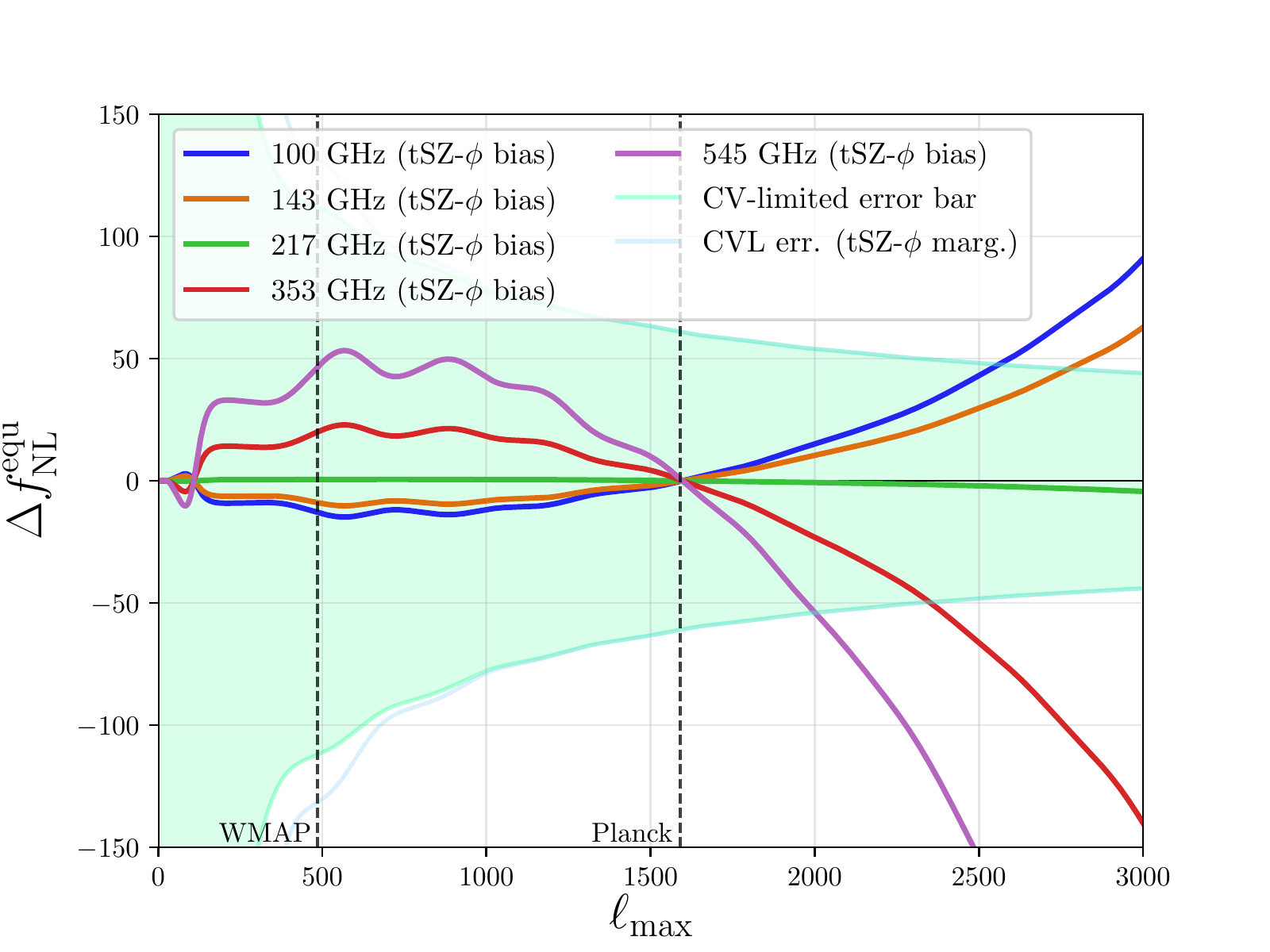} 
\caption{\label{fig:tSZxphi_bias_equ} Bias on $f_{\rm NL}^{\rm equ}$ from the lensing-tSZ bispectrum for an experiment that is CV-limited to a maximum multipole $\ell_{\rm max}$, analogous to Fig.~\ref{fig:tSZxphi_bias} for $f_{\rm NL}^{\rm loc}$.  All curves and shaded regions are identical in meaning to those in Fig.~\ref{fig:tSZxphi_bias}.}
\end{figure}

Here, we repeat all of the calculations presented in \S\ref{sec:lensing} and~\S\ref{sec:ISW}, but with the replacement $B^{\rm loc} \rightarrow B^{\rm equ}$, i.e., the local bispectrum is replaced by the equilateral bispectrum.  We calculate $B^{\rm equ}$ via the explicit formula given in Eq.~22 of Ref.~\cite{Creminellietal2006}.  We follow the guidance of Ref.~\cite{Senatoreetal2010} in this calculation, extending the integral over comoving distance to $\chi_* + 500 \, {\rm Mpc}/h$ (rather than the usual upper limit of $\chi_*$) in order to obtain convergence.

We implement the formalism of \S\ref{sec:fNL} to compute foreground biases and foreground-marginalized error bars on $f_{\rm NL}^{\rm equ}$ using the models described in \S\ref{sec:lensing} and \S\ref{sec:ISW}.  The general conclusion of these calculations is that for the current {\em Planck} analysis, none of these foreground bispectra are a major concern; however, the lensing-tSZ and lensing-CIB biases are potentially large for future experiments with $\ell_{\rm max}=3000$.  Moreover, it is important to note that we are only considering the foreground contributions that are likely to dominate in the squeezed limit in this paper, and these are generally not those expected to dominate in equilateral configurations.  In particular, the tSZ-tSZ-tSZ and CIB-CIB-CIB (and ISW-ISW-ISW) auto-bispectra are the foreground bispectra that peak in equilateral configurations (as well as point source bispectra).  However, these are the foreground bispectra that are most heavily suppressed by the multifrequency component separation algorithms (since they include three ``cleaning'' factors).  Of course, the ISW-ISW-ISW bispectrum is not removed by multifrequency cleaning, but it is only important at low $\ell$ and is unlikely to be a major contaminant to the {\em Planck} analysis.  However, note that we have also not computed the tSZ-kSZ-kSZ and CIB-kSZ-kSZ bispectra (or ISW-ISW-tSZ or ISW-ISW-CIB bispectra), which are the contributions least suppressed by the foreground cleaning.  These foreground terms could have non-negligible equilateral contributions.  We leave a calculation of these signals for future work.

In the remainder of this appendix, we briefly comment on the foreground biases on $f_{\rm NL}^{\rm equ}$ due to the seven bispectra considered in this paper.  Fig.~\ref{fig:ISWxphi_bias_equ} shows the $f_{\rm NL}^{\rm equ}$ bias and foreground-marginalized error bar for the lensing-ISW bispectrum.  It is apparent that the bias is always much less than the statistical error, and furthermore that marginalization over the lensing-ISW amplitude has no effect on $\sigma(f_{\rm NL}^{\rm equ})$.  Fig.~\ref{fig:ISWxphi_bias_equ} also includes an additional lensing-ISW bias prediction (beyond those shown in Fig.~\ref{fig:ISWxphi_bias}), in which non-linear theory is used to compute the lensing-ISW bispectrum, via the halo model.  This calculation is a test as to whether the use of linear theory is sufficient for predicting the $f_{\rm NL}^{\rm equ}$ bias due to the lensing-ISW bispectrum (see Appendix~\ref{app:orth} for an analogous calculation for orthogonal NG).  As noted earlier, Ref.~\cite{Junk-Komatsu2012} performed this check for the lensing-ISW bias on $f_{\rm NL}^{\rm loc}$, but we are not aware of a similar check in the literature for $f_{\rm NL}^{\rm equ}$ (or $f_{\rm NL}^{\rm orth}$).  Fig.~\ref{fig:ISWxphi_bias_equ} shows that the fractional change in the bias prediction is non-negligible at high multipoles, but since the bias itself is small compared to the statistical error bar, this change is nevertheless not important for NG analyses.

\begin{figure}
\includegraphics[width=0.5\textwidth]{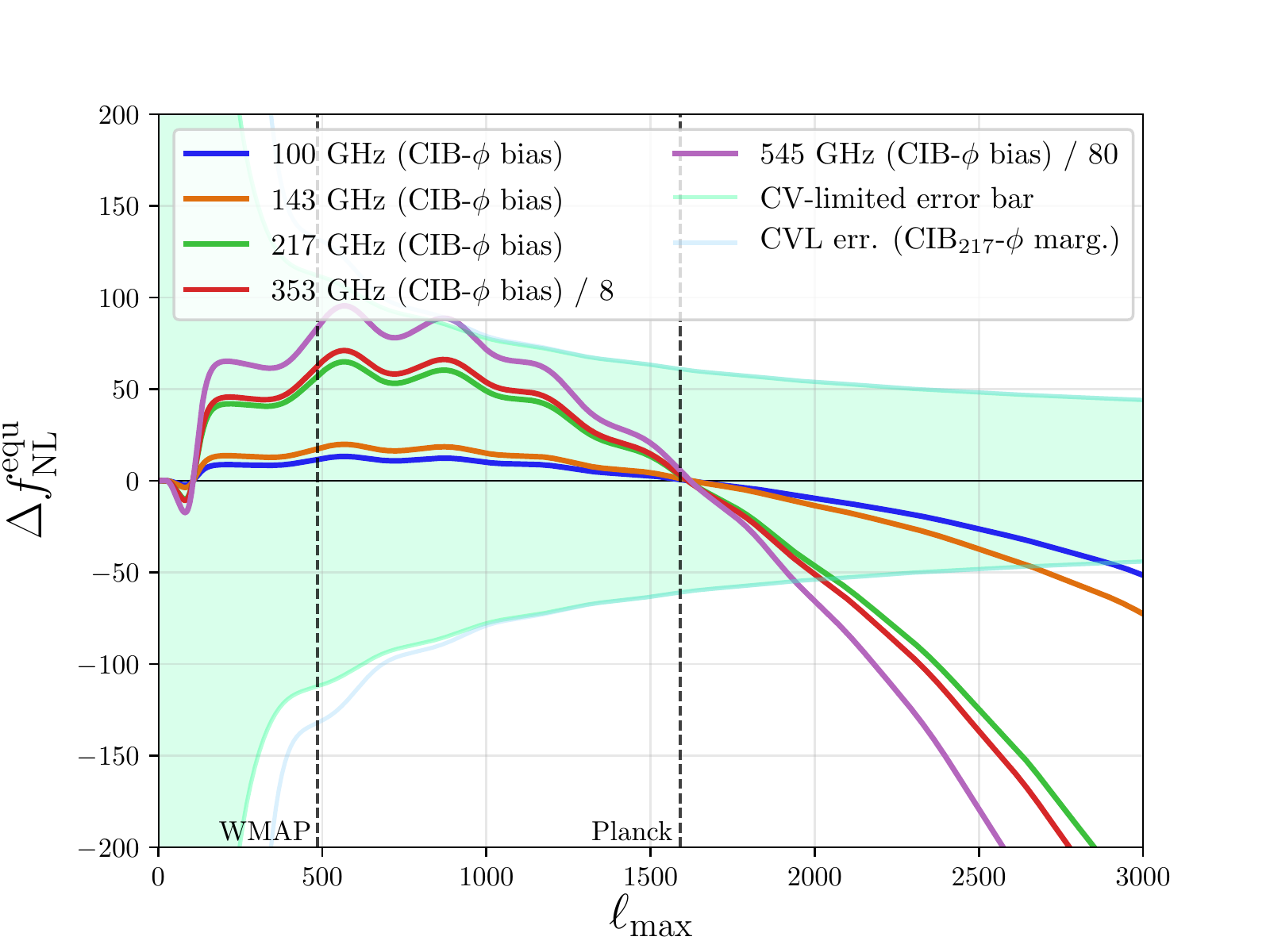} 
\caption{\label{fig:CIBxphi_bias_equ} Bias on $f_{\rm NL}^{\rm equ}$ from the lensing-CIB bispectrum for an experiment that is CV-limited to a maximum multipole $\ell_{\rm max}$, analogous to Fig.~\ref{fig:CIBxphi_bias} for $f_{\rm NL}^{\rm loc}$.  All curves and shaded regions are identical in meaning to those in Fig.~\ref{fig:CIBxphi_bias}.}
\end{figure}

Figs.~\ref{fig:tSZxphi_bias_equ} and~\ref{fig:CIBxphi_bias_equ} show the $f_{\rm NL}^{\rm equ}$ biases and foreground-marginalized error bars for the lensing-tSZ and lensing-CIB bispectra, respectively.  While these biases have a fortuitous zero-crossing in exactly the neighborhood of the {\em Planck} value of $\ell_{\rm max}$, they subsequently become larger than the statistical error bar on $f_{\rm NL}^{\rm equ}$ at for higher-sensitivity experiments.  In particular, the lensing-tSZ bias at 100 or 143 GHz is larger than $\sigma(f_{\rm NL}^{\rm equ})$ for $\ell_{\rm max} \gtrsim 2500$.  The lensing-CIB bias at 217 GHz is larger than $\sigma(f_{\rm NL}^{\rm equ})$ for $\ell_{\rm max} \gtrsim 2000$, and even at 100 GHz, it is larger than $\sigma(f_{\rm NL}^{\rm equ})$ at $\ell_{\rm max}=3000$.  Given that some residual CIB signal will always persist in multifrequency-cleaned CMB maps, the latter bias is perhaps the most concerning for ongoing and future experiments.  However, Figs.~\ref{fig:tSZxphi_bias_equ} and~\ref{fig:CIBxphi_bias_equ} also show that the amplitudes of these foreground bispectra can be marginalized over with little increase in $\sigma(f_{\rm NL}^{\rm equ})$ (except for low-sensitivity experiments).

\begin{figure}
\includegraphics[width=0.5\textwidth]{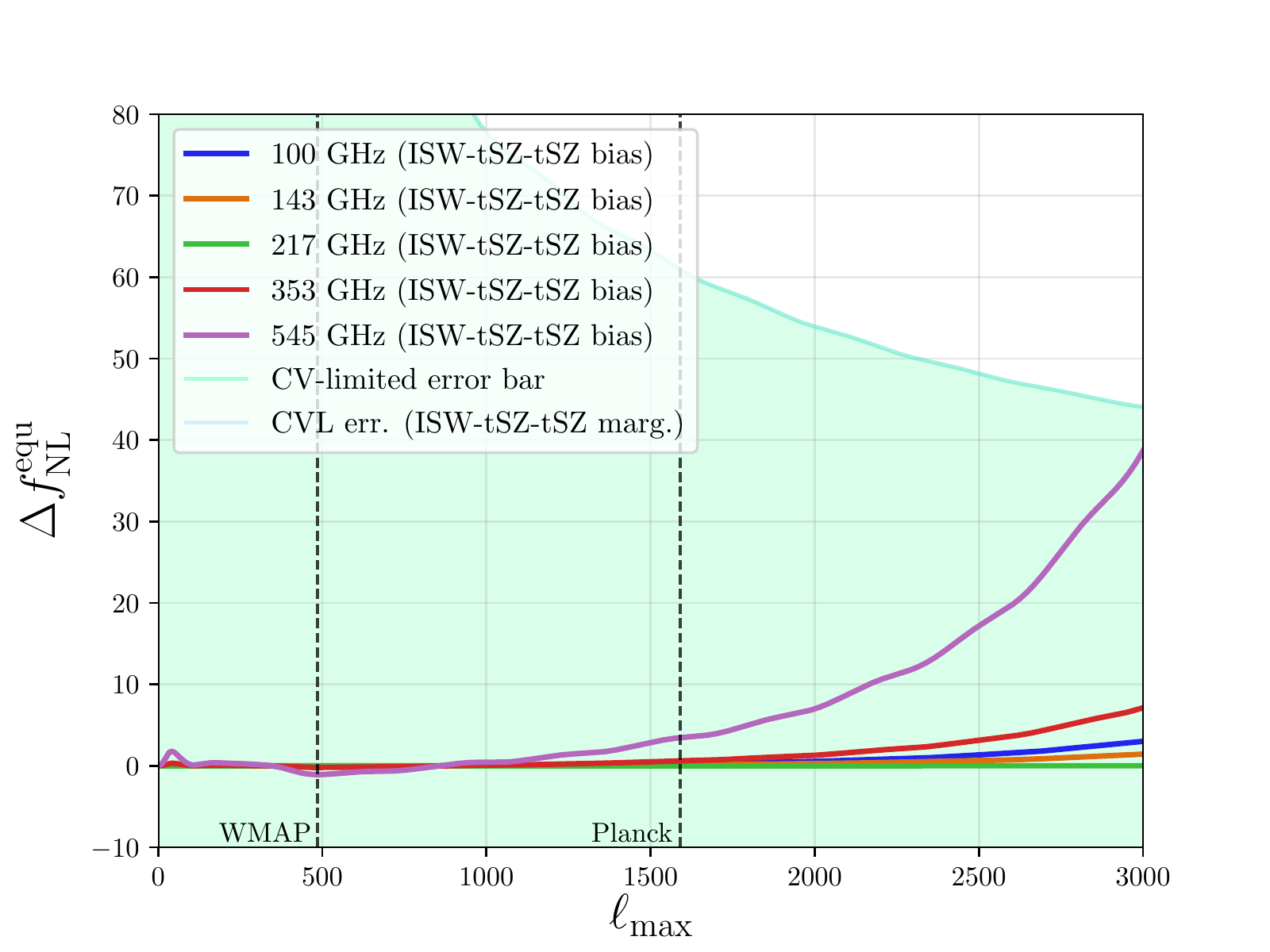} 
\caption{\label{fig:ISWyy_bias_equ} Bias on $f_{\rm NL}^{\rm equ}$ from the ISW-tSZ-tSZ bispectrum for an experiment that is CV-limited to a maximum multipole $\ell_{\rm max}$, analogous to Fig.~\ref{fig:ISWyy_bias} for $f_{\rm NL}^{\rm loc}$.  All curves and shaded regions are identical in meaning to those in Fig.~\ref{fig:ISWyy_bias}.}
\end{figure}

\begin{figure}
\includegraphics[width=0.5\textwidth]{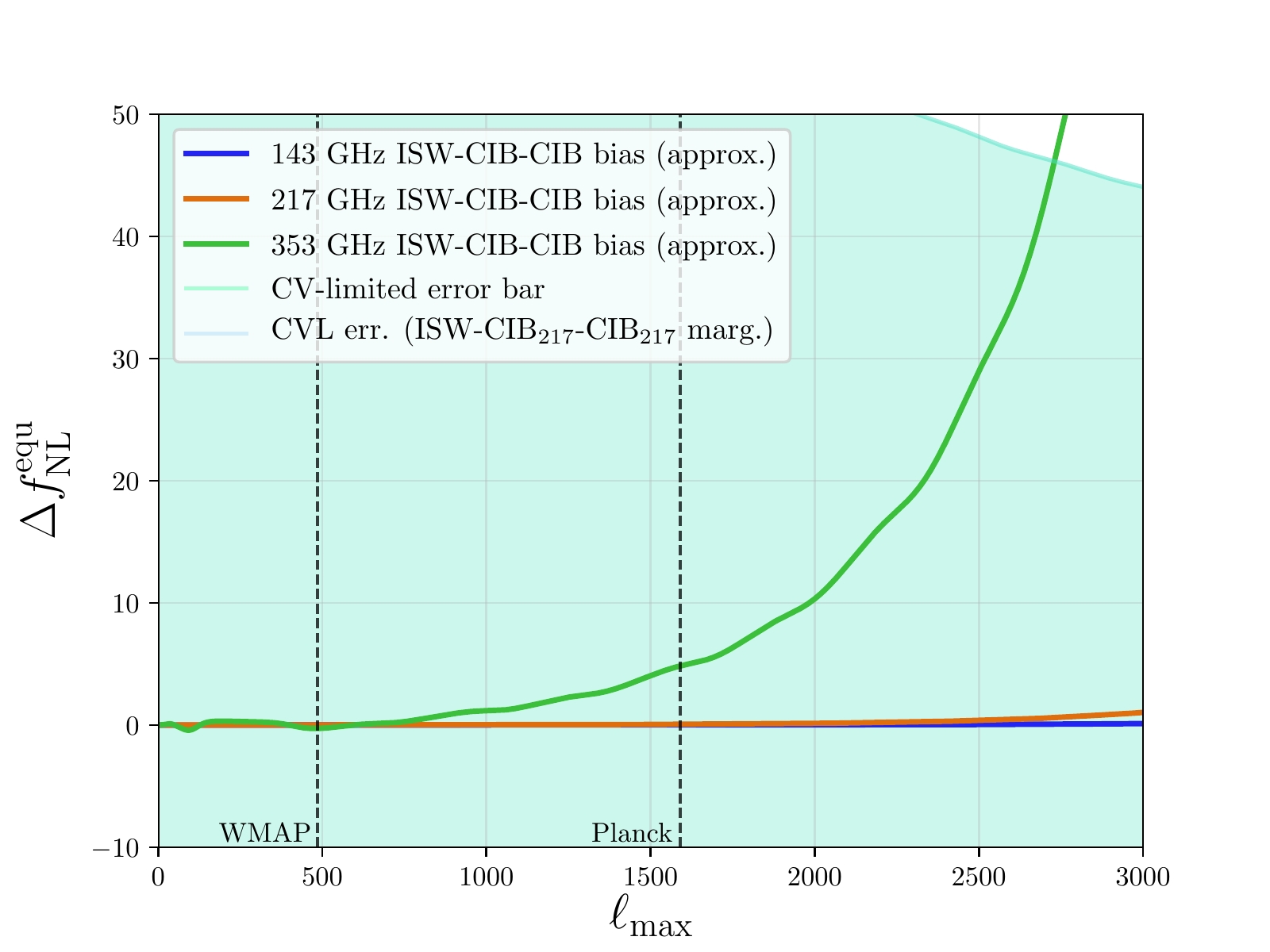} 
\caption{\label{fig:ISWCIBCIB_bias_equ} Bias on $f_{\rm NL}^{\rm equ}$ from the (approximate) ISW-CIB-CIB bispectrum for an experiment that is CV-limited to a maximum multipole $\ell_{\rm max}$, analogous to Fig.~\ref{fig:ISWCIBCIB_bias} for $f_{\rm NL}^{\rm loc}$.  All curves and shaded regions are identical in meaning to those in Fig.~\ref{fig:ISWCIBCIB_bias}.}
\end{figure}

\begin{figure}
\includegraphics[width=0.5\textwidth]{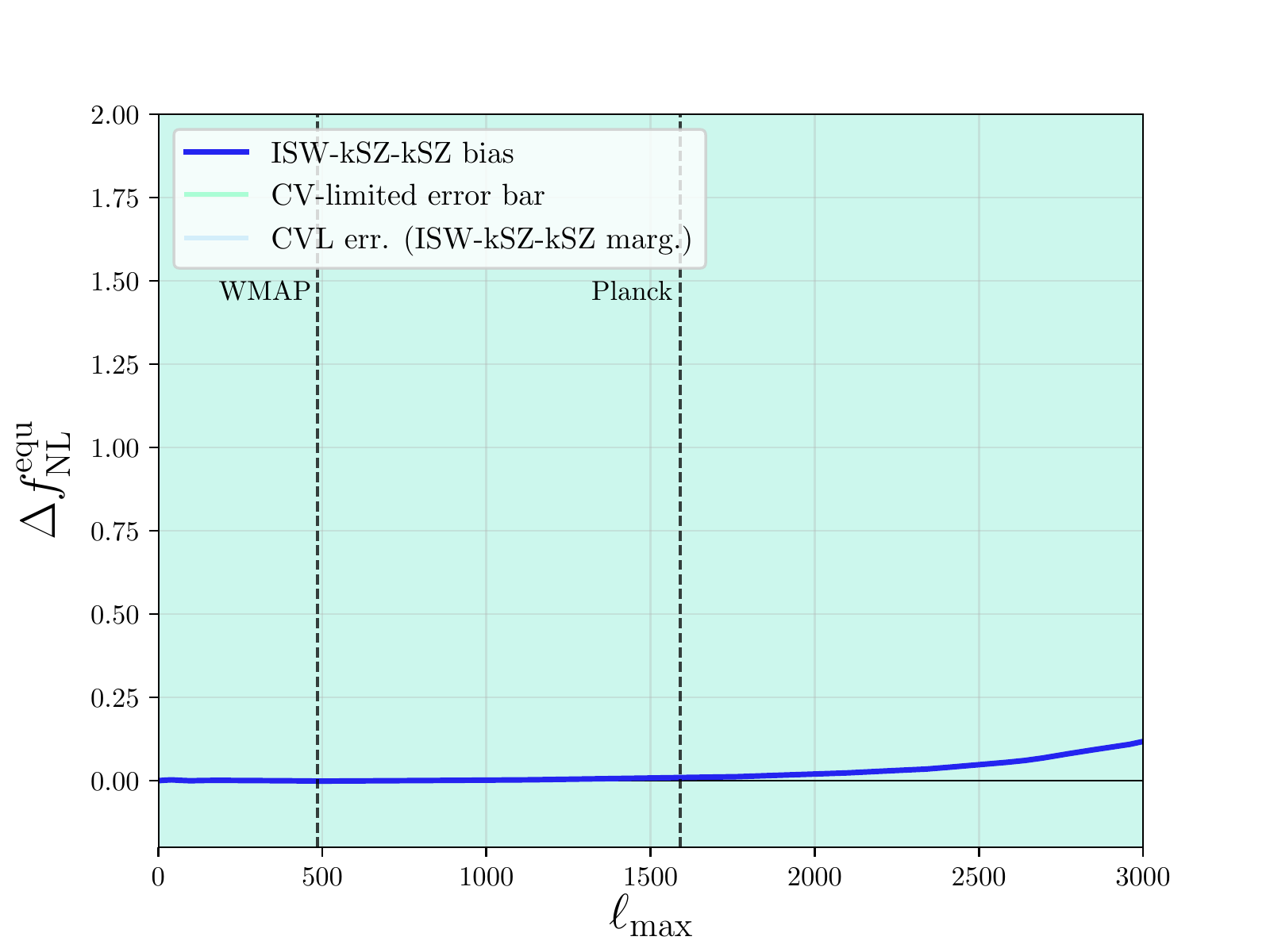} 
\caption{\label{fig:ISWkSZkSZ_bias_equ} Bias on $f_{\rm NL}^{\rm equ}$ from the ISW-kSZ-kSZ bispectrum for an experiment that is CV-limited to a maximum multipole $\ell_{\rm max}$, analogous to Fig.~\ref{fig:ISWkSZkSZ_bias} for $f_{\rm NL}^{\rm loc}$.  All curves and shaded regions are identical in meaning to those in Fig.~\ref{fig:ISWkSZkSZ_bias}.}
\end{figure}

Figs.~\ref{fig:ISWyy_bias_equ},~\ref{fig:ISWCIBCIB_bias_equ}, and~\ref{fig:ISWkSZkSZ_bias_equ} show the $f_{\rm NL}^{\rm equ}$ biases and foreground-marginalized error bars for the ISW-tSZ-tSZ, ISW-CIB-CIB, and ISW-kSZ-kSZ bispectra, respectively.  We do not include a plot for the ISW-tSZ-CIB bispectrum, as the biases and effects of marginalization in this case are even smaller than those shown in these figures (e.g., for the CIB at 353 GHz and tSZ signal at 100 GHz, the ISW-tSZ-CIB bias is $\Delta f_{\rm NL}^{\rm equ} = -0.05$ for the {\em Planck} value of $\ell_{\rm max}$).  For all of these bispectra, the associated biases on $f_{\rm NL}^{\rm equ}$ are far smaller than the statistical error bar, except when considering foreground-dominated channels at high sensitivity (e.g., 545 GHz at $\ell_{\rm max}=3000$).  In addition, the amplitudes of these bispectra can be marginalized over with no increase in $\sigma(f_{\rm NL}^{\rm equ}$).  These results are expected due to the fact that none of these bispectra peak in equilateral configurations.  As mentioned earlier, other bispectra that are not considered here will likely lead to higher levels of bias for $f_{\rm NL}^{\rm equ}$ (e.g., the tSZ-tSZ-tSZ or tSZ-kSZ-kSZ bispectra).

We conclude that the seven foreground bispectra considered in this paper do not present serious problems for the {\em Planck} analysis of equilateral NG.  However, the lensing-tSZ and lensing-CIB bispectra could be an issue for temperature-based $f_{\rm NL}^{\rm equ}$ constraints from high-sensitivity experiments.  Finally, we emphasize that a complete calculation including the other foreground terms not considered here is necessary before a fully robust conclusion can be reached.

\section{Foreground Biases on Orthogonal-Type Primordial non-Gaussianity}
\label{app:orth}

In this appendix, we provide foreground bias results for orthogonal-type primordial NG analogous to those given in the main text for local-type NG and in Appendix~\ref{app:equ} for equilateral-type NG.  Orthogonal NG was identified in the context of the effective field theory of inflation as an additional shape that is naturally generated by operators in the Lagrangian, but which is effectively orthogonal to the local and equilateral shapes~\cite{Senatoreetal2010}.  The orthogonal bispectrum signal peaks in both equilateral and ``flattened'' triangle configurations in momentum space (but with opposite signs), where the latter refers to triangles where the two shortest sides are exactly half of the longest side.  As for local and equilateral NG, the {\em Planck} 2015 CMB anisotropy data yield the tightest current constraint on the amplitude of orthogonal NG, $f_{\rm NL}^{\rm orth}$: $f_{\rm NL}^{\rm orth} = -34 \pm 33$ (temperature data only) or $f_{\rm NL}^{\rm orth} = -26 \pm 21$ (temperature and polarization data)~\cite{Planck2015NG}.  As for equilateral NG, the polarization data (particularly mixed bispectra of temperature and $E$-modes) are somewhat more constraining here than in the local NG analysis, yielding a comparable constraint to that derived from temperature alone.  In the following, we compute foreground biases (and foreground-marginalized error bars) for $f_{\rm NL}^{\rm orth}$ inferred from the temperature bispectrum alone.

Here, we repeat all of the calculations presented in \S\ref{sec:lensing} and~\S\ref{sec:ISW} (and in Appendix~\ref{app:equ}), but with the replacement $B^{\rm loc} \rightarrow B^{\rm orth}$, i.e., the local bispectrum is replaced by the orthogonal bispectrum.  We calculate $B^{\rm orth}$ following Sec.~4 of Ref.~\cite{Senatoreetal2010}, including their guidance on the upper limit in the integral over comoving distance (as mentioned in Appendix~\ref{app:equ}).

We use the formalism of \S\ref{sec:fNL} to compute foreground biases and foreground-marginalized error bars on $f_{\rm NL}^{\rm orth}$ using the models described in \S\ref{sec:lensing} and \S\ref{sec:ISW}.  As for $f_{\rm NL}^{\rm loc}$ (but unlike $f_{\rm NL}^{\rm equ}$), we find foreground biases that are potentially concerning for current ({\em Planck}) analyses and clearly an issue for future measurements (considering CMB temperature only).  Furthermore, we again emphasize that we are only considering the foreground contributions that are likely to dominate in the squeezed limit in this paper, which are generally not those expected to dominate in configurations relevant to orthogonal NG.  The most concerning contributions that we have not computed are those due to the tSZ-kSZ-kSZ and CIB-kSZ-kSZ bispectra (or possibly the ISW-ISW-tSZ or ISW-ISW-CIB bispectra), which are the contributions least suppressed by multifrequency foreground cleaning.  These foreground terms could have non-negligible orthogonal-type contributions.  We defer a calculation of these signals to future work.

\begin{figure}
\includegraphics[width=0.5\textwidth]{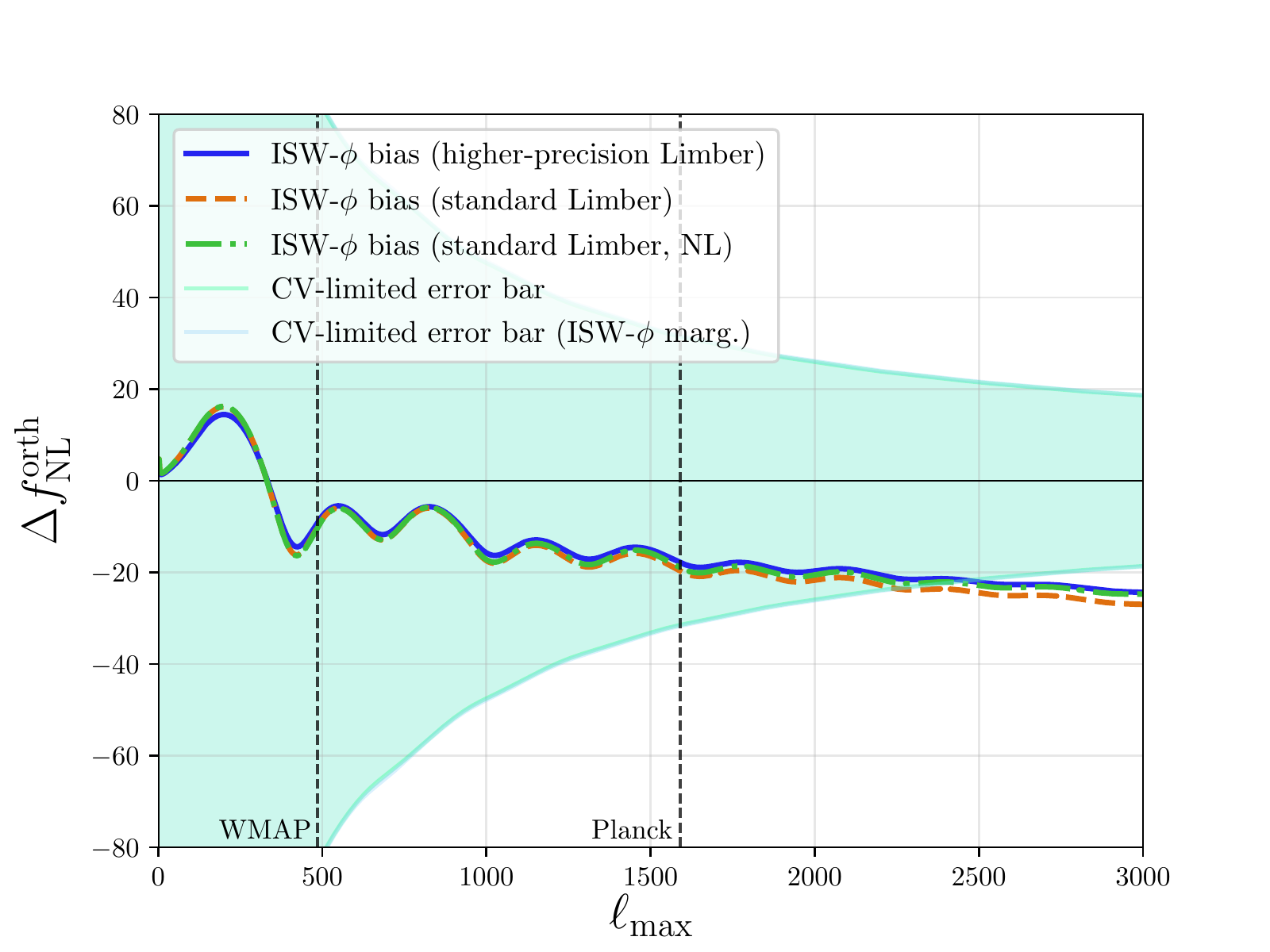} 
\caption{\label{fig:ISWxphi_bias_orth} Bias on $f_{\rm NL}^{\rm orth}$ from the lensing-ISW bispectrum for an experiment that is CV-limited to a maximum multipole $\ell_{\rm max}$, analogous to Fig.~\ref{fig:ISWxphi_bias} for $f_{\rm NL}^{\rm loc}$ and Fig.~\ref{fig:ISWxphi_bias_equ} for $f_{\rm NL}^{\rm equ}$.  All curves and shaded regions are identical in meaning to those in Fig.~\ref{fig:ISWxphi_bias_equ}.}
\end{figure}

In the remainder of this appendix, we briefly comment on the foreground biases on $f_{\rm NL}^{\rm orth}$ due to the seven bispectra considered in this paper.  Fig.~\ref{fig:ISWxphi_bias_orth} shows the $f_{\rm NL}^{\rm orth}$ bias and foreground-marginalized error bar for the lensing-ISW bispectrum.  In agreement with the {\em Planck} 2015 NG analysis~\cite{Planck2015NG}, we find that the lensing-ISW bias is non-negligible.  For {\em Planck}, the bias is $\approx 2/3$ of the statistical error bar on $f_{\rm NL}^{\rm orth}$ and cannot be neglected; for future experiments with $\ell_{\rm max} = 3000$, the bias is $\approx 1.5\sigma(f_{\rm NL}^{\rm orth})$.  In addition, the exact value depends on the accuracy of the Limber approximation used (compare the solid blue and dashed orange curves in Fig.~\ref{fig:ISWxphi_bias_orth}, corresponding to the use of $k=(\ell+1/2)/\chi$ or $k=\ell/\chi$ in the Limber approximation, respectively), as well as on the use of linear or non-linear theory in the calculation (see the dash-dotted green curve in the figure).  As for $f_{\rm NL}^{\rm equ}$ in Fig.~\ref{fig:ISWxphi_bias_equ}, the non-linear theory calculation is a test as to whether the use of linear theory is sufficient for predicting the $f_{\rm NL}^{\rm orth}$ bias due to the lensing-ISW bispectrum.  For $f_{\rm NL}^{\rm loc}$, linear theory is known to suffice~\cite{Junk-Komatsu2012}, but for orthogonal NG, Fig.~\ref{fig:ISWxphi_bias_orth} indicates that non-linear theory should be used, although the fractional correction is relatively small.  Finally, Fig.~\ref{fig:ISWxphi_bias_orth} shows that marginalization over the lensing-ISW bispectrum amplitude has no effect on $\sigma(f_{\rm NL}^{\rm orth})$, indicating that the correlation coefficient between these bispectra is not large.

\begin{figure}
\includegraphics[width=0.5\textwidth]{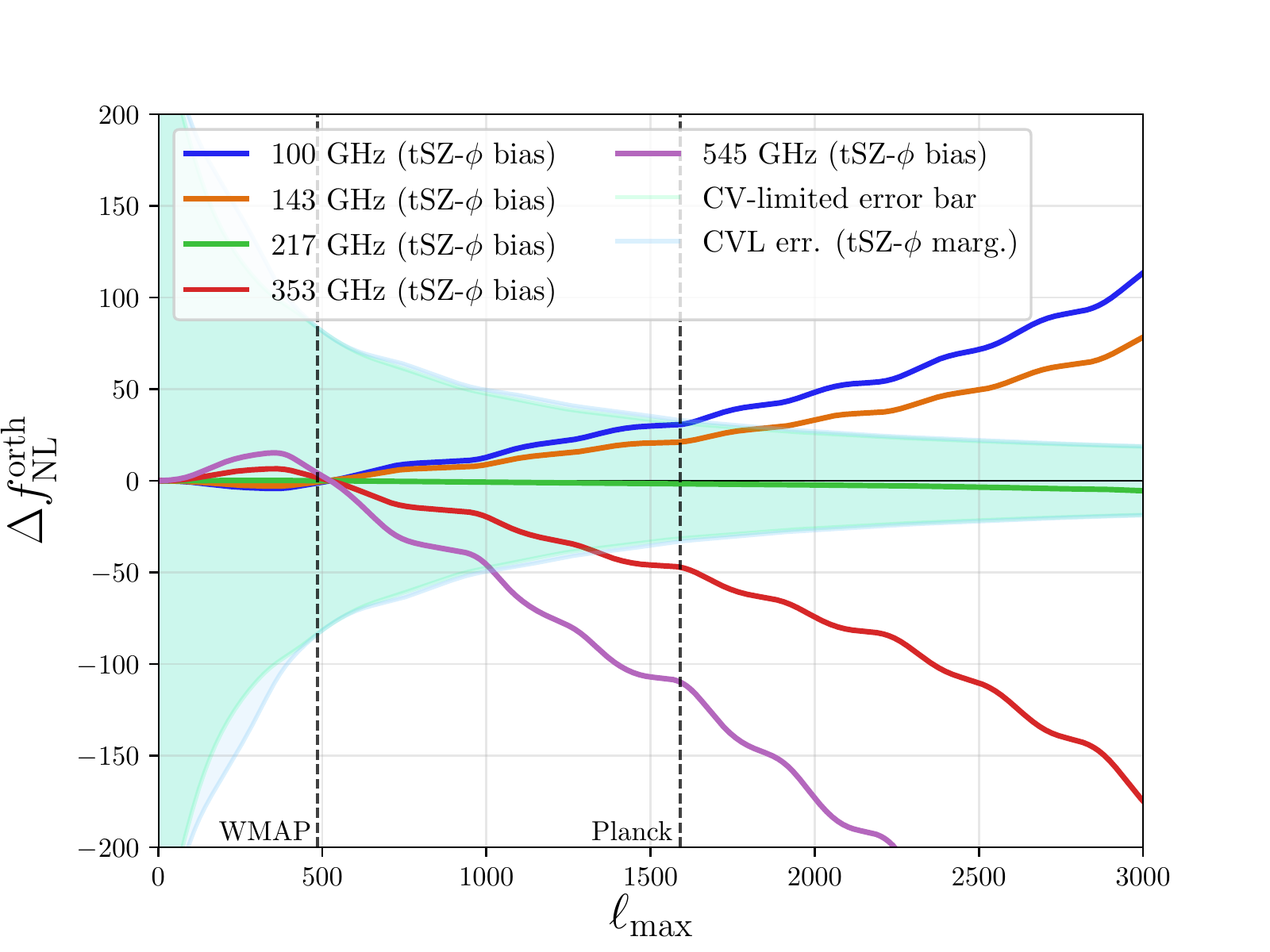} 
\caption{\label{fig:tSZxphi_bias_orth} Bias on $f_{\rm NL}^{\rm orth}$ from the lensing-tSZ bispectrum for an experiment that is CV-limited to a maximum multipole $\ell_{\rm max}$, analogous to Fig.~\ref{fig:tSZxphi_bias} for $f_{\rm NL}^{\rm loc}$ and Fig.~\ref{fig:tSZxphi_bias_equ} for $f_{\rm NL}^{\rm equ}$.  All curves and shaded regions are identical in meaning to those in Fig.~\ref{fig:tSZxphi_bias}.}
\end{figure}

\begin{figure}
\includegraphics[width=0.5\textwidth]{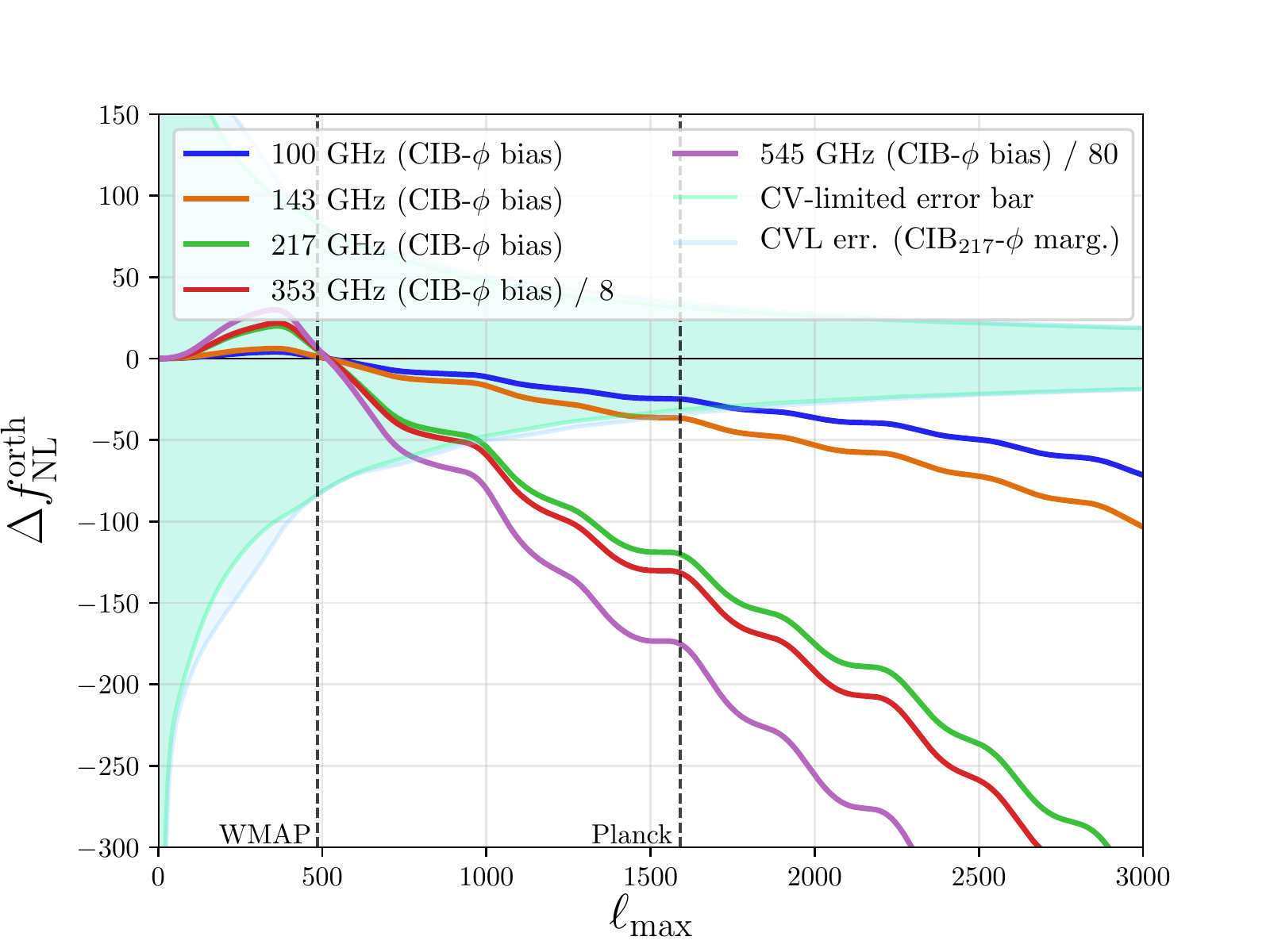} 
\caption{\label{fig:CIBxphi_bias_orth} Bias on $f_{\rm NL}^{\rm orth}$ from the lensing-CIB bispectrum for an experiment that is CV-limited to a maximum multipole $\ell_{\rm max}$, analogous to Fig.~\ref{fig:CIBxphi_bias} for $f_{\rm NL}^{\rm loc}$ and Fig.~\ref{fig:CIBxphi_bias_equ} for $f_{\rm NL}^{\rm loc}$.  All curves and shaded regions are identical in meaning to those in Fig.~\ref{fig:CIBxphi_bias}.}
\end{figure}

Figs.~\ref{fig:tSZxphi_bias_orth} and~\ref{fig:CIBxphi_bias_orth} show the $f_{\rm NL}^{\rm orth}$ biases and foreground-marginalized error bars for the lensing-tSZ and lensing-CIB bispectra, respectively.  In both cases, we find that the bias on $f_{\rm NL}^{\rm orth}$ can be substantial, even for {\em Planck}.  For the lensing-tSZ bispectrum, Fig.~\ref{fig:tSZxphi_bias_orth} shows that the bias on $f_{\rm NL}^{\rm orth}$ at 100 GHz is comparable to the $1\sigma$ error bar for {\em Planck}, and is nearly this large at 143 GHz.  For higher values of $\ell_{\rm max}$, the bias is significantly larger than $\sigma(f_{\rm NL}^{\rm orth})$.  These results strongly motivate the use of tSZ-nulled CMB maps in orthogonal NG analyses.  However, Fig.~\ref{fig:tSZxphi_bias_orth} also shows that the lensing-tSZ bispectrum amplitude can be marginalized over with effectively no increase in $\sigma(f_{\rm NL}^{\rm orth})$ (except for experiments with very low $\ell_{\rm max}$).

Fig.~\ref{fig:CIBxphi_bias_orth} shows that the bias on $f_{\rm NL}^{\rm orth}$ due to the lensing-CIB bispectrum is also significant.  For {\em Planck}, the bias is larger than the $1\sigma$ error bar at all HFI frequencies, except for 100 GHz, although it is still non-negligible at this frequency.  At 217 GHz, the bias is roughly four times larger than the {\em Planck} error bar on $f_{\rm NL}^{\rm orth}$.  Even for a relatively low level of CIB leakage into the component-separated CMB map used in the {\em Planck} NG analysis, this bias could be quite important.  For future experiments with $\ell_{\rm max}=3000$, the bias is many times larger than the statistical error bar, and will necessitate very accurate CIB cleaning.  Alternatively, as seen in Fig.~\ref{fig:CIBxphi_bias_orth} (considering the CIB at 217 GHz as an example), the lensing-CIB bispectrum amplitude can be marginalized over with little penalty on $\sigma(f_{\rm NL}^{\rm orth})$.  To guarantee robustness, marginalizing over such a template in NG analyses is likely to be advantageous for ongoing and upcoming CMB experiments.  Finally, we speculate that a combination of the lensing-tSZ and lensing-CIB biases seen in Figs.~\ref{fig:tSZxphi_bias_orth} and~\ref{fig:CIBxphi_bias_orth} could be responsible for the weak ($\approx 1\sigma$) preference for negative $f_{\rm NL}^{\rm orth}$ in the {\em Planck} 2015 NG analysis, due to the amplitude and sign of the results presented here.

\begin{figure}
\includegraphics[width=0.5\textwidth]{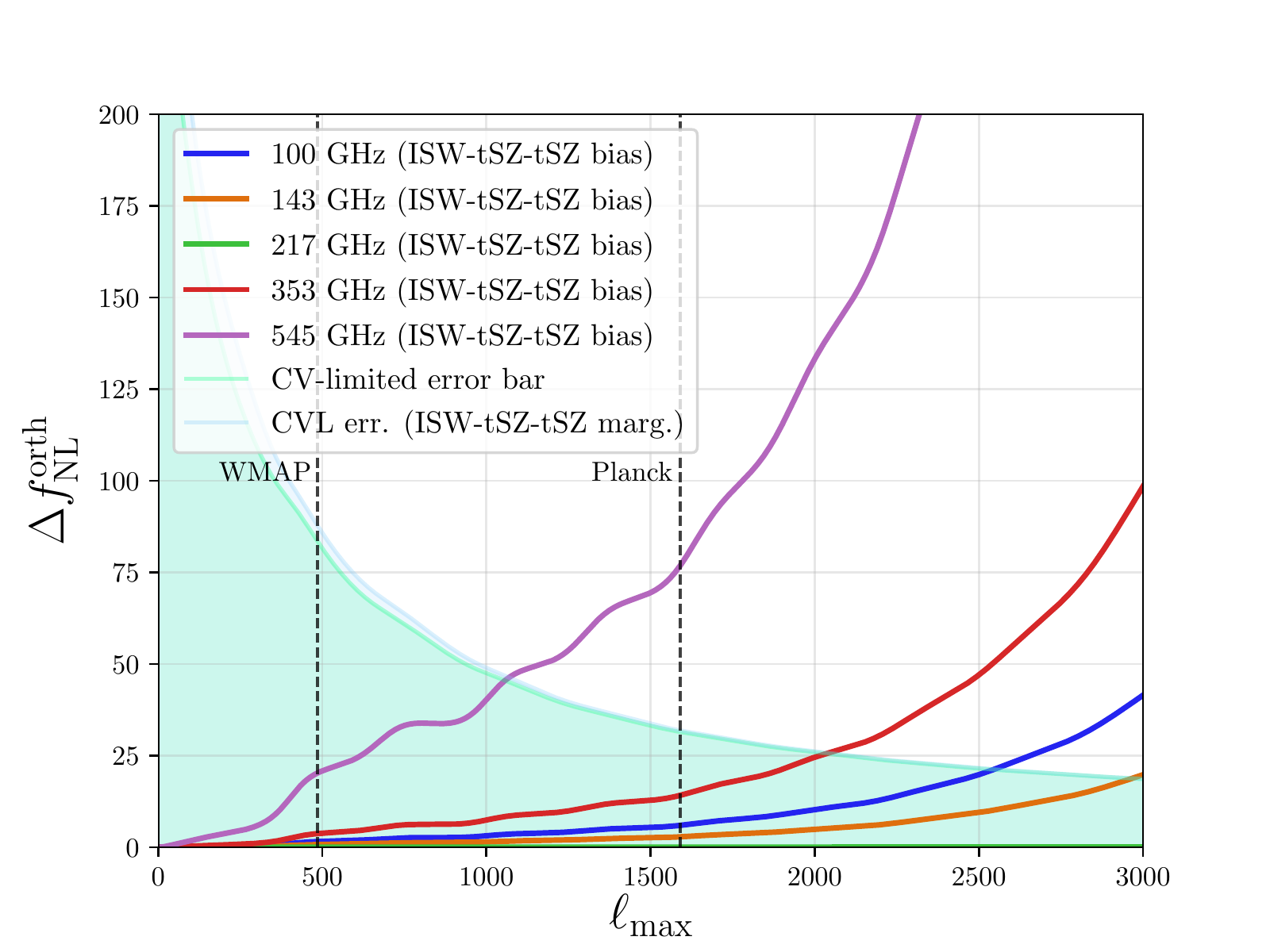} 
\caption{\label{fig:ISWyy_bias_orth} Bias on $f_{\rm NL}^{\rm orth}$ from the ISW-tSZ-tSZ bispectrum for an experiment that is CV-limited to a maximum multipole $\ell_{\rm max}$, analogous to Fig.~\ref{fig:ISWyy_bias} for $f_{\rm NL}^{\rm loc}$ and Fig.~\ref{fig:ISWyy_bias_equ} for $f_{\rm NL}^{\rm equ}$.  All curves and shaded regions are identical in meaning to those in Fig.~\ref{fig:ISWyy_bias}.}
\end{figure}

\begin{figure}
\includegraphics[width=0.5\textwidth]{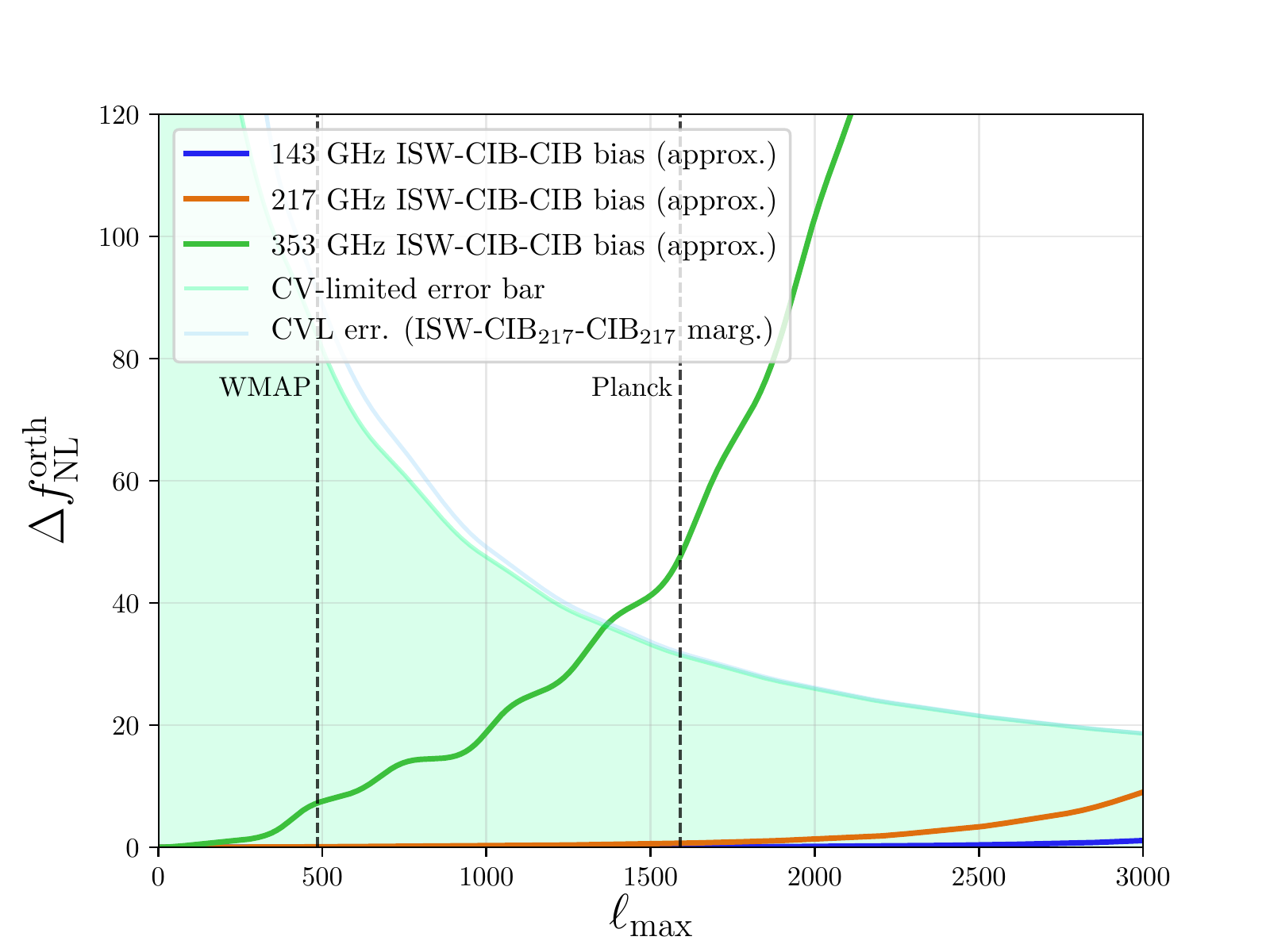} 
\caption{\label{fig:ISWCIBCIB_bias_orth} Bias on $f_{\rm NL}^{\rm orth}$ from the (approximate) ISW-CIB-CIB bispectrum for an experiment that is CV-limited to a maximum multipole $\ell_{\rm max}$, analogous to Fig.~\ref{fig:ISWCIBCIB_bias} for $f_{\rm NL}^{\rm loc}$ and Fig.~\ref{fig:ISWCIBCIB_bias_equ} for $f_{\rm NL}^{\rm equ}$.  All curves and shaded regions are identical in meaning to those in Fig.~\ref{fig:ISWCIBCIB_bias}.}
\end{figure}

\begin{figure}
\includegraphics[width=0.5\textwidth]{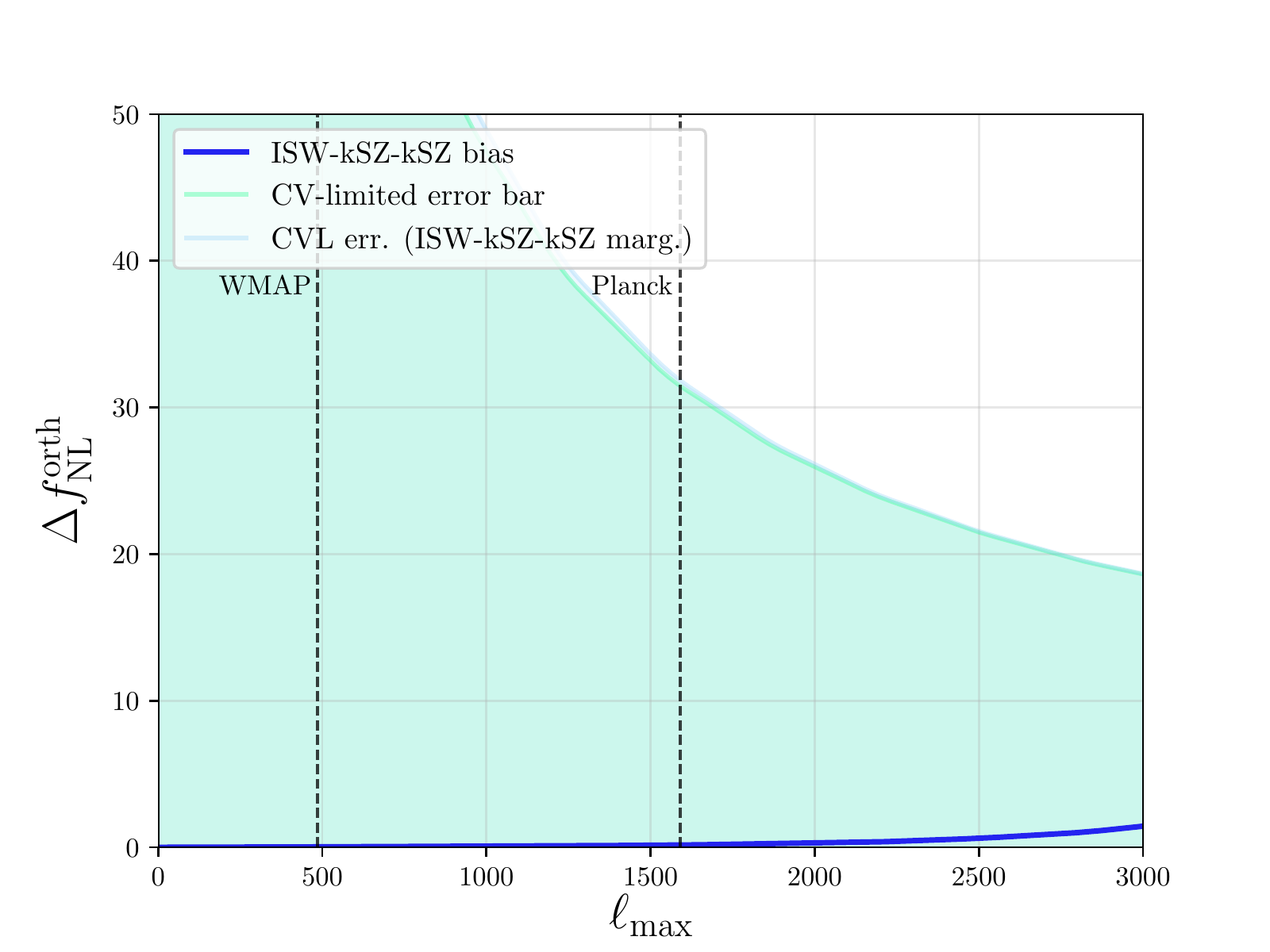} 
\caption{\label{fig:ISWkSZkSZ_bias_orth} Bias on $f_{\rm NL}^{\rm orth}$ from the ISW-kSZ-kSZ bispectrum for an experiment that is CV-limited to a maximum multipole $\ell_{\rm max}$, analogous to Fig.~\ref{fig:ISWkSZkSZ_bias} for $f_{\rm NL}^{\rm loc}$ and Fig.~\ref{fig:ISWkSZkSZ_bias_equ} for $f_{\rm NL}^{\rm equ}$.  All curves and shaded regions are identical in meaning to those in Fig.~\ref{fig:ISWkSZkSZ_bias}.}
\end{figure}

Figs.~\ref{fig:ISWyy_bias_orth},~\ref{fig:ISWCIBCIB_bias_orth}, and~\ref{fig:ISWkSZkSZ_bias_orth} show the $f_{\rm NL}^{\rm orth}$ biases and foreground-marginalized error bars for the ISW-tSZ-tSZ, ISW-CIB-CIB, and ISW-kSZ-kSZ bispectra, respectively.  We do not include a plot for the ISW-tSZ-CIB bispectrum, as the biases and effects of marginalization in this case are extremely small (e.g., for the CIB at 353 GHz and tSZ signal at 100 GHz, the ISW-tSZ-CIB bias is $\Delta f_{\rm NL}^{\rm orth} = -0.45$ for the {\em Planck} value of $\ell_{\rm max}$).  Of the ISW-related contributions, only the ISW-tSZ-tSZ bispectrum appears to present a serious concern.  For {\em Planck}, the ISW-tSZ-tSZ bias is much smaller than $\sigma(f_{\rm NL}^{\rm orth})$ (except for the foreground-dominated 545 GHz channel); however, for future experiments with $\ell_{\rm max}=3000$, the bias is comparable to or larger the statistical error bar, even at 143 GHz.  Thus, similar to the lensing-tSZ bias, this result motivates the use of tSZ-nulled CMB maps in future $f_{\rm NL}^{\rm orth}$ analyses.  Even for {\em Planck}, this is likely to be worthwhile simply for the purpose of robustness, since the statistical penalty for explicitly removing the tSZ signal is not very large with {\em Planck} (see the discussion in \S\ref{sec:discussion}).  Alternatively, Fig.~\ref{fig:ISWyy_bias_orth} also shows that the ISW-tSZ-tSZ bispectrum amplitude can be marginalized over with no increase in the error bar on $f_{\rm NL}^{\rm orth}$, presenting a mitigation option even for single-frequency measurements.

Figs.~\ref{fig:ISWCIBCIB_bias_orth} and~\ref{fig:ISWkSZkSZ_bias_orth} show that the ISW-CIB-CIB and ISW-kSZ-kSZ bispectra generally yield small biases on $f_{\rm NL}^{\rm orth}$ (except for the ISW-CIB-CIB bispectrum at 353 GHz).  For a future experiment with $\ell_{\rm max}=3000$, the ISW-CIB-CIB bias at 217 GHz is $\approx 0.5\sigma$.  However, in both the ISW-CIB-CIB and ISW-kSZ-kSZ cases, the foreground bispectrum amplitudes can be marginalized over with no penalty in the statistical error on $f_{\rm NL}^{\rm orth}$.  Overall, we conclude that the ISW-related bispectra considered here are generally not a major problem for $f_{\rm NL}^{\rm orth}$ analyses from the CMB temperature bispectrum.  However, as mentioned earlier, other bispectra that are not considered here may lead to higher levels of bias for $f_{\rm NL}^{\rm orth}$ (e.g., the tSZ-kSZ-kSZ or CIB-kSZ-kSZ bispectra).

We conclude that the lensing-related bispectra considered in this paper yield potentially serious biases on $f_{\rm NL}^{\rm orth}$, even at the {\em Planck} sensitivity level.  Orthogonal NG analyses using the CMB temperature bispectrum should utilize tSZ- and CIB-nulled maps, while also accounting for the blackbody lensing-ISW bias (this bias has been included in the {\em Planck} 2015 NG analysis).  In contrast, we find that the ISW-related bispectra do not generally yield significant biases on $f_{\rm NL}^{\rm orth}$.  In all cases, the amplitudes of the foreground bispectra can be marginalized over with little increase in the error bar on $f_{\rm NL}^{\rm orth}$.  Finally, we emphasize that a complete calculation including the other foreground terms not considered here is necessary before a fully robust conclusion can be reached regarding foregrounds in orthogonal NG analyses.

\end{appendix}



\end{document}